\documentclass[twoside,11pt,preprint]{article}
\usepackage[abbrvbib]{jmlr2e}

\usepackage{bbm}
\usepackage[utf8]{inputenc}

\newtheorem{assumption}[theorem]{Assumption}
\hypersetup{hidelinks}
\usepackage{enumitem}
\usepackage[T1]{fontenc}
\usepackage{rotating}

\usepackage{url}
\usepackage{booktabs}
\usepackage{amsfonts}
\usepackage{nicefrac}
\usepackage{microtype}
\usepackage{xcolor}
\usepackage{graphicx}
\graphicspath{ {./figures/} }
\usepackage{array}
\usepackage{amsmath}
\usepackage[normalem]{ulem}
\usepackage{dirtytalk}
\usepackage{csquotes}
\usepackage{makecell}
\usepackage{float}
\restylefloat{table}
\usepackage{mathrsfs}
\usepackage{algorithm}
\usepackage{algpseudocode}

\usepackage{cleveref}

\newcommand{\R}{\mathbb{R}}

\newcommand{\f}{\mathscr{F}}
\newcommand{\g}{\mathscr{G}}

\algnewcommand{\NOSTATE}{\State\unskip}
\newcommand{\bs}{\boldsymbol}

\usepackage{lastpage}
\jmlrheading{27}{2026}{1-\pageref{LastPage}}{9/25; Revised 6/26}{6/26}{25-2146}{Guglielmo Gattiglio, Lyudmila Grigoryeva, and Massimiliano Tamborrino}

\ShortHeadings{Prob-GParareal: A Probabilistic PinT Solver}
{Gattiglio, Grigoryeva, and Tamborrino}
\firstpageno{1}

\begin{document}

\title{Prob-GParareal: A Probabilistic Numerical Parallel-in-Time Solver for Differential Equations}

\author{\name Guglielmo Gattiglio \email g.gattiglio@gmail.com \\
       \addr Department of Statistics\\
       University of Warwick\\
       Coventry, CV4 7AL, UK
       \AND
       \name Lyudmila Grigoryeva \email Lyudmila.Grigoryeva@unisg.ch \\
       \addr  Division of Mathematics and Statistics\\
       University of St.~Gallen\\
       St.~Gallen, CH-9000,~Switzerland
       \AND
       \name Massimiliano Tamborrino \email Massimiliano.Tamborrino@warwick.ac.uk \\
       \addr Department of Statistics\\
       University of Warwick\\
       Coventry, CV4 7AL, UK}

\editor{Samuel Kaski}

\maketitle

\begin{abstract}
We introduce Prob-GParareal, a probabilistic extension of the GParareal algorithm designed to provide uncertainty quantification for the Parallel-in-Time (PinT) solution of (ordinary and partial) differential equations (ODEs, PDEs). The method employs Gaussian processes (GPs) to model the Parareal correction function, in line with GParareal, further enabling the propagation of numerical uncertainty across time and yielding probabilistic forecasts of the system's evolution. Furthermore, Prob-GParareal accommodates probabilistic initial conditions and maintains compatibility with classical numerical solvers,  ensuring its straightforward integration into existing Parareal frameworks. Here, we first conduct a theoretical analysis of the computational complexity and derive error bounds of Prob-GParareal. Then, we numerically demonstrate the accuracy and robustness of the proposed algorithm on five benchmark ODE systems, including chaotic, stiff, and bifurcation problems.  To showcase the flexibility and potential scalability of the proposed algorithm, we also consider Prob-nnGParareal, a variant obtained by replacing the GPs in Parareal with the nearest-neighbors GPs, illustrating its improved computational performance on an additional PDE example.
This work bridges a critical gap in the development of probabilistic counterparts to established PinT methods.
\end{abstract}


\section{Introduction}
\label{sec:intro}

Efficient numerical methods for solving differential equations (DEs) are a cornerstone of modern simulation and modeling techniques, with applications ranging from climate, medical, and financial modeling to aerospace engineering and many other fields where understanding dynamic processes is crucial. Among many advances in this field, Parallel-in-Time (PinT) methods have emerged as a powerful approach to accelerate the solution of large-scale, high-dimensional DEs where traditional parallelization strategies, such as spatial decomposition, reach saturation, preventing full use of available computational resources~\citep{samaddar2019application}. PinT techniques address the limitations of conventional sequential solvers by enabling concurrent computations over the time domain, which is especially useful when applied to problems with a long simulation horizon, such as, for example, in the case of molecular dynamics simulations~\citep{gorynina2022combining}.

Over the past two decades, the PinT field has grown substantially, with several algorithmic classes introduced. Following~\cite{gander201550}, these can be categorized according to how they discretize the space-time domain to implement parallelization: multiple shooting, space-time domain decomposition, multigrid methods, and direct solvers. Notable examples include Parareal~\citep{lions2001resolution}, the Parallel Full Approximation Scheme in Space and Time~\citep{emmett2012toward, minion2011hybrid}, and Multigrid Reduction in Time~\citep{Falgout2014, Friedhoff2012}, see~\cite{gander2023unified} for an overview. Among these approaches, Parareal has received the most attention due to its simplicity, flexibility, and demonstrated success in a wide range of applications~\citep{ reynolds2012mechanisms,samaddar2010parallelization,samaddar2019application,bal2002parareal,pages2018parareal,philippi2022parareal,philippi2023micro}. Extensive work on theoretical analysis~\citep{bal2005convergence,gander2007analysis,gander2008nonlinear,ruprecht2018wave,staff2005stability,pentland2023error}, and algorithmic extensions~\citep{haut2014asymptotic,peddle2019parareal,legoll2013micro,pentland2022stochastic,pentland2023gparareal,nngparareal,randnet_parareal} have further established Parareal as a foundational method in PinT research. This algorithm uses a computationally cheap coarse numerical solver to obtain an approximate sequential solution of the DE system. The intermediate solution is iteratively refined by executing a precise, although expensive, numerical solver in parallel, and correcting the coarse evaluation by accounting for the estimated error between these two solvers.

One of the recent PinT advances, the so-called GParareal, introduced in~\cite{pentland2023gparareal}, is especially relevant for this work. By using Gaussian processes (GPs) to learn the solver's discrepancy from Parareal's past iteration data, GParareal approximates the Parareal correction function using the GP posterior mean, yielding substantial speed-ups over the canonical Parareal algorithm. This approach was later extended to utilize nearest-neighbor GPs (nnGPs) in~\cite{nngparareal}, resulting in the nnGParareal algorithm, offering improved scalability for higher-dimensional systems and an increased number of discretization points in the time domain. High-quality and numerically efficient approximation of the Parareal correction function has been further pursued using other families of models with universal approximation properties. For example, RandNet-Parareal, proposed in~\cite{randnet_parareal}, uses shallow random weights neural networks to learn the solvers' discrepancy, allowing its application to partial differential equations (PDEs),  accommodating up to $10^5$ spatial discretization points. It is worth noting that these algorithms yield deterministic solvers. However, while the RandNet-Parareal method is deterministic by construction, GParareal and its extension have probabilistic foundations. Indeed, although GParareal~\citep{pentland2023gparareal} relies solely on the prediction with the posterior mean of the GP, its posterior covariance structure contains important uncertainty information that the resulting solver could potentially leverage.

During the last two decades, the field of probabilistic numerics has also experienced rapid development. Although its origins can be traced back to the pioneering works of~\cite{Suldin1959} and~\cite{Larkin1972} in the second half of the 20th century (see~\citealt{oates2019modern} for a review), it is in recent years that it has experienced a surge of interest, advancing on a broad front: quadrature methods~\citep{minka2000deriving,sarkka2014gaussian,xi2018bayesian}, linear algebra~\citep{selig2012improving,fitzsimons2017bayesian}, global~\citep{hennig2012entropy} and local~\citep{mahsereci2017probabilistic} optimization, and DEs~\citep{kersting2020convergence,kramer2024stable,
 tronarp2019probabilistic,  teymur2018implicit,abdulle2020random}.
  Probabilistic numerics seeks to quantify epistemic uncertainty arising from intractable or incomplete numerical computation~\citep{hennig2022probabilistic}. For instance, quadrature methods and DE solvers rely on finite evaluations of the integrand and vector field, respectively, while the exact solution would theoretically require infinitely many of those.  By framing numerical problems as inference tasks, probabilistic numerics provides a principled approach to uncertainty quantification (UQ), yielding uncertainty-aware computation tools.

  In the specific context of ordinary differential equations (ODEs), probabilistic numerical methods can be classified into \textit{ODE filters and smoothers}, based on GP regression, and \textit{perturbative solvers}, which characterize uncertainty through perturbation of classic numerical methods~\citep{hennig2022probabilistic}. The cost of the former is cubic in the number of time steps, and, although an approximate computation of GP regression using Bayesian filters can be achieved in linear time~\citep{kersting2016active,kersting2020convergence,schober2019probabilistic,kramer2024stable}, these methods underperform in representing more complex dynamics (for example, for chaotic systems, see~\citealt{hennig2022probabilistic, tronarp2019probabilistic}). At the same time, perturbative solvers are more expressive, though more numerically expensive, requiring multiple simulations of the ODE.

\paragraph{Contribution.} In this work, we bridge the fields of PinT computation and UQ by introducing Prob-GParareal, which, to the best of our knowledge, is the first probabilistic extension of the (G)Parareal algorithm. The main idea consists in modeling the uncertainty in the Parareal update rule using GPs, and propagating it nonlinearly through time via sampling. Unlike GParareal~\citep{pentland2023gparareal}, which exclusively exploits the posterior mean of GPs, Prob-GParareal also effectively uses additional probabilistic information provided by their posterior covariance. Our proposed probabilistic solver, Prob-GParareal, offers several key advantages over its deterministic counterparts:
 \begin{itemize}
     \item \textit{Uncertainty quantification}. Prob-GParareal produces a probabilistic forecast of the system's evolution, explicitly quantifying the dynamics of the error across both the temporal domain and the algorithm's iterations. We demonstrate the accuracy of our method for both non-chaotic and chaotic systems, addressing the well-documented challenges of ODE filters in accurately representing chaotic behavior~\citep{hennig2022probabilistic, tronarp2019probabilistic}.
    \item \textit{Support for random initial conditions}. Prob-GParareal generalizes deterministic initial value problems (IVPs) to random initial conditions under non-restrictive assumptions. This extension enables its use for systems where precise information about the initial condition is not available.
    \item \textit{Solver compatibility}. Our method is agnostic to the choice of the numerical solver used, and it can be integrated with any existing (deterministic) Parareal implementation without major modifications to the simulation specifics. This flexibility is an important feature, as stability guarantees in Parareal applications often require problem-specific solvers~\citep{de2024parallel,ruprecht2018wave}. Since these can be directly embedded in our approach, Prob-GParareal naturally inherits enhanced stability properties.
    \item \textit{Flexible and controlled resource allocation}. Prob-GParareal enables flexible control over computational resources by supporting early termination, either based on predefined solution variance thresholds or computational budget constraints, without requiring full convergence of the algorithm. Empirical results suggest that probabilistic forecasts remain well-calibrated under early termination, with the predictive variance accurately reflecting the uncertainty induced by incomplete convergence.
    \item \textit{Scalability}. To achieve scalable performance under an increasing number of processors and DE dimensions, we build upon~\cite{nngparareal} and extend our Prob-GParareal framework to Prob-nnGParareal using nnGPs, improving the computational efficiency of our probabilistic solver.
 \end{itemize}
While probabilistic numerical methods have seen extensive development in other domains, their application to PinT methods remains unexplored, with the notable exceptions of~\cite{bosch2024parallel, iqbal2024parallel}. The authors leverage the associativity of the Bayesian smoothing operator~\citep{sarkka2020temporal} to achieve temporal parallelization. Their method differs from Prob-GParareal in several key aspects. First, their algorithm relies on extended Kalman filtering and smoothing. This technique provides exact solutions for affine ODEs, but requires first-order Taylor approximations of nonlinear vector fields. Second, their approach to UQ is intrinsically tied to the Bayesian smoothing procedure. Our proposed Prob-GParareal does not have these constraints, and can be applied to nonlinear and/or chaotic ODEs/PDEs in combination with any classical solver.

\medskip

The paper is organized as follows. In~\Cref{sec:background}, we review the Parareal and (nn)GParareal algorithms.
In~\Cref{sec:prob_parareal}, we introduce our novel method, Prob-GParareal, present its probabilistic formulation,  algorithmic structure, and computational complexity. In~\Cref{sec:th_convergence}, we provide its theoretical and error bound properties. \Cref{sec:empirics} validates our novel framework through numerical experiments on five benchmark ODE systems that exhibit stiff behavior, bifurcations, and chaotic dynamics. The extension of the Prob-GParareal algorithm to systems with probabilistic initial conditions is contained in~\Cref{sec:prob_init_cond}. In~\Cref{sec:nnprobpara}, we demonstrate the scalability improvements and the numerical efficiency increase achieved with Prob-nnGParareal. Conclusions and future directions are presented in~\Cref{sec:conclusion}.

\section{Background: Parareal and (nn)GParareal}
\label{sec:background}
Without loss of generality, to simplify the notation, we focus on autonomous IVPs (a non-autonomous PDE is considered in Section \ref{sec:nnprobpara}), described by a system of $d$ ODEs, $d \in \mathbb{N}$,
\begin{equation}
\label{eq:ode}
\frac{d\bs{u}}{dt} = h(\bs{u}(t)), \quad t \in \left[t_0,t_N\right], \quad \bs{u}(t_0)=\bs{u}_{(0)},
\end{equation}
where $\bs{u}:\left[t_0, t_N\right]\rightarrow \mathbb{R}^d$, $N\in\mathbb{N}$, is the time-dependent vector solution, $h: \mathbb{R}^d \rightarrow \mathbb{R}^d$ is a smooth multivariate function, and $\bs{u}_{(0)} \in \mathbb{R}^d$ is the known initial value at time $t_0$, which we assume deterministic, unless otherwise stated. As an exact solution to~\eqref{eq:ode} is generally  not available, one typically relies on a numerical solver $\mathscr{F}$ to obtain a high-accuracy numerical solution.
Depending on the system~\eqref{eq:ode} and the length of the interval over which it is
integrated, the sequential application of $\mathscr{F}$ may be computationally infeasible. The Parareal algorithm offers a remedy to this by partitioning the time domain into $N$ sub-intervals (usually of equal length), so that the problem can be split into $N$ IVPs given by:
\begin{equation}
\label{eq:2}
\frac{d\bs{u}_i}{dt} = h\left(\bs{u}_i (t)\right), \quad t \in \left[t_i,t_{i+1}\right], \quad \bs{u}_i(t_i)=\bs{u}_{(i)}, \quad i=0,\ldots,N-1,
\end{equation}
with $\bs{u}_{(i)} = \varphi_{\Delta t_i}(\bs{u}_{(i-1)})$, $i=1,\ldots, N-1$, where $\Delta t_{i}=t_{i} - t_{i-1}$ is the $i$th time step and ${\varphi}_{\Delta t_i}:\mathbb{R}^d\rightarrow \mathbb{R}^d$ denotes the $\Delta t_i$-time flow map (i.e. the solution) of the $i$th  IVP with initial condition $\bs{u}_{(i-1)}$ after time $\Delta t_i$.
Since only $\bs{u}_{(0)}$ is known, the other initial conditions $\bs{u}_{(i)}$, $ i=1,\ldots, N-1$, would need to be estimated. To do this, Parareal relies on an iterative scheme using a faster but less accurate coarse solver $\mathscr{G}$. At iteration $0$, the initial conditions $\bs{u}_{(i)}$ are approximated as $\bs{u}_{i,0}=\mathscr{G}(\bs{u}_{i-1,0})$, $i=1,\ldots,N-1$, where $\bs{u}_{i,k}$ denotes the Parareal solution at time $t_i$ and iteration $k$. These approximations at iteration $k=0$ are then updated \textit{sequentially}, for iteration $k \geq 1$, using the Parareal predictor-corrector rule
\begin{equation}
\label{eq:update_rule_para}
\bs{u}_{i,k} = \mathscr{G}(\bs{u}_{i-1,k}) + \left(\mathscr{F}-\mathscr{G}\right)(\bs{u}_{i-1,k-1}), \quad i=1,\ldots,N,
\end{equation}
where $\f(\bs{u}_{i-1,k-1})$ is computed \textit{in parallel} over $N$ processors. Parareal is said to $\epsilon$-converge at iteration $k$ for some chosen accuracy level $\epsilon>0$ whenever
\begin{equation}
    \label{eq:para_stp_rule}
\max_{1\leq i \leq N-1}\|\bs{u}_{i,k}-\bs{u}_{i,k-1}\|_\infty < \epsilon.
\end{equation}
Recent contributions modify~\eqref{eq:update_rule_para} by using the current $k$th iteration data $\bs{u}_{i-1,k}$, instead of $\bs{u}_{i-1,k-1}$ at iteration $k-1$, and by modeling the correction (discrepancy) function
\begin{equation}\label{discrepancy}
f_c := \left(\mathscr{F}-\mathscr{G}\right): \mathbb{R}^d\rightarrow \mathbb{R}^d,
\end{equation}
using alternative techniques. In particular,
GParareal~\citep{pentland2023gparareal} and nnGParareal~\citep{nngparareal} approximate this function using $d$ independent scalar GPs and nnGPs, respectively.
More precisely, in GParareal, each $s$th coordinate of the correction function $f_c$, is modeled as
\begin{equation}
f_c^{(s)}= (\f-\g)^{(s)} \sim GP(0, {K}_{\rm GP}), \enspace s=1,\ldots, d,
    \label{eq:gp_prior}
\end{equation}
that is, a one-dimensional GP with zero mean and variance kernel function ${K}_{\rm GP}: \R^d \times \R^d \rightarrow \R$.
The GPs are then trained on the accumulated dataset $\mathcal{D}_k$ (or a subset for nnGParareal) defined as
\[
\mathcal{D}_k = \left\{ \left(\bs{u}_{i-1,j},f_c(\bs{u}_{i-1,j})\right) | \, i=1,\ldots,N,\; j=0,\ldots,k-1 \right\}, \quad k\in \mathbb{N},
\]
leading to the following posterior distribution for the $s$th coordinate of a point $\bs{u}' \in \R^d$,
\begin{equation}
    f_c^{(s)}(\bs{u}')|\mathcal{D}_k \sim \; \mathcal{N}(\boldsymbol{\mu}_{\mathcal{D}_k}^{(s)}(\bs{u}'), {\sigma^{(s)}_{\mathcal{D}_k}}(\bs{u}')^2),\label{eq:gp_posterior}
\end{equation}
with posterior mean $\boldsymbol{\mu}_{\mathcal{D}_k}^{(s)}(\bs{u}') \in \mathbb{R}$ and posterior variance ${\sigma^{(s)}_{\mathcal{D}_k}}(\bs{u}')^2 \in \mathbb{R}^+$, whose expressions are given in~\Cref{app:GP_details}, when $K_{\rm{GP}}$ is a Gaussian kernel, also known as the radial basis function or square exponential kernel (an alternative kernel, the Mat\'ern kernel, popular in spatial statistics analysis~\citep{Matern1986}, is also presented there).
The Parareal update rule~\eqref{eq:update_rule_para} for GParareal is then constructed by taking the posterior means of the GPs as predictions of the discrepancy as follows:
\begin{equation}
\label{eq:update_rule_gpara}
\bs{u}_{i,k} = \mathscr{G}(\bs{u}_{i-1,k}) + \widehat{f}_{\rm GPara}(\bs{u}_{i-1,k}),
\end{equation}
where $\widehat{f}_{\rm GPara}(\bs{u}_{i-1,k}) = \left(\widehat{f}^{(1)}_{\rm GPara}(\bs{u}_{i-1,k}),\ldots,\widehat{f}^{(d)}_{\rm GPara}(\bs{u}_{i-1,k})\right)^\top \in \mathbb{R}^d$ is the vector of posterior means, with $\widehat{f}^{(s)}_{\rm GPara}(\bs{u}_{i-1,k}):= \boldsymbol{\mu}_{\mathcal{D}_k}^{(s)}(\bs{u}_{i-1,k})$, $s=1,\ldots, d$.

The framework is unchanged for nnGParareal, except that the nearest neighbors are recomputed for each test point $\bs{u}'\in \mathbb{R}^d$, and nnGPs are trained on these neighbors instead of the entire $\mathcal{D}_k$ before making a prediction $\widehat f_\text{nnGPara}$. Although GParareal (and similarly nnGParareal) uses only the posterior mean in~\eqref{eq:update_rule_gpara}, the GP framework naturally provides uncertainty estimates through its posterior variance in~\eqref{eq:gp_posterior}. This feature motivates our proposed probabilistic extension described in Section~\ref{sec:prob_parareal}.

\section{Prob-GParareal: a probabilistic Parareal framework}
\label{sec:prob_parareal}
The primary source of error in the Parareal algorithm is the coarse solver $\g$, which is inaccurate by design. In the context of sequential solvers of deterministic DEs, the true solution is recovered in the limit as the time step $\Delta t\rightarrow 0$ for convergent numerical schemes. However, infinitesimal time steps are not computable, and, in practice, finite $\Delta t$ leads to arbitrarily small errors consistent with the scheme's order of accuracy. Given an autonomous ODE with a unique solution and equidistant time steps $\Delta t=t_i-t_{i-1}, i=1,\ldots, N$, the coarse solver solution to the IVP starting in $\bs{u}_{i-1,k}$ can be written as
\[
\g(\bs{u}_{i-1,k}) = {\varphi}_{\Delta t}(\bs{u}_{i-1,k}) + \epsilon_\g^{ep}(\bs{u}_{i-1,k}), \quad i=1,\ldots,N,
\]
where $\epsilon_\g^{ep}(\bs{u}_{i-1,k})$ is the numerical error associated with $\bs{u}_{i-1,k}$ and $\varphi$ is the $\Delta t$-flow map defined as in~\eqref{eq:2}. Taking $\f$ as $\varphi$ is equivalent to assuming that $\f$ is sufficiently accurate to represent the true solution to~\eqref{eq:ode}, a common assumption in the (theoretical and applied) Parareal literature~\citep{gander2008nonlinear, pentland2023gparareal}.
Under this assumption, the correction function $f_c$ in~\eqref{discrepancy} accounts for all sources of uncertainty. While a direct application of $f_c$ is unfeasible, as it would require \textit{sequential} runs of $\f$, GPs could be used instead, as described in \Cref{sec:background}. In particular, the GParareal predictor-corrector rule~\eqref{eq:update_rule_gpara} can be readily extended by including the posterior distribution of the GP, to model the error incurred by approximating $f_c$. This leads to random solutions $\{\bs{U}_{i,k}\}_{1\leq i\leq N,k\in \mathbb{N}}$ (with $\bs{u}_{i,k}$ representing one of their possible outcomes/draws/realizations) that define a process, Markov conditionally on the training data $\mathcal{D}_k$, with the law of $\bs{U}_{i,k}$ conditionally independent of $\bs{U}_{j,\ell}$, given $\bs{U}_{i-1,k}$, for  $j\leq i-2$, $\ell\neq k$. In particular, the probabilistic update rule for $\bs{U}_{i,k}$ is then given by
\begin{equation}
\bs{U}_{i, k}=\g\left(\bs{U}_{i-1, k}\right)+\boldsymbol{Z}_{i, k}, \quad i=1,\ldots,N, \quad k \geq 1,\label{eq:update_rule_probpara}
\end{equation}
with
\begin{equation}
\label{eq:cond normal}
\boldsymbol{Z}_{i,k}|(\bs{U}_{i-1, k}=\bs{u}_{i-1,k},\mathcal{D}_k) \enspace \sim \mathcal{N}_d(\boldsymbol{\mu}_{\mathcal{D}_k}\left(\bs{u}_{i-1, k}\right), \Sigma_{\mathcal{D}_k}(\bs{u}_{i-1,k})),
\end{equation}
where $\mathcal{N}_d(\boldsymbol{a},B)$ denotes a $d$-dimensional normal distribution with mean vector $\boldsymbol{a} \in \mathbb{R}^d$ and $d\times d$-dimensional covariance matrix $B$,
and $\boldsymbol{Z}_{i, k}$ is conditionally independent
of $ \bs{U}_{j, \ell}$ given $\bs{U}_{i-1,k}$ for all $j\leq i-2$ and $\ell \neq k$. Once a realization $\bs{u}_{i-1,k}$ of $\bs{U}_{i-1, k}$ is given, $\boldsymbol{Z}_{i, k}$ represents the only source of uncertainty in~\eqref{eq:update_rule_probpara}.

From the update rule~\eqref{eq:update_rule_probpara}, it is immediate to see that the conditional distribution of $\bs{U}_{i,k}$, given $\bs{U}_{i-1,k}=\bs{u}_{i-1,k}$, follows a $d$-dimensional normal distribution
\begin{equation}
    \label{eq:pp_cond_distr}
    \bs{U}_{i,k}|\bs{U}_{i-1,k}=\bs{u}_{i-1,k} \enspace  \sim \mathcal{N}_d\left( \g(\bs{u}_{i-1,k}) + \boldsymbol{\mu}_{\mathcal{D}_k}(\bs{u}_{i-1,k}), \Sigma_{\mathcal{D}_k}(\bs{u}_{i-1,k}) \right).
\end{equation}
Instead, the unconditional distribution of $\bs{U}_{i,k}$ is unknown. Even if it were known and Gaussian, determining the distribution of $\bs{U}_{i+1,k}$ via~\eqref{eq:update_rule_probpara} using $\g$ would be neither theoretically possible nor computationally feasible for nonlinear IVPs~\citep{pentland2023gparareal}.
 This is why Bayesian/probabilistic numerics ODE filters and smoothers linearize the vector field of nonlinear ODEs, resulting in approximate solutions (see, e.g.~\citealt{hennig2022probabilistic}). In the following subsections, we propose a sampling scheme overcoming these limitations, circumventing the need for linearization.

\subsection{Derivation of Prob-GParareal}
\label{sec:theoretical_construction}
The update rule~\eqref{eq:pp_cond_distr} allows  modeling the correlation between the components of $\bs{U}_{i,k}|\bs{U}_{i-1,k}$ with multi-output GPs. However, the computational costs associated with such a flexible approach may be significantly high, with marginal, if any, improvements in terms of accuracy and the number of iterations until the algorithm converges, as discussed in~\cite{pentland2023gparareal} for GParareal. Moreover, for systems of DEs, one of the simplest multi-output GP extensions, the intrinsic coregionalization model~\citep{goovaerts1997geostatistics}, has been shown to be equivalent to independent GPs' predictions over each coordinate, with each GP trained on the same dataset~\citep{alvarez2012kernels, wackernagel2003multivariate}. Hence, we adopt the same strategy here for Prob-GParareal and, similarly to GParareal, assume that the entries of $\bs{U}_{i,k}|\bs{U}_{i-1,k}$ are uncorrelated and, thus, independent, given the underlying multivariate Gaussian distribution. More precisely, the covariance matrix $\Sigma_{\mathcal{D}_k}$ in~\eqref{eq:pp_cond_distr} is assumed to be diagonal with diagonal components $\sigma^{(s)}_{\mathcal{D}_k}(\bs{u}_{i-1,k})^2$, $s=1,\ldots, d$ as in~\eqref{eq:gp_posterior}.

We now formalize the Prob-GParareal sampling procedure to obtain realizations of $\bs{U}_{i,k}$.
Let $P_{\bs{U}_{i,k}\mid\bs{U}_{i-1,k}}$, $i=1,\ldots, N$, $k\in \mathbb{N}$, denote the conditional distribution of $\bs{U}_{i,k}|\bs{U}_{i-1,k}$ given in~\eqref{eq:pp_cond_distr} (in some cases, we explicitly write $P_{\bs{U}_{i,k}\mid\bs{U}_{i-1,k}=\bs{u}_{i-1,k}}$ to specify a particular realization $\bs{u}_{i-1,k}$), and let $P_{\bs{U}_{i,k}}$ be the unconditional marginal distribution of $\bs{U}_{i,k}$, given by
\begin{equation}
P_{\bs{U}_{i,k}}(\bs{u}_{i,k}) = \int_{\R^d} P_{\bs{U}_{i,k}\mid\bs{U}_{i-1,k}}(\bs{u}_{i,k}\mid\bs{u}_{i-1,k}) \; P_{\bs{U}_{i-1,k}}(\bs{u}_{i-1,k}) d\bs{u}_{i-1,k},
\label{eq:cont_mixture}
\end{equation}
which, in general, cannot be solved analytically. However, it can be interpreted as a continuous Gaussian mixture, for which sampling is feasible via many available procedures. Here, we employ a two-step procedure known as ancestral sampling~\cite[Chapter~11]{deisenroth2020mathematics}, designed to draw $n\in \mathbb{N}$
{\it observed} samples, or samples of {\it observations/realizations},
$\mathcal{U}_{i,k} = \{ \bs{u}_{i,k}^{(j)}\}_{j=1}^n$, from $ P_{\bs{U}_{i,k}}$, $i=1,\ldots, N$, $k\in \mathbb{N}$,  in~\eqref{eq:cont_mixture} using~\eqref{eq:pp_cond_distr}, as follows.\\
For $i=1,\ldots, N$, $k\in \mathbb{N}$:
\begin{enumerate}
    \item
    Sample independently $n$ observations $\bs{u}_{i-1,k}^{(1)}, \ldots, \bs{u}_{i-1,k}^{(n)}$ from the marginal distribution $P_{\bs{U}_{i-1,k}}$.
    \item Given $\bs{U}_{i-1,k}=\bs{u}_{i-1,k}^{(j)}$, sample $\bs{u}_{i,k}^{(j)}$ from $ P_{\bs{U}_{i,k}|\bs{U}_{i-1,k}= \bs{u}_{i-1,k}^{(j)}}$ in~\eqref{eq:pp_cond_distr},  $j=1,\ldots, n$.
\end{enumerate}
This sampling procedure is then embedded within the GParareal framework, as described in the following Subsection.
\subsection{Algorithm}
\label{sec:probpara_algo}
\begin{algorithm}[ht!]
\caption{Prob-GParareal Algorithm}\label{alg:Prob-GParareal}
\textbf{Input}: initial distribution $P_{\bs{U}_{0,0}}$, number of samples $n\in\mathbb{N}$, coarse solver $\g$, fine solver $\f$, tolerance $\epsilon>0$, exit condition,
statistic $g:(\mathbb{R}^d)^n\rightarrow \mathcal{X}$, distance metric $d_{\mathcal{U}}:\mathcal{X}\times \mathcal{X}\rightarrow \mathbb{R}^+$.
\begin{algorithmic}[1]
\Statex \textit{Initialization of the Algorithm}
\State Initialize the dataset $\mathcal{D}_0=\emptyset$, the number of converged intervals $L=0$, the iteration $k=0$ and $K_{\rm conv}=K_{\rm stop}=0$.
\State Sample $n$ observations $\bs{u}_{0,0}^{(j)}$, $j=1,\ldots, n$ from $P_{\bs{U}_{0,0}}$, set $\mathcal{U}_{0,0}=\{\bs{u}_{0,0}^{(j)}\}_{j=1}^n$.
\For{$i=1,\ldots, N$}
\For{$j=1,\ldots, n$} \textit{(in parallel)}
\State Compute $\bs{u}_{i,0}^{(j)}=\g(\bs{u}_{i-1,0}^{(j)})$.
\EndFor
\State Set $\mathcal{U}_{i,0}=\{\bs{u}_{i,0}^{(j)}\}_{j=1}^n$.
\EndFor
\Statex \textit{Execution of the recursive algorithm}
\While{$L<N$ {\bf and} exit condition not met}
\State Set $k=k+1$
\State Set $\mathcal U_{i,k}=\mathcal U_{i,k-1}$ for $i=0,\ldots,L$.
\For{$i=L,\ldots, N-1$
}  \textit{(in parallel)}
\State   Compute the sample means $\overline{\bs{u}}_{i,k-1} = \frac{1}{n} \sum\limits_{j=1}^n \bs{u}_{i,k-1}^{(j)}$ for observation samples $\mathcal{U}_{i,k-1}$.
\State Evaluate $\f(\overline{\bs{u}}_{i,k-1})$ and $\g(\overline{\bs{u}}_{i,k-1})$.
\EndFor
\State Update the dataset $\mathcal{D}_{k} = \mathcal{D}_{k-1} \cup \{\left(\overline{\bs{u}}_{i,k-1}, f_c (\overline{\bs{u}}_{i,k-1})\right)\}_{i=L}^{N-1}$, with $f_c$ defined in~\eqref{discrepancy}.
\State \mbox{Train the $d$ scalar GPs on $\mathcal{D}_k$ to compute the  posterior mean $\bs{\mu}_{\mathcal{D}_k}(\cdot)$ and covariance $\Sigma_{\mathcal{D}_k}(\cdot)$.}
\For {$i=L+1,\ldots, N$}
\For{$j=1,\ldots, n$}  \textit{(in parallel)}
\State \mbox{Draw $\bs{z}_{i,k}^{(j)}$  from $\mathcal{N}_d \left(\boldsymbol{\mu}_{\mathcal{D}_k}(\bs{u}_{i-1,k}^{(j)}), \Sigma_{\mathcal{D}_k}(\bs{u}_{i-1,k}^{(j)}) \right)$, i.e.~\eqref{eq:cond normal} conditioned on  $\bs{U}_{i-1,k}=\bs{u}_{i-1,k}^{(j)}$.}
\State \mbox{Compute $\bs{u}_{i,k}^{(j)} = \g(\bs{u}_{i-1,k}^{(j)}) + \bs{z}_{i,k}^{(j)}$ by the Prob-GParareal predictor-corrector rule~\eqref{eq:update_rule_probpara}.}
\EndFor
\State Set $\mathcal{U}_{i,k}=\{ \bs{u}_{i,k}^{(j)}\}_{j=1}^n$.
\EndFor
\For{$i=L+1,\ldots, N$}
\If{$d_{\mathcal{U}}(g({\mathcal{U}_{i,k}}),g({\mathcal{U}_{i,k-1}}))<\epsilon$}
\State Set $L=i$.
\Else
\State \textbf{Break}
\EndIf
\EndFor
\EndWhile
\If{$L=N$}
\State Set $K_{\rm end}=K_{\rm conv}=k$.
\Else
\State Set $K_{\rm end}=K_{\rm stop}=k$.
\EndIf
\end{algorithmic}
\textbf{Output}: Set of solution samples $\{\mathcal{U}_{i,K_{\rm end}}\}_{i=0}^N$, $K_{\rm end}, K_{\rm conv}$, and $K_\textrm{stop}$.
\end{algorithm}

The schematic description of Prob-GParareal is provided in Algorithm \ref{alg:Prob-GParareal}. The algorithm is initialized at iteration $k=0$ by drawing $n$ initial values from the initial distribution $P_{\bs{U}_{0,0}}$ (Line 2). When the initial condition $\bs{u}_{(0)}$ is deterministic, as considered in the experiment results in Section~\ref{sec:empirics} (other than~\Cref{sec:prob_init_cond}) and~\Cref{sec:nnprobpara}, we set $P_{\bs{U}_{0,0}}=\delta_{\bs{u}_{0,0}}$, a Dirac measure centered at $\bs{u}_{0,0}=\bs{u}_{(0)}$, leading to $\bs{u}_{0,0}^{(j)}=\bs{u}_{(0)}$ and $\bs{u}_{i,0}^{(j)}=\g(\bs{u}_{i-1,0}^{(j)}), i=1,\ldots, N$, $j=1,\ldots, n$, otherwise the $n$ sampled values are propagated sequentially via $\g$ (Line 5), leading to $\mathcal{U}_{i,0}=\{\bs{u}_{i,0}^{(j)}\}_{j=1}^n, i=0,\ldots, N$, collection of observed samples at iteration $k=0$.

After the initialization phase, a recursive execution consisting of a five-step procedure is launched, running until either the algorithm converges or an early stopping criterion is met, as described below.

Let $L$ be the number of converged intervals, which is initially set to $0$. At iteration $k$ and interval $i=L,\ldots, N$, the sample means $\overline {\bs{u}}_{i,k-1}$ of the observation samples $\mathcal{U}_{i,k-1}$ at iteration $k-1$ are first computed and then propagated through $\g$ and $\f$ \textit{in parallel} (Lines 12-13, {\bf Step 1}). The pairs $\left(\overline{\bs u}_{i,k-1},f_c(\overline{\bs u}_{i,k-1})\right)$, $i=L,\ldots, N$, are then added to the dataset $\mathcal{D}_{k-1}$, yielding the updated dataset $\mathcal{D}_k$ (Line 15, {\bf Step 2}), which is then used to train $d$ scalar GPs to obtain the estimates of the posterior mean $\bs{\mu}_{\mathcal{D}_k}(\cdot)$ and posterior covariance $\Sigma_{\mathcal{D}_k}(\cdot)$ functions (Line 16, {\bf Step 3}). After that, the Prob-GParareal predictor-corrector rule $\bs{U}_{i,k}=\g(\bs{U}_{i-1,k})+\bs Z_{i,k}$ in~\eqref{eq:update_rule_probpara} is applied by drawing, \textit{in parallel} over $j=1,\ldots,n$,  a realization $\bs z_{i,k}^{(j)}$ from the multivariate conditional Gaussian distribution $\bs U_{i,k}|\bs{U}_{i-1,k}=\bs{u}_{i-1,k}^{(j)}$\footnote{Note that, given the chosen GP implementation consisting of $d$ independent scalar GPs, see~\eqref{discrepancy}-\eqref{eq:gp_posterior}, we sample the $s$th coordinate of $\bs{z}_{i,k}^{(j)}$ from $\mathcal{N} \left(\boldsymbol{\mu}^{(s)}_{\mathcal{D}_k}(\bs{u}_{i-1,k}^{(j)}), \sigma^{(s)}_{\mathcal{D}_k}(\bs{u}_{i-1,k}^{(j)})^2 \right)$.
}, and then computing, sequentially over $i=L+1,\ldots, N$, the resulting sampled predictor-corrector values $\bs{u}_{i,k}^{(j)}=\g(\bs{u}_{i-1,k}^{(j)})+\bs z_{i,k}^{(j)}$ (Lines 17-23, {\bf Step 4}). This defines the resulting sampled values $\mathcal{U}_{i,k}=\{\bs{u}_{i,k}^{(j)}\}_{j=1}^n$ at iteration $k$ (Line 22). At this point, ({\bf Step 5}):
\begin{itemize}
\item We check whether the Prob-GParareal solutions have converged up to some interval $l$, $L+1\leq l\leq N$. Intuitively, the Prob-GParareal solutions
$\mathcal{U}_{i,k}$, $i=L+1,\ldots, N$ have converged up to time $t_l\leq t_N$, with $i\leq  l\leq N$, if the sampled points in $\mathcal{U}_{i,k-1}$ and $\mathcal{U}_{i,k}$ at interval $i=L+1,\ldots, l$ and iteration $k-1$ and $k$, respectively, are \lq\lq close enough\rq\rq.
To formalize this, we choose an accuracy threshold $\epsilon>0$, a statistic $g:(\mathbb{R}^d)^n\to\mathcal{X}$, for some metric space $\mathcal{X}$, applied to both $\mathcal{U}_{i,k-1}$, and $\mathcal{U}_{i,k}$, and a distance metric $d_{\mathcal{U}}:\mathcal{X}\times\mathcal{X}\to \mathbb{R}_+$, which measures their similarity. In our implementation and theoretical analysis, $g$ is defined as the empirical measure with $\widehat{P}_{\mathcal{U}_{i,k} }:=g(\mathcal{U}_{i,k})=\dfrac{1}{n} \sum_{j=1}^n\delta_{\bs{u}_{i,k}^{(j)}}$. Additionally, $d_{\mathcal{U}}$ is chosen as the $p$-power Wasserstein-$p$ ($p\in\mathbb{N}$) distance~\citep{villani2008optimal} between these empirical measures. We say that the Prob-GParareal solution has $\epsilon$-converged up to iteration $l$ when
      \begin{equation}
            \label{eq:pp_stp_rule}
        W_p(\widehat{P}_{\mathcal{U}_{i,k} }, \widehat{P}_{\mathcal{U}_{i,k-1} })^{p}= \min_{ \pi \in \Pi } \left( \frac{1}{n} \sum_{j=1}^n \left \| \bs{u}_{i,k}^{(j)} - \bs{u}_{i,k-1}^{( \pi(j) )} \right\|^p \right) < \epsilon, \quad 0\leq L<i\leq l\leq N,
        \end{equation}
        where $ \Pi $ is the set of all permutations of $ \{1, 2, \dots, n\} $. If this happens, we set $L=l$ and, if $L<N$, continue the execution run.
        Here, the tolerance $\epsilon$ is expressed on the $p$th-power scale.
\item We verify whether an early termination rule, based on practical constraints, such as maximum solution variance or computational budget limits, is met. If so, the algorithm terminates and the solutions $\mathcal{U}_{i,k}$, $i=0,\ldots, N$, are returned.
\end{itemize}
\begin{remark}
The Prob-GParareal algorithm can be easily modified to evaluate $\f$ and $\g$ in parallel on other summaries (e.g., median or a certain quantile of $\mathcal{U}_{i,k-1}$) than the sample mean $\overline{\bs u}_{i,k-1}$ (Lines 12-13 of Algorithm~\ref{alg:Prob-GParareal}).
One could also potentially propagate all $n$ samples through $\f$ for each $i$th time interval (as done in SParareal by~\citealt{pentland2022stochastic}), but this would require $nN$ processors, which is a prohibitively expensive requirement in practice. Despite the limitation of considering only a summary of the sample, we derive theoretical error bounds
in~\Cref{sec:th_convergence}, and numerically illustrate the algorithm's convergence on several models in Section~\ref{sec:empirics}.
\end{remark}
\begin{remark}\label{remark2}
In~\eqref{eq:pp_stp_rule}, we used empirical measures and the Wasserstein-$p$ distance between them. This distance, computed with the Hungarian (Kuhn-Munkres) algorithm, typically has an $O(n^3)$ complexity, although certain geometric assumptions or approximation techniques can mitigate this cost (see, e.g.,~\citealt{bernton2019approximate} and~\citealt{NIPS2013_af21d0c9}). In our implementation, rather than computing the exact Wasserstein distance between the empirical measures in \eqref{eq:pp_stp_rule}, we approximate each empirical distribution by a multivariate Gaussian with its sample mean and covariance matrix and compute the squared Wasserstein-2 distance between these Gaussian approximations (cost of $O(nd)$ is achieved under diagonal covariance structure). Other statistics and distance metrics could be considered. For instance, multivariate statistics $g:(\mathbb{R}^d)^n\rightarrow \mathbb{R}^q$, $q\in \mathbb{N}$, paired with an appropriate metric $d_{\mathcal{U}}: \mathbb{R}^q \times \mathbb{R}^q\rightarrow \mathbb{R}^+$ represent a viable alternative. Among other probability metrics for empirical measures (see~\citealt{sriperumbudur2009integral} for an overview), the Maximum Mean Discrepancy (MMD)~\citep{gretton2006kernel} stands out due to its computational efficiency, with an $O(n^2)$ complexity~\citep{sriperumbudur2010hilbert}, which can be further reduced via approximation methods. Despite the computational advantages and the availability of simple upper and lower bounds linking MMD to the Wasserstein-$p$ distance~\citep{sriperumbudur2010hilbert}, its performance is more sensitive to ad-hoc tuning of the kernel bandwidth, which is why it is not considered here.

\end{remark}
\begin{remark}
    The Prob-GParareal algorithm introduces two complementary stopping mechanisms in Step 5; the $\epsilon$-convergence, controlling the accuracy of the desired solution (Line 25 in Algorithm~\ref{alg:Prob-GParareal}); an independent termination rule (that we call the exit condition, see Line 28 in  Algorithm~\ref{alg:Prob-GParareal}). By explicitly accounting for
the uncertainty of the solution throughout its execution, Prob-GParareal allows these two stopping criteria to operate simultaneously. We explore the impact of early termination of the algorithm in~\Cref{sec:early_stop}. \end{remark}

\subsection{Computational complexity}
\label{sec:comp_complx}

The computational cost of Prob-GParareal, denoted as $T_\textrm{Prob-GPara}$, can be divided into that of running $\f$ and $\g$ over one interval $[t_i, t_{i+1}]$, $i=0,\ldots, N-1$, denoted by $T_\f$ and $T_\g$, respectively, and that of computing the correction function $f_c$ in~\eqref{discrepancy} during iteration $k\in \mathbb{N}$, denoted as $T_f^{\textrm{Prob-GPara}}(k)$. At iteration $k$, the $d$ GPs are trained in parallel on the dataset $\mathcal{D}_k$, containing up to $Nk$ observations, and are used to sample $\bs{z}_{i,k}^{(j)}$, $j=1,\ldots,n$, for up to $N$ intervals, which yields the following complexity:
\[
\begin{aligned}
    T^{\text{Prob-GPara}}_f(k) &= \underbrace{(\tfrac{d}{N} \vee 1) \; O\left(d (Nk)^2 + (Nk)^3\right)}_{\text{Training}}
    +
\underbrace{N \; (\tfrac{dn}{N} \vee 1) \;  O\left( dNk + (Nk)^2    \right)}_{\text{Sampling of } \{\bs{z}_{i,k}^{(j)}\}_{j=1}^n},
\end{aligned}
\]
where $\vee$ denotes the maximum operator.
The factors $(\tfrac{d}{N} \vee 1)$ and $(\tfrac{dn}{N} \vee 1)$ result from parallelizing computations across the $d$ GPs. The training term complexity $O(d (Nk)^2)$ corresponds to constructing the $Nk$-dimensional symmetric kernel matrix, while the $O((Nk)^3)$ term accounts for its inversion, with both operations carried out once per GP. In the sampling term, the complexities $O(dNk)$ and $O((Nk)^2)$ represent the cost of evaluating the posterior mean and variance, respectively, repeated across each GP and each sample $j=1,\ldots, n$  \textit{in parallel}. Moreover, the
 $N$ factor in the sampling term accounts for \textit{sequential} repetition of sampling for up to $N$ intervals.

At iteration $k$, the training cost of Prob-GParareal is the same as that of GParareal. Instead, the sampling cost of GParareal is lower, being $(N-k) \; (\tfrac{d}{N} \vee 1) \; O(dNk)$~\citep{pentland2023gparareal} as it does not need to evaluate the posterior variance.
This is due to the fact that GParareal applies the discrepancy function only once per interval, with no need to
estimate the variance. The factor $(N-k)$ follows from the application of $\f$ directly on the solution $\bs{u}_{i-1,k}$, rather than on the mean $\bar{\bs{u}}_{i-1,k}$ as in Prob-GParareal. For moderate values of $d$ and when $\tfrac{n}{N} \in O(k)$, the training and sampling steps in Prob-GParareal share the same cubic complexity, indicating no additional overhead for uncertainty estimation compared to the deterministic GParareal.

To mitigate the cubic inversion cost of GPs, we follow the approach proposed by~\cite{nngparareal} in nnGParareal, using nnGPs trained on a reduced dataset consisting of the $m$ observations (typically $  m\approx 15\ll Nk$) that are nearest neighbors in terms of Euclidean distance to the prediction point $\bs{u}_{i-1,k}^{(j)}, j=1,\ldots, n$
. Although this method would require re-training the nnGPs for each of the $n$ new prediction points, as the data subset may change, we empirically observe that most of the $n$ observations belonging to the sample $\mathcal{U}_{i,k}$ at interval $i$ share the same nearest neighbors. For this reason, we train the nnGP once per interval $i$, obtaining the following worst-case cost per iteration for the Prob-nnGParareal algorithm:
\[
\begin{aligned}
    T^{\text{Prob-nnGPara}}_f(k) &=
    \underbrace{N  }_{\text{For each $i$}}
    \underbrace{(\tfrac{d}{N} \vee 1) \; O\left(d m^2 + m^3\right)}_{\text{Training once}}
     +
\underbrace{N \; (\tfrac{dn}{N} \vee 1) \; O\left( dm + m^2 + \log(Nk)   \right)}_{\text{Sampling of } \{\bs{z}_{i,k}^{(j)}\}_{j=1}^n},
\end{aligned}
\]
where $\log(Nk)$ is the cost associated with maintaining a kd-tree structure~\citep{bentley1975multidimensional}  for fast nearest neighbor computation~\citep{nngparareal}. Similar considerations of the training costs of Prob-GParareal and GParareal hold when comparing Prob-nnGParareal and nnGParareal.

Specifically, the cost of Prob-nnGParareal includes the additional term $m^2$ to compute the covariance matrix, and the parallelizable cost of making $n$ posterior evaluations per interval $i$. Overall, Prob-nnGParareal is considerably faster  than Prob-GParareal, leading to similar results, albeit with slightly higher uncertainty in the algorithm solution due to the additional approximation, as shown in~\Cref{sec:nnprobpara}.

As mentioned above, the total computational cost $T_\textrm{Prob-GPara}$ of Prob-GParareal combines the cost of running $\f$ and $\g$, with the cost of computing the correction function. A validation of the cost model against observed wall-clock time, including scaling and speedup results with respect to the number of time intervals \(N\), is reported in \Cref{app:cost_model_validation}. Let $K_{\rm conv}$ be the number of iterations required for the algorithm to converge. Prob-GParareal performs up to $N$ parallel evaluations of $\f$ per iteration $k =1, \ldots, K_{\rm conv}$, and $n$ parallel evaluations of $\g$ per interval $i$, yielding an approximate worst-case total computational cost, ignoring serial overheads, of
\[
T_{\text{Prob-GPara}} \approx K_{\rm conv} T_{\f} + (\tfrac{n}{N} \vee 1)(K_{\rm conv}+1)NT_{\g}+\sum_{k=1}^{K_{\rm conv}}T^{\text{Prob-GPara}}_f(k) + NK_{\rm conv} C_{\rm dist},
\]
where $C_{\rm dist}$ is the general cost of computing the statistics in~\eqref{eq:pp_stp_rule}, which is $O(n^3)$ for the Wasserstein-$p$ case, and $O(nd)$ for the special case $p=2$ under Gaussian assumption, as discussed in Remark \ref{remark2}.

The overall Prob-GParareal cost is slightly higher than that of GParareal (i.e., $T_{\rm GPara} =K_{\rm conv} T_{\f} + (K_{\rm conv}+1)(N-K_{\rm conv}/2)T_{\g}+T_f$,~\citealt{pentland2023gparareal}), due to the execution of $n$  applications of $\g$ per interval during the sequential update procedure. With sufficient computational resources, the $n$ runs of $\g$ can be parallelized, reducing the total cost of Prob-GParareal to that of GParareal.

\section{Theoretical analysis}
\label{sec:th_convergence}
In this section, we derive two main theoretical results for Prob-GParareal: an error bound  of its solution to the true one in terms of $W_2^{2}$
(\Cref{prop:convergence}), and an error bound between the Prob-GParareal mean behavior and the GParareal solution (\Cref{prop:meanbound}), which we use to provide a theoretical comparison between the two algorithms.

\subsection{Notation}
Here, we adopt the notation introduced, for example, in~\cite{Zhu08}. Given $p \in \mathbb{N}, \alpha=\left(\alpha_1, \ldots, \alpha_d\right) \in \mathbb{N}^d$ and $|\alpha|=\sum_{s=1}^d \alpha_s$, we define the index set $I_p:=\left\{\alpha \in \mathbb{N}^d:|\alpha| \leq p\right\}$. For a function $f: \mathbb{R}^d \rightarrow \mathbb{R}^d$, we denote the $\alpha$th partial derivative of its $s$th coordinate, if it exists, as
$$
D^\alpha f^{(s)}(\bs{u})=\frac{\partial^{|\alpha|}}{\partial u_1^{\alpha_1} \partial u_2^{\alpha_2} \cdots \partial u_d^{\alpha_d}} f^{(s)}(\bs{u}), \quad \text { for all } \bs{u} \in \mathbb{R}^d.
$$
Given the set $C^p\left(\mathbb{R}^d, \mathbb{R}^d\right)$ of $p$-continuously differentiable functions from $\mathbb{R}^d$ to $\mathbb{R}^d$, let $C_b^p\left(\mathbb{R}^d, \mathbb{R}^d\right)$ be the set of the corresponding bounded $p$-continuously differentiable functions with bounded derivatives given by
$$
C_b^p(\mathbb{R}^d, \mathbb{R}^d)=\left\{f \in C^p(\mathbb{R}^d, \mathbb{R}^d): \|f\|_{C_b^p}:=\max_{1\leq s\leq d}\|f^{(s)}\|_{C_b^p}=\max_{1\leq s\leq d}\left\{\sup _{\alpha \in I_p}\left\|D^\alpha f^{(s)}\right\|_{\infty}\right\}<\infty\right\},
$$
where $\|g\|_{\infty}=\sup _{\bs{u} \in \mathbb{R}^d}|g (\bs{u})|$ for any $g:\mathbb{R}^d\rightarrow \mathbb{R}$.
Unless otherwise specified, $\|\cdot \|$ denotes the Euclidean norm for vectors, and its induced matrix (spectral) norm if applied to matrices.

\subsection{Kernels and induced RKHS}

For our Prob-GParareal error bound analysis, the reproducing kernel Hilbert space (RKHS) structure is especially advantageous. We start by recalling standard facts and some recent results in the literature, which are convenient for our derivations.

\begin{definition}[RKHS,~\cite{Steinwart2008}]Let $K: \mathcal{U} \times \mathcal{U} \rightarrow \mathbb{R}$ be a Mercer (symmetric positive semi-definite) kernel on a nonempty set $\mathcal{U} \subset \mathbb{R}^d$. A Hilbert space $\mathcal{H}_K$ of real-valued functions on $\mathcal{U}$ endowed with the pointwise sum and pointwise scalar multiplication, and with inner product $\langle\cdot, \cdot\rangle_{\mathcal{H}_K}$ is called a RKHS associated to $K$ if the following properties hold:
\begin{description}
    \item[(i)]For all $\bs{u} \in \mathcal{U}$, the function $K(\bs{u}, \cdot) \in \mathcal{H}_K$.
    \item[(ii)]For all $\bs{u} \in \mathcal{U}$ and for all $f \in \mathcal{H}_K$, the  reproducing property $f(\bs{u})=\langle f, K(\bs{u}, \cdot)\rangle_{\mathcal{H}_K}
$ holds.
\end{description}
\end{definition}
Note that two commonly used Mercer kernels are the Gaussian and the Mat\'ern kernels defined in~\Cref{app:GP_details}. Given a Mercer kernel $K$ on a set $\mathcal{U}$, by Moore-Aronszajn Theorem~\citep{aronszajn1950theory}, there exists a unique Hilbert space of real-valued functions for which $K$ is a reproducing kernel. The RKHS $\mathcal{H}_K$ induced by $K$ is given by
\begin{equation}
    \mathcal{H}_K=\left\{f=\sum_{i=1}^{\infty} c_i K\left(\bs{u}_i, \cdot\right) \mid c_i \in \mathbb{R}, \bs{u}_i \in \mathcal{U},\|f\|_{\mathcal{H}_K}^2=\sum_{i, j=1}^{\infty} c_i K\left(\bs{u}_i, \bs{u}_j\right) c_j<\infty\right\},
\end{equation}
see~\cite{Steinwart2008} for more details.

\subsection{
Preparatory assumptions}
Here, we introduce the assumptions on the numerical coarse solvers and on the properties of the GP posterior variance, which will be needed to prove the theoretical results of \Cref{theoProbGP}.
Following~\cite{gander2008nonlinear}, we assume that $h:\mathbb{R}^d\rightarrow \mathbb{R}^d$ in~\eqref{eq:ode} is of appropriate regularity, the time steps are uniform with $\Delta t_i=\Delta t:=(t_N-t_0)/N$, and that $\f$ yields an exact solution, that is $\bs{u}(t_i) = \f(\bs{u}(t_{i-1})) = \varphi_{\Delta t}(\bs{u}(t_{i-1}))$, for all $i=1, \ldots, N$.
\begin{assumption}[Order of the one-step coarse solver $\g$]
\label{ass:1_beg}
$\g$ is a one-step numerical solver with uniform local truncation error $O(\Delta t^{p+1})$ for $p \geq 1$. More precisely, for all $\bs{u} \in \R^d$, it holds that
\begin{equation}
\label{eq:decomp}
\f(\bs{u}) - \g(\bs{u})
=
c^{(p+1)}(\bs u)\Delta t^{p+1}
+
R_{p+2}(\bs u,\Delta t),
\end{equation}
where $c^{(p+1)}\in C_b^1(\mathbb{R}^d,\mathbb{R}^d)$, and there exist constants
$C_R>0$ and $\Delta t_0>0$ such that, for all $0<\Delta t\leq \Delta t_0$, $
\|R_{p+2}(\cdot,\Delta t)\|_{C_b^1}
\leq
C_R\Delta t^{p+2}.$
\end{assumption}
Note that, the one-step truncation error in~\eqref{eq:decomp} corresponds by definition~\eqref{discrepancy} to the correction function $f_c$, so Assumption \ref{ass:1_beg} can also be seen as an assumption on $f_c$ rather than on $\g$.

\begin{assumption}[The correction function is Lipschitz]
\label{ass:2_beg}
    The correction function $f_c$ in~\eqref{discrepancy} is Lipschitz continuous, that is, for all $\bs{u},\bs{u}' \in \R^d$, there exists some constant $L_c>0$, such that
    \begin{equation}
    \label{ass:2}
    \| f_c(\bs{u}) - f_c(\bs{u}') \| \leq L_c \| \bs{u} - \bs{u}' \|.
    \end{equation}
\end{assumption}

\begin{assumption}[$\g$ is Lipschitz]
\label{ass:3_beg}
    $\g$ is Lipschitz continuous, that is, for all $\bs{u},\bs{u}' \in \R^d$, there exists some constant $L_{\g} >0$, such that
    \begin{equation}
        \|\g(\bs{u})-\g(\bs{u}')\| \leq L_{\g} \|\bs{u}-\bs{u}'\|.
        \label{eq:ass_3}
    \end{equation}
\end{assumption}
\begin{remark}
The vector field $h:\mathbb{R}^d\rightarrow \mathbb{R}^d$ in~\eqref{eq:ode} must be at least $C_b^{p+2}(\mathbb{R}^d,\mathbb{R}^d)$ for Assumption~\ref{ass:1_beg}  to hold. If $h\in C_b^{p+2}(\mathbb{R}^d,\mathbb{R}^d)$, the flow is Lipschitz by Gr\"onwall's lemma and so is the exact solver $\f$, meaning that Assumptions~\ref{ass:1_beg}-\ref{ass:3_beg} immediately hold.
\end{remark}
\begin{remark}
\label{common_remark}
Under Assumption~\ref{ass:1_beg}, let
$
c_1
:=
\max_{1\leq s\leq d}
\|D(c^{(p+1)})^{(s)}\|_{\infty}.
$
Then the correction function $f_c$ is Lipschitz continuous
(Assumption~\ref{ass:2_beg}) with Lipschitz constant satisfying
$
L_c
\leq
\sqrt d\,\left(c_1 + C_R\,\Delta t\right)\Delta t^{p+1},
$
where $C_R$ is as in Assumption~\ref{ass:1_beg}.
In particular, for sufficiently small $\Delta t$, $L_c$ is of order
$\Delta t^{p+1}$.
\end{remark}
Our final assumption is stated in terms of the fill distance, which we now define.

\begin{definition}[Fill distance or covering radius]
\label{def:fill distance}
Let $\mathcal{D}=\left\{\bs{u}_1, \ldots, \bs{u}_n\right\}   \subset \mathbb{R}^d$. The global fill distance for $\mathcal{D}\subset \mathcal{U}\subset \mathbb{R}^d$, that is, the largest distance from any point in $\mathcal{U}$ to its nearest point in $\mathcal{D}$, is defined as
\begin{equation}\label{globalfill}
h_{\mathcal{U},\mathcal{D}}:=\sup _{\bs{u} \in \mathcal{U}} \min _{\bs{u}_i \in \mathcal{D}}\left\|\bs{u}-\bs{u}_i\right\|.
\end{equation}
For a constant $\rho>0$, the local fill distance  at $\bs{u}' \in \mathcal{U}$ is defined as
\begin{equation}\label{localfill}
h_{\rho, \mathcal{D}}(\bs{u}'):=\sup _{\bs{u} \in B_{\rho}(\bs{u}')\cap \mathcal{U}} \min _{\bs{u}_i \in \mathcal{D}}\left\|\bs{u}-\bs{u}_i\right\|,
\end{equation}
where $B_\rho(\bs{u}')\subset \mathbb{R}^d$ denotes the ball of radius $\rho>0$
centered at $\bs{u}'$.
\end{definition}

\begin{assumption}[Posterior variance decay]
\label{ass:Posterior variance decay}
Let $K$ be a kernel on $\mathbb{R}^d$ and let ${\mathcal{H}}_{{K}}$ be the RKHS induced by it. For $s=1,\ldots,d$, let the $s$th coordinate of the correction function $f_c^{(s)}=(\f-\g)^{(s)} \in {\mathcal{H}}_{{K}}$, and let ${\sigma^{(s)}_{\mathcal{D}}}(\bs{u}')^2 \in \R$ be the posterior variance of a scalar-output GP built on a dataset $\mathcal{D}$ to approximate $f_c^{(s)}
\in {\mathcal{H}}_{{K}}$. We assume that there exist some constants $\beta>0$, $h_0>0$, and $\{C_{\beta, s}\}_{s=1}^d$, $C_{\beta, s}>0$ such that for any $\bs{u}' \in \mathbb{R}^d$ and any set of points $\mathcal{D}=\left\{\bs{u}_1, \ldots, \bs{u}_n\right\} \subset \mathbb{R}^d$ satisfying $h_{\rho, \mathcal{D}}(\bs{u}') \leq h_0$,  the GP regression posterior variance satisfies $\sigma_{\mathcal{D}}^{(s)}(\bs{u}')^2 \leq C_{\beta, s} h_{\rho, \mathcal{D}}(\bs{u}')^\beta$.
\end{assumption}
Additionally, it is convenient to introduce the following definition.
\begin{definition}[Maximum norm for vector-valued functions]
\label{def:max RKHS norm}
Let \((\mathcal{H}, \|\cdot\|_\mathcal{H})\) be a nor\-med vector space of scalar-valued functions. For a vector-valued function $f$ with the  $s$th coordinate \( f^{(s)} \in\mathcal{H} \) for all \(s = 1,\ldots,d \), define the maximum norm of \( f \) induced by \( \|\cdot\|_\mathcal{H} \) as $\| f \|_{\infty,\mathcal{H}} := \max_{1 \leq s \leq d} \| f^{(s)} \|_\mathcal{H}.$
\end{definition}
All assumptions are stated globally on $\mathbb{R}^d$ for notational simplicity, but may need to be imposed, instead, locally on a compact convex set $\Omega \subset \mathbb{R}^d$ with nonempty interior chosen large enough to contain the true trajectory $\{\bs u(t_i)\}$, the GParareal trajectory $\{\bs u^{\mathrm{GPara}}_{i,k}\}$, and the Prob-GParareal means $\{\bs \mu_{i,k}=\mathbb{E}[\bs U_{i,k}]\}$. The constants appearing in the bounds are then interpreted as local constants on $\Omega$, with the posterior variance and fill-distance assumptions imposed on a fixed neighborhood $\Omega_\rho:=\{\bs v\in\mathbb R^d \mid \exists \bs u\in\Omega
\text{ such that }\|\bs v - \bs u\|\le \rho\}$ of $\Omega$,  where $\rho > 0$ and one takes $\mathcal{U} = \Omega_\rho$ in  Definition~\ref{def:fill distance}. Note that for every evaluation point \(\bs u'\in\Omega\) appearing in the bounds, the ball \(B_\rho(\bs u')\) entering the local fill distance is contained
in \(\Omega_\rho\). The condition $f_c^{(s)}\in \mathcal{H}_K$ is then understood locally, by requiring the restrictions $f_c^{(s)}|_{\Omega_\rho}$ to admit extensions belonging to the corresponding RKHS on $\mathbb{R}^d$. 

\subsection{Prob-GParareal error bounds}\label{theoProbGP}

We are now ready to study the behavior of the probabilistic Prob-GParareal solution.
Consider $W_2( \delta_{\bs{u}(t_i)}, P_{\bs{U}_{i,k}} )^2$, the squared Wasserstein-2 distance between $\bs{u}(t_i)$, the true solution of~\eqref{eq:ode} at time $t_i$, and $P_{\bs{U}_{i,k}}$, the distribution of the Prob-GParareal solution at interval $i$ and iteration $k$. This quantity can be bounded as a function of the expected local fill distance, as stated below.
\begin{theorem}[Prob-GParareal error bound]
    \label{prop:convergence}

    Let $P_{\bs{U}_{0,k}}$ be an initial distribution satisfying
$\mathbb{E}[\bs{U}_{0,k}]=\bs{u}_{0,0}=\bs{u}_{(0)}$ and
$\operatorname{Var}(\bs{U}_{0,k})=\Sigma_{0,k}$, 
    $P_{\boldsymbol{U}_{i,k}}$ be the distribution of the Prob-GParareal solution at interval $i\in \{1,\ldots,N\}$ and iteration $k\in \mathbb{N}$, and $\delta_{\boldsymbol{u}(t_i)}$ be a Dirac measure centered at ${\boldsymbol{u}(t_i)}$, the true solution of the system \eqref{eq:ode} at time $t_i$. Let $K$ be a kernel on $\mathbb{R}^d$ and let ${\mathcal{H}}_{{K}}$ be the RKHS induced by this kernel. For $s=1,\ldots,d$, let the $s$th coordinate of the correction function $f_c^{(s)}=(\f-\g)^{(s)} \in {\mathcal{H}}_{{K}}$, and let ${\sigma^{(s)}_{\mathcal{D}_k}}^2$, be the posterior variance function of a scalar-output GP built on a dataset $\mathcal{D}_k$ to approximate $f_c^{(s)} \in {\mathcal{H}}_{{K}}$. Further, assume that one of the following holds:
    \begin{description}
    \item[(i)] {\bf Differentiability:} $K \in C_b^{2p
    +1}(\mathbb{R}^d \times \mathbb{R}^d)$, $
    p\geq 1$, and Assumptions~\ref{ass:1_beg}, \ref{ass:3_beg}, and  \ref{ass:Posterior variance decay} (with some $\beta>0$) hold.
\item[(ii)] {\bf Sobolev norm-equivalence:} $K$ is such that its induced RKHS ${\mathcal{H}}_{{K}}$ is norm-equivalent\footnote{Vector spaces $(\mathcal{H}_1,\| \cdot \|_{\mathcal{H}_1})$ and $(\mathcal{H}_2,\|\cdot \|_{\mathcal{H}_2})$ are called norm-equivalent, if $\mathcal{H}_1=\mathcal{H}_2$ as a set, and if there are constants $c_1, c_2>0$ such that $c_1\|f\|_{\mathcal{H}_2} \leq\|f\|_{\mathcal{H}_1} \leq c_2\|f\|_{\mathcal{H}_2}$ holds for all $f \in \mathcal{H}_1=\mathcal{H}_2$.} to $(W^q_2(\mathbb{R}^d), \|\cdot\|_{W^q_2})$, the Sobolev space of order $q>d/2$, and Assumptions~\ref{ass:1_beg} and~\ref{ass:3_beg} hold. 
\item[(iii)] {\bf Smoothness:} $K$ is infinitely smooth and Assumptions~\ref{ass:1_beg} and \ref{ass:3_beg} hold.
\end{description}
Additionally, define \begin{equation}
    \label{a def}
    a := 4\left(L_{\g}^2 + L_c^2 \right).
\end{equation}
Then, for all $i=1,\ldots, N$, and  $k \in \mathbb{N}$, it holds that
\begin{equation}
\label{eq:mse_w2}
    W_2( \delta_{\bs{u}(t_i)}, P_{\bs{U}_{i,k}} )^2 = \mathbb{E} \left[ \| \bs{u}(t_i) - \bs{U}_{i,k} \|^2 \right],
\end{equation}
i.e., the $W_2^2$ distance equals the mean squared error (MSE) of $\bs{U}_{i,k}$, and
\begin{align}
\label{bound error}
     W_2( \delta_{\bs{u}(t_i)}, P_{\bs{U}_{i,k}} )^2 \leq
     a^i \operatorname{tr}(\Sigma_{0,k})
     +
     \sum_{j=1}^{i} a^{i-j}b_{j-1,k},
\end{align}
where $b_{l,k}$, $l=0,\ldots,N-1$, is defined separately for each of the cases {\bf (i)}-{\bf (iii)} as follows:
    \begin{description}
    \item[(i)] {\bf Differentiability:}
\begin{align*}
b_{l,k} &= b_{l,k}(\beta) := 4d \, C_{\beta} (1+
              \| f_c \|_{\infty,{\mathcal{H}_{{K}}}}^2) \mathbb{E} \left[h_{\rho, \mathcal{D}_k}(\bs{U}_{l,k})^\beta\right],
\end{align*}
with $C_{\beta}=\max_{1\leq s\leq d}C_{\beta,s}$
and $\{C_{\beta,s}\}_{s=1}^d$, $C_{\beta,s}>0$, as in Assumption~\ref{ass:Posterior variance decay}.
\item[(ii)] {\bf Sobolev norm-equivalence:}
\begin{align*}
b_{l,k} &= 4d \, C (1+\| f_c \|_{\infty,W^q_2}^2) \mathbb{E} \left[h_{\rho, \mathcal{D}_k}(\bs{U}_{l,k})^{2q-d}\right],
\end{align*}
with
$C=\max_{1\leq s\leq d}C_s$ and
$\{C_s\}_{s=1}^d$, $C_s>0$, defined in Theorem~5.4 in~\cite{kanagawa2018gaussian} (reported in part {\bf (i)} in~\Cref{th:posterior variance p smooth} in Appendix \ref{pretheory} for convenience).
\item[(iii)] {\bf Smoothness:}
\begin{align*}
b_{l,k} &= b_{l,k}(\alpha) := 4d \, C_{\alpha} (1+
              \| f_c \|_{\infty,{\mathcal{H}_{{K}}}}^2) \mathbb{E} \left[h_{\rho, \mathcal{D}_k}(\bs{U}_{l,k})^{\alpha} \right],
\end{align*}
with
$C_{\alpha}=\max_{1\leq s\leq d}C_{\alpha,s}$
and $\{C_{\alpha,s}\}_{s=1}^d$, $C_{\alpha,s}>0$, defined in Theorem 5.14 in~\cite{wu1993local} (reported in part {\bf (ii)} in~\Cref{th:posterior variance p smooth} in Appendix \ref{pretheory} for convenience).
\end{description}
In all cases, $h_{\rho, \mathcal{D}_k}$ is the local fill distance defined in~\eqref{localfill}.

\end{theorem}

\begin{remark}
\label{sobolev_remark}
In {\bf (ii)}, whenever $q>d/2+1$, the Sobolev embedding
$W^q_2(\mathbb{R}^d)\hookrightarrow C^1_b(\mathbb{R}^d)$ holds.
Since $f_c^{(s)}\in\mathcal{H}_K\simeq W^q_2(\mathbb{R}^d)$
for each $s=1,\ldots,d$, it follows that
$f_c^{(s)}\in C^1_b(\mathbb{R}^d,\mathbb{R})$ for each
coordinate. Hence $f_c$ is globally Lipschitz,
Assumption~\ref{ass:2_beg} is automatically satisfied with
$L_c \leq \sqrt{d}\,\max_{1\leq s\leq d}\|Df_c^{(s)}\|_{\infty}$, and Assumption~\ref{ass:1_beg} is not needed to obtain
Lipschitz continuity of $f_c$.
\end{remark}
\begin{remark}
Assumption~\ref{ass:1_beg} is listed in cases {\bf (i)}-{\bf (iii)} to obtain
Lipschitz continuity of $f_c$ via Remark~\ref{common_remark} and could be replaced by
Assumption~\ref{ass:2_beg}. However, Assumption~\ref{ass:1_beg} additionally
yields the explicit rate
$
L_c \leq \sqrt{d}\,(c_1 + C_R\,\Delta t)\,\Delta t^{p+1},
$
which connects the bound via \eqref{a def} to the order
$p$ of the coarse solver $\mathscr{G}$.
\end{remark}
\begin{remark}[RKHS membership and kernel choice]
\label{rem:rkhs}
The condition $f_c^{(s)}\in \mathcal{H}_K$ is used to construct the error bounds and is not required to run Prob-GParareal. Assumption~\ref{ass:1_beg} implies global Lipschitz continuity of $f_c$ (see Remark \ref{common_remark}), but not RKHS membership, which requires additional regularity of $h$ and of the one-step solvers. For Mat\'ern-$\nu$ kernels, membership reduces to
$f_c^{(s)}\in W_2^{\nu+d/2}(\Omega)$ on the relevant compact region $\Omega$ of the state space, and can be verified through standard regularity assumptions on $h$ and on the one-step solvers. For Gaussian kernels, the condition is strictly stronger than analyticity, requiring Gaussian-weighted Fourier integrability of $f_c^{(s)}$. Hence, the Gaussian kernel should be understood as a practical high-regularity modeling choice. In either case, kernel hyperparameters can be selected by marginal likelihood optimization and assessed empirically through the observed decay of the posterior variance.
\end{remark}

The proof of~\Cref{prop:convergence} is provided in~\Cref{app:pr:conv}.

\begin{corollary}[Uniform Prob-GParareal error bound]
\label{cor:uniform}
Under the same Assumptions of Theorem\,\ref{prop:convergence}, fix an iteration $k \geq 1$ and suppose there exists a constant $B_k\ge0$ such that $b_{l,k}\;\le\;B_k$, for all $l=0,\dots,N-1$.
Then,
it holds that
\begin{equation}
\label{eq:uniform-bound}
W_2\bigl(\delta_{\bs{u}(t_i)},P_{\bs{U}_{i,k}}\bigr)^2
\;\le\;
a^i\operatorname{tr}(\Sigma_{0,k})+
B_k\;\frac{1 - a^i}{1 - a},
\end{equation}
where $a$ is defined in~\eqref{a def} and satisfies $a \neq 1$.
\end{corollary}

\begin{remark}
The bound~\eqref{bound error} depends on the initial condition variance, $\Sigma_{0,k}$, with $\Sigma_{0,k}$ equal to the zero matrix when the initial condition $\bs{u}_{(0)}$ is deterministic ($P_{\bs{U}_{0,k}}=\delta_{\bs{u}_{0,k}}$), and on the expected local fill distance $h_{\rho,\mathcal{D}_k}$ through the coefficients $b_{l,k}$. Similar fill-distance bounds appear in the deterministic (nn)GParareal literature~\citep{pentland2023gparareal,nngparareal}. In the contractive regime, i.e. $a<1$, the first term decays exponentially in $i$ and $\;W_2(\delta_{\bs{u}(t_i)},P_{\bs{U}_{i,k}})^2$ is controlled by a geometric sum.  When instead $a>1$, the worst case bound would grow exponentially in $i$. However, empirical evidence shows that the coefficients $b_{l,k}$ decay exponentially with the iteration number $k$. This behavior is driven by the progressive refinement of the Prob-GParareal solution, which increases the informativeness of the training dataset $\mathcal{D}_k$ and induces a corresponding decay in the expected local fill distance $h_{\rho,\mathcal{D}_k}$. Although the theoretical analysis of this decay as a function of $k$ is challenging, \Cref{app:fill_dist} presents empirical evidence of its exponential decay, mitigating the exponential growth term. An empirical assessment of the sharpness of the bounds in \Cref{prop:convergence} and Corollary~\ref{cor:uniform} is provided in Appendix~\ref{AppendixH} for the deterministic initial condition case, where we compare the Wasserstein error with the corresponding fill-distance terms in the coefficients \(b_{i,k}\).
\end{remark}

After having bounded the Prob-GParareal error with respect to the true solution, we now draw
a connection between the Prob-GParareal solution (in particular, its mean) and its deterministic counterpart, the GParareal solution $\bs{u}_{i,k}^{\textrm{GPara}}$
obtained by sequential applications of $\g$ corrected by the GP posterior mean $\widehat{f}_{\rm GPara}(\cdot)=\bs{\mu}_{\mathcal{D}_k}(\cdot)$ as in~\eqref{eq:update_rule_gpara}, namely
\[
\bs{u}_{i,k}^{\rm GPara} := \underbrace{\left((\g + \bs{\mu}_{\mathcal{D}_k}){\circ}(\g + \bs{\mu}_{\mathcal{D}_k})\circ\cdots \circ (\g + \bs{\mu}_{\mathcal{D}_k})\right)}_{i \text{ times}}(\bs{u}_{0,k}^{\rm GPara}), \quad i = 1,\ldots,N, \enspace k\in \mathbb{N},
\]
with $\bs{u}_{0,k}^{\rm GPara} = \bs{u}_{(0)}$.
In particular, in~\Cref{prop:meanbound}, we derive an error bound between the Prob-GParareal mean and the GParareal solution as a function of the maximum variance of $\bs{U}_{i,k}$ over the $d$ dimensions. While such variance

is generally unknown, we provide an explicit bound in the following \Cref{th:probpara:var:bound}.
 \begin{theorem}[Variance bound for $\bs{U}_{i,k}$]
 \label{th:probpara:var:bound}
 Let the conditions of~\Cref{prop:convergence} hold. Let $\bs{U}_{i,k}$ be the Prob-GParareal solution at the $i$th interval, $i\in\{1, \ldots, N\}$, and $k$th iteration, $k\in \mathbb{N}$, with mean $\boldsymbol{\mu}^{(s)}_{i,k}
=\mathbb{E}[\bs{U}_{i,k}^{(s)}]$ and variance $\sigma^{(s),2}_{i,k}={\rm Var}(\bs{U}_{i,k}^{(s)})$ of its  $s$th coordinate, $s=1,\ldots, d$. Denote the maximum variance of $\bs{U}_{i,k}$ as
 \begin{equation}
     \label{max var}
\sigma^{{\rm max},2}_{{i,k}}:=\max_{1\leq s\leq d} \sigma_{i,k}^{(s),2}, \quad i=1,\ldots,N, \enspace k\in \mathbb{N}.
 \end{equation}

Assume that one of the {\bf (i)}-{\bf (iii)} cases of~\Cref{prop:convergence} holds, and define
\begin{equation}
    \label{a def var}
    a := 2 \; d(L_\g^2 + 3L_c^2).
\end{equation}
Then, for all $i=1,\ldots, N$, and  $k \in \mathbb{N}$, it holds that
\begin{equation}
\label{bound variance}
\sigma^{{\rm max},2}_{i,k} \le a^i\,\sigma^{{\rm max},2}_{0,k} + \sum_{j=1}^{\,i}a^{\,i-j}\,b_{j-1,k},
\end{equation}
where $b_{l,k}$, $l=0,\ldots,N-1$, is defined separately for each of the cases {\bf (i)}-{\bf (iii)} in~\Cref{prop:convergence} as follows:
\begin{description}
\item[(i)] {\bf Differentiability:}
\begin{align*}
b_{l,k} &= C_{\beta} \mathbb{E}\left[ h_{\rho,\mathcal{D}_{k}}(\bs{U}_{l,k})^\beta \right] + 6 C_{\beta} \| f_c \|^2_{\infty,\mathcal{H}_{{K}}} \left\{\mathbb{E}\left[ h_{\rho,\mathcal{D}_{k}}(\bs{U}_{l,k})^{\beta} \right]  +  h_{\rho,\mathcal{D}_{k}}(\boldsymbol{\mu}_{l,k})^{\beta}\right\},
\end{align*}
with
$C_{\beta}$
defined as in
part {\bf(i)} of~\Cref{prop:convergence}.
\item[(ii)] {\bf Sobolev norm-equivalence:}
\begin{align*}
b_{l,k} &= C \mathbb{E}\left[ h_{\rho,\mathcal{D}_{k}}(\bs{U}_{l,k})^{2q-d} \right] + 6 C \| f_c \|^2_{\infty,W^q_2} \left\{\mathbb{E}\left[ h_{\rho,\mathcal{D}_{k}}(\bs{U}_{l,k})^{2q-d} \right]  +  h_{\rho,\mathcal{D}_{k}}(\boldsymbol{\mu}_{l,k})^{2q-d}\right\}
\end{align*}
with $C$
defined as in part {\bf(ii)} of~\Cref{prop:convergence}.
    \item[(iii)] {\bf Smoothness:}
\begin{align*}
b_{l,k} &= C_{\alpha} \mathbb{E}\left[ h_{\rho,\mathcal{D}_{k}}(\bs{U}_{l,k})^\alpha \right] + 6 C_{\alpha} \| f_c \|^2_{\infty,\mathcal{H}_{{K}}} \left\{\mathbb{E}\left[ h_{\rho,\mathcal{D}_{k}}(\bs{U}_{l,k})^{\alpha} \right]  +  h_{\rho,\mathcal{D}_{k}}(\boldsymbol{\mu}_{l,k})^{\alpha}\right\},
\end{align*}
with $C_{\alpha}$
defined as in part {\bf(iii)} of~\Cref{prop:convergence}.
\end{description}
In addition, Remark \ref{common_remark} holds under the assumptions of this theorem.
  \end{theorem}
 The proof of this theorem is provided in~\Cref{app:pr:var_bound}.
\begin{corollary}[Uniform variance bound for $\bs{U}_{i,k}$]
\label{cor:uniform_var}
Under the same assumptions as in Theorem~\ref{th:probpara:var:bound}, fix an iteration $k \in \mathbb{N}$, and suppose there exists a constant $B_k \ge 0$ such that  $b_{l,k} \le B_k$ for all  $l=0,\dots,N-1$.
Then, it holds that
\begin{equation}
\label{eq:uniform-bound-var}
\sigma^{{\rm max},2}_{{i,k}} \leq B_k\frac{1-a^i}{1- a } + a^i \sigma^{{\rm max},2}_{{0,k}}, \quad i=1,\ldots,N, \enspace k\in \mathbb{N},
\end{equation}
where $a$ is defined in~\eqref{a def var} and satisfies $a \neq 1$.
\end{corollary}

 After having bounded $\sigma_{i,k}^{\rm{max},2}$, we now present a bound on the difference between the mean of the Prob-GParareal solution and its deterministic counterpart, the GParareal solution, as a function of such quantity. An error bound between the mean of Prob-GParareal and the true solution will then follow.

\begin{theorem}[Mean error bound with respect to GParareal]
\label{prop:meanbound}
    Let $\sigma^{{\rm max},2}_{{i,k}}$  be the maximum coordinate-wise variance of the Prob-GParareal solution $\bs{U}_{i,k}$ defined as in~\eqref{max var} and let $\bs{\mu}_{\mathcal{D}_k}$ be the GP posterior mean computed on the dataset $\mathcal{D}_k$ at interval $i\in \{1, \ldots, N\}$ and iteration $k\in \mathbb{N}$. Let $(\g+\bs{\mu}_{\mathcal{D}_k})\in C^2(\mathbb{R}^d,\mathbb{R}^d)$ and $H_{(\g + \bs{\mu}_{\mathcal{D}_k})^{(s)}}(\bs{U})$ be the Hessian matrix of $(\g + \bs{\mu}_{\mathcal{D}_k})^{(s)}$, $s=1,\ldots,d$, evaluated at $\bs{U}\in \R^d$. Assume there exists a constant $M_s>0$ such that
$\|H_{(\g + \bs{\mu}_{\mathcal{D}_k})^{(s)}}(\bs{U})\| \leq M_s$
for all $\bs{U}\in \R^d$. Define $M := \max_{1\leq s \leq d} M_s$.  Let either of the cases {\bf (i)}, {\bf (ii)}, or {\bf (iii)} of~\Cref{prop:convergence} holds, and define
\begin{equation}
    \label{a def mean}
    a := L_\g + L_c.
\end{equation}
Then, for all $i=1,\ldots, N$, and  $k \in \mathbb{N}$ it holds that
\begin{equation}
\label{mean bound}
\|\bs{\mu}_{i,k}- \bs{u}_{i,k}^{\rm GPara}\| \leq \sum_{j=1}^{i} a^{i-j} b_{j-1, k},\end{equation}
where $b_{l,k}$, $l=0,\ldots,N-1$, is defined separately for each of the cases {\bf (i)}-{\bf (iii)} in~\Cref{prop:convergence} as follows:
\begin{description}
\item[(i)] {\bf Differentiability:}
\begin{align*}b_{l,k} = \frac{M \, d^{3/2}}{2} \, \sigma^{{\rm max},2}_{{l,k}}
+ \sqrt{C_\beta d} \| f_c \|_{\infty,{\mathcal{H}_{{K}}}}
\left(h_{\rho, \mathcal{D}_k}(\boldsymbol{\mu}_{l,k}) ^{\beta/2}
+ h_{\rho, \mathcal{D}_k}(\bs{u}_{l,k}^{\rm GPara})^{\beta/2}\right)\end{align*}
with $C_{\beta}$
defined as in part {\bf(i)} of~\Cref{prop:convergence} for $\beta>0$.
\item[(ii)] {\bf Sobolev norm-equivalence:}
\begin{align*}
b_{l,k} = \frac{M \, d^{3/2}}{2} \, \sigma^{{\rm max},2}_{{l,k}}  + \sqrt{C d} \| f_c \|_{\infty,W^q_2} \left(h_{\rho, \mathcal{D}_k}(\boldsymbol{\mu}_{l,k}) ^{q-d/2} + h_{\rho, \mathcal{D}_k}(\bs{u}_{l,k}^{\rm GPara})^{q-d/2}\right).
\end{align*}
with $C$
defined as in part {\bf(ii)} of~\Cref{prop:convergence}.
\item[(iii)] {\bf Smoothness:}
\begin{align*}
b_{l,k} = \frac{M \, d^{3/2}}{2} \, \sigma^{{\rm max},2}_{{l,k}} + \sqrt{C_\alpha d} \| f_c \|_{\infty,{\mathcal{H}_{{K}}}} \left(h_{\rho, \mathcal{D}_k}(\boldsymbol{\mu}_{l,k}) ^{\alpha/2} + h_{\rho, \mathcal{D}_k}(\bs{u}_{l,k}^{\rm GPara})^{\alpha/2}\right),
\end{align*}
with $C_\alpha$ defined as in part {\bf (iii)} of~\Cref{prop:convergence} for $\alpha>0$.
\end{description}
In addition, Remark \ref{common_remark} holds under the assumptions of this theorem.
 \end{theorem}

The proof of this theorem is provided in~\Cref{app:pr:mean_bound}. Following the same approach as in~Corollary~\ref{cor:uniform}, we simplify~\eqref{mean bound} by assuming the existence of a constant $B_k \ge 0$ such that $b_{l,k} \le B_k$; this is presented in~Corollary~\ref{cor:uniform_mean} below. Alternatively, $\sigma^{{\rm max},2}_{{l,k}}$ in $b_{l,k}$ can be replaced by the variance bound established in~\Cref{th:probpara:var:bound}, allowing~\eqref{mean bound} to be expressed in terms of the variance of the initial condition, $\sigma^{\max,2}_{0,k}$, which is typically known.

\begin{corollary}[Uniform mean error bound]
\label{cor:uniform_mean}
Under the same assumptions as in~\Cref{prop:meanbound}, fix an iteration $k \geq 1$ and suppose there exists a constant $B_k \ge 0$ such that $b_{l,k} \le B_k$, for all $l=0,\dots,N-1$.
Then, it holds that
\begin{equation}
\label{eq:uniform-bound-mean}
\left\| \bs{\mu}_{i,k}- \bs{u}_{i,k}^{\rm GPara} \right\| \leq B_k \frac{1- a^i}{1-a}, \quad i=1,\ldots,N, \enspace k\in \mathbb{N},
\end{equation}
where $a$ is defined in~\eqref{a def mean} and satisfies $a \neq 1$.
\end{corollary}

\begin{corollary}[Explicit mean-error bound]
\label{cor:mean_full_expr}
Under the assumptions of \Cref{th:probpara:var:bound} and  \Cref{prop:meanbound}, let
$$
a \;=\;2\,d\,(L_{\g}^{2}+3L_{c}^{2}),
\quad
\tilde a \;=\;L_{\g}+L_{c},
$$
and for each of the cases {\bf (i)}-{\bf (iii)}, let $b_{l,k}$, $l=0,\ldots,N-1$, be as in~\Cref{th:probpara:var:bound}. If $a \neq \tilde{a}$,
then for every $i=1,\dots,N$ and $k\in\mathbb N$, the result of  \Cref{prop:meanbound} simplifies to
\[
\|\boldsymbol{\mu}_{i,k}-\bs{u}_{i,k}^{\rm GPara}\|
\le
 \frac{M  d^{3/2}}{2({\tilde a - a})}\left(\sigma^{\max,2}_{0,k}\;
  \;(\tilde a ^i - a^i)
+ \sum_{j=1}^{i-1} b_{j-1,k} (\tilde a^{\,i-j}-a^{\,i-j})\right)
+r_{i,k}.
\]
If $a=\tilde a$, the same bound holds with the first term replaced by its limiting value, namely
\[
\|\boldsymbol{\mu}_{i,k}-\bs{u}_{i,k}^{\rm GPara}\|
\le
 \frac{M d^{3/2}}{2}\left(
 \sigma^{\max,2}_{0,k}\, i a^{i-1}
+ \sum_{j=1}^{i-1} b_{j-1,k} (i-j)a^{i-j-1}
\right)
+r_{i,k}.
\]
 In these bounds $r_{i,k}$ is defined separately for each of the cases {\bf (i)}-{\bf (iii)} in~\Cref{prop:convergence} as follows:
\begin{description}
\item[(i)] {\bf Differentiability:}
\[
r_{i,k} = \sqrt{C_\beta d} \| f_c \|_{\infty,{\mathcal{H}_{{K}}}}\sum_{j=1}^i \tilde a^{\,i-j}
   \bigl(h_{\rho,\mathcal{D}_k}(\boldsymbol{\mu}_{j-1,k})^{\beta/2} + h_{\rho,\mathcal{D}_k}(\bs{u}_{j-1,k}^{\rm GPara})^{\beta/2}\bigr),
\]
with $C_{\beta}$
defined as in part {\bf(i)} of~\Cref{prop:convergence} for $\beta>0$.
\item[(ii)] {\bf Sobolev norm-equivalence:}
\[
r_{i,k} = \sqrt{C d} \| f_c \|_{\infty,W^q_2}\sum_{j=1}^i \tilde a^{\,i-j}
   \bigl(h_{\rho,\mathcal{D}_k}(\boldsymbol{\mu}_{j-1,k})^{q-d/2} + h_{\rho,\mathcal{D}_k}(\bs{u}_{j-1,k}^{\rm GPara})^{q-d/2}\bigr),
\]
with
$C$
defined as in part {\bf(ii)} of~\Cref{prop:convergence}.
    \item[(iii)] {\bf Smoothness:}
\begin{align*}
r_{i,k} &=  \sqrt{C_\alpha d}\| f_c \|_{\infty,{\mathcal{H}_{{K}}}}\sum_{j=1}^i \tilde a^{\,i-j}
   \bigl(h_{\rho,\mathcal{D}_k}(\boldsymbol{\mu}_{j-1,k})^{\alpha/2} + h_{\rho,\mathcal{D}_k}(\bs{u}_{j-1,k}^{\rm GPara})^{\alpha/2}\bigr),
\end{align*}
with
$C_{\alpha}$
defined as in part {\bf(iii)} of~\Cref{prop:convergence} for $\alpha>0$.
\end{description}
\end{corollary}
 Proof is provided in~\Cref{app:pr:mean_bound}.
 \begin{remark}
 \label{remark:bias}
We can bound the error between the mean $\bs{\mu}_{i,k}$ of the Prob-GParareal solution $\bs{U}_{i,k}$ and the true solution $\bs{u}(t_i)$, that is  the bias of $\bs{U}_{i,k}$, using the triangular inequality
\begin{equation}
    \label{eq:bias}
     \|\bs{\mu}_{i,k}- \bs{u}(t_i)\| \leq \|\bs{\mu}_{i,k}- \bs{u}_{i,k}^{\rm GPara}\|
 + \|\bs{u}_{i,k}^{\rm GPara} - \bs{u}(t_i)\|,
\end{equation}
where the first term is bounded by \Cref{prop:meanbound} or Corollaries~\ref{cor:uniform_mean} and \ref{cor:mean_full_expr}, and the second term by Theorem 3.4 in~\cite{pentland2023gparareal}. Combining \eqref{eq:bias} (squared) with the variance bound in Theorem~\ref{th:probpara:var:bound} gives an upper bound of the MSE of $\bs{U}_{i,k}$, or, equivalently, by \eqref{eq:mse_w2} the upper bound of distance $W_2^2$.
 \end{remark}

\section{Empirical results for Prob-GParareal}
\label{sec:empirics}

In this section, we illustrate Prob-GParareal and its performance on five different ODE systems with deterministic initial conditions (except in~\Cref{sec:prob_init_cond}, with random initial conditions), four of which were considered for GParareal~\citep{pentland2023gparareal}, allowing for a direct comparison. These include the FitzHugh-Nagumo (FHN) model, a two-dimensional representation of an animal nerve axon~\citep{nagumo1962active}; the R\"ossler system, a three-dimensional chaotic model for turbulence~\citep{rossler1976equation}; the nonlinear Hopf bifurcation model, a two-dimensional non-autonomous ODE~\citep{seydel2009practical}; and the double pendulum, a four-dimensional system describing the dynamics of two connected pendula~\citep{pettet1999computer}. The fifth added model is the Lorenz system, a commonly studied chaotic system used in simplified weather modeling~\citep{lorenz1963deterministic}. Overall, these systems offer a wide range of complexities, from stiff to non-autonomous and chaotic dynamics, to assess the performance of Prob-GParareal. For a detailed description of the models, we refer to~\cite{pentland2023gparareal},~\cite{nngparareal}, and the original references. A summary of the simulation setup, including the coarse and fine solvers, initial conditions, evolution timespans, and other relevant parameters required to replicate the results is reported in~\Cref{tab:simsetup} in~\Cref{app:setup}. Finally, a comparison with the parallel-in-time probabilistic ODE solver of \citet{bosch2024parallel} is reported in \Cref{app:bosch_comparison}. As previously done in~\cite{nngparareal,randnet_parareal}, we rescale each ODE by a change of variables such that each coordinate lies within $[-1,1]$. This normalization step does not alter the system's properties, helps stabilize the GP learning, and facilitates comparisons across systems with different scales.

\subsection{
Prob-GParareal accuracy and probabilistic forecast}
\label{sec:prob_forecast}

Prob-GParareal produces a \textit{probabilistic} forecast for the evolution of a DE system - a sequence of predicted distributions $P_{\bs{U}_{i,K_{\rm end}}}$ for the target $\bs{u}(t_i)$, $i=1,\ldots,N$, - as opposed to the single-point forecast of classical deterministic DE solvers. These distributions are then approximated using their empirical counterparts $\widehat{P}_{\bs{U}_{i,K_{\rm end}}}$, based on the $n$ drawn samples in $\mathcal{U}_{i,K_{\rm end}}$, where, $K_{\rm end}$ denotes the final Prob-GParareal iteration, when the algorithm either converges (so $K_{\rm end}=K_{\rm conv})$ or is stopped due to early termination (so $K_{\rm end}=K_{\rm stop})$.

Evaluating probabilistic forecasts requires metrics that capture the similarity of the predicted distribution with empirical data in terms of both \textit{calibration} and \textit{sharpness}~\citep{gneiting2007strictly}.
 Calibration refers to the statistical consistency  between predicted probabilities and observed event frequencies, indicating the reliability of a forecast. Sharpness measures the concentration of the predictive distribution, with sharper (more concentrated) distributions being preferable when well-calibrated.
Proper scoring rules, extensively studied in the literature~\citep{gneiting2007strictly, brocker2007scoring, gneiting2014probabilistic} and widely adopted across disciplines~\citep{galbraith2012assessing, dumas2022deep, lerch2017forecaster, bogner2016post}, are a natural choice for evaluating probabilistic forecasts, as they balance sharpness and calibration. Comparative studies~\citep{pic2024proper,alexander2024evaluating,ziel2019multivariate} emphasize the importance of employing multiple scoring rules, as each metric highlights distinct distributional characteristics. For instance, for $i=1,\ldots,N$, $k \geq 1$, and $s=1,\ldots,d$, the energy score (ES)~\citep{gneiting2007probabilistic}, defined as
\[
\text{ES}(\mathcal{U}_{i,k}, \bs{u}(t_i))
= \frac{1}{n} \sum_{j=1}^{n} \|\bs{u}_{i,k}^{(j)} - \bs{u}(t_i)\|
- \frac{1}{2n^2}\sum_{j=1}^{n}\sum_{l=1}^{n}\|\bs{u}_{i,k}^{(j)} - \bs{u}_{i,k}^{(l)}\|,
\]
prioritizes mean accuracy over capturing covariance structure, while the variogram score (VS)~\citep{scheuerer2015variogram}
\[
\text{VS}(\mathcal{U}_{i,k}, \bs{u}(t_i))
= \sum_{s_1,s_2=1}^{d} w_{s_1, s_2} \left(\frac{1}{n}\sum_{j=1}^{n}\left|\bs{u}_{i,k}^{(j)(s_1)}-\bs{u}_{i,k}^{(j)(s_2)}\right|^{p}-\left|\bs{u}(t_i)^{(s_1)}-\bs{u}(t_i)^{(s_2)}\right|^{p}\right)^2
\]
addresses dependence structures more effectively. Here, $\bs{u}_{i,k}^{(j)(s)}$ denotes the $s$th coordinate of the $j$th forecast sample, $\bs{u}(t_i)^{(s)}$ the $s$th coordinate of the observed solution at time $t_i$, and $w_{s_1,s_2}\ge 0$ are user-specified weights that determine the contribution of each coordinate pair. In our setting, we take $w_{s_1,s_2}=1$ for all pairs. Following common practice~\citep{alexander2024evaluating}, we set $p=0.5$, which has been shown to discriminate between forecasts of different dependence structures effectively. Henceforth, we omit the subscript $p$ for simplicity.
To capture the temporal performance of probabilistic forecasts, both ES and VS are averaged over time intervals as
\[
\text{ES}_k = \frac{1}{N}\sum_{i=1}^{N}\text{ES}(\mathcal{U}_{i,k}, \bs{u}(t_i)) \qquad \textrm{and} \qquad
\text{VS}_k = \frac{1}{N}\sum_{i=1}^{N}\text{VS}(\mathcal{U}_{i,k}, \bs{u}(t_i)).
\]
Although ES and VS quantify forecast quality, their interpretation can be challenging. Therefore, we additionally employ the mean absolute deviation (MAD) of the observed solution from forecasted confidence intervals (CIs)~\citep{winkler1972decision}, defined as:
\[
\text{MAD}_{k}=\frac{1}{N}\sum_{i=1}^N \text{MAD}(\mathcal{U}_{i,k}, \bs{u}(t_i))
\]
with
\[    \text{MAD}(\mathcal{U}_{i,k}, \bs{u}(t_i))
    =\frac{1}{d} \sum_{s=1}^{d}
    \left( \max\,\bigl(\bs{u}(t_i)^{(s)} - \bs{B}_{i,k}^{(s)}, 0\bigr)
          + \max\,\bigl(\bs{A}_{i,k}^{(s)} - \bs{u}(t_i)^{(s)}, 0\bigr) \right),
\]
where $\bs{A}_{i,k}^{(s)}$ and $\bs{B}_{i,k}^{(s)}$ represent the lower and upper bounds, respectively, of a one-dimensional two-sided symmetric 95\% CI of the $s$th coordinate, thus neglecting the dependence structure.
Although MAD captures forecast calibration in a straightforward manner, it does not explicitly consider sharpness, and loses distributional information by summarizing forecasts with CIs. Nonetheless, it serves as a baseline measure complementary to ES and VS, offering more intuitive interpretability.

Finally, given the equality between $W_2( \delta_{\bs{u}(t_i)}, P_{\bs{U}_{i,k}} )^2$ and the MSE of $\bs{U}_{i,k}$ shown in \Cref{prop:convergence},  we include the MSE as evaluation metric alongside those presented above. We use a similar notation where ${\rm MSE}_{i,k}$ denotes the MSE of $\bs{U}_{i,k}$ computed using the intermediate solution at iteration $k$ (e.g. $\mathcal{U}_{i,k}$ for Prob-GParareal), and ${\rm MSE}_k$ is its average over the intervals. Moreover, in \Cref{fig:rmse_comparison} and in \Cref{tab:burges}, we additionally show the bias, which we compute as $ {\rm Bias}_{i,k} = \|\bs{\bar{u}}_{i,k}- \bs{u}(t_i)\| $ where $\bs{\bar{u}}_{i,k}$ is the sample mean of $\mathcal{U}_{i,k}$ (i.e. the sample estimate of $\mathbb{E}[\bs{U}_{i,k}]$).
Note that the MSE and the bias squared coincide for all algorithms but Prob-(nn)GParareal, as they yield a deterministic solution.
\begin{table}[t]
\centering
{
\footnotesize
\begin{tabular}{lcccc}
\toprule
System & ${\rm MAD}_{K_{\rm conv}}$ & ${\rm VS}_{K_{\rm conv}}$ & ${\rm ES}_{K_{\rm conv}}$ & ${\rm MSE}_{K_{\rm conv}}$ \\
\midrule
FHN & 0 (0) & $2.9e^{-13}$ ($1.5e^{-13}$) & $7.9e^{-8}$ ($6.6e^{-9}$) & $2.1e^{-13}$ ($4.0e^{-14}$) \\
Hopf & $2.8e^{-7}$ ($3.9e^{-8}$) & $6.9e^{-9}$ ($8.5e^{-9}$) & $2.3e^{-5}$ ($9.2e^{-6}$) & $9.1e^{-11}$ ($4.8e^{-11}$) \\
\makecell{Dbl Pend.} & $3.9e^{-7}$ ($4.6e^{-7}$) & $2.3e^{-8}$ ($2.2e^{-8}$) & $1.2e^{-5}$ ($7.3e^{-6}$) & $1.3e^{-11}$ ($5.0e^{-12}$) \\
Lorenz & 0 (0) & $2.9e^{-6}$ ($4.6e^{-6}$) & $2.1e^{-4}$ ($4.6e^{-5}$) & $3.1e^{-11}$ ($1.5e^{-12}$) \\
R\"ossler & 0 (0) & $7.0e^{-8}$ ($2.6e^{-8}$) & $2.7e^{-5}$ ($2.7e^{-6}$) & $3.7e^{-11}$ ($1.4e^{-12}$) \\
\bottomrule
\end{tabular}

\vspace{1em}

\begin{tabular}{lcccc}
\toprule
System & $T_{\text{Prob-GPara}}$ & $T_{\text{GPara}}$ & $K_{\text{Prob-GPara}}$ & $K_{\rm GPara}$ \\
\midrule
FHN & 6s (0.3s) & 5s & 5 (0) & 5 \\
Hopf & 21s (1.0s) & 22s & 9 (0) & 10 \\
\makecell{Dbl Pend.} & 30s (1.4s) & 29s & 7.5 (0.5) & 10 \\
Lorenz & 80s (4.1s) & 46s & 13.6 (0.67) & 12 \\
R\"ossler & 41s (0.8s) & 37s & 6 (0) & 7 \\
\bottomrule
\end{tabular}
}
\caption
{Evaluation of the Prob-GParareal probabilistic forecast across various DEs. The reported metrics are averaged over ten independent runs of the simulations (launched with a different random seed to account for algorithmic randomness), with standard deviation in parentheses.
$T_\cdot$ and $K_\cdot$ denote the runtime and iterations to converge, respectively, for Prob-GParareal (Prob-GPara) and GParareal (GPara).
For Prob-GParareal (GParareal) an accuracy threshold of $\epsilon_{\text{Prob-GPara}} = 1e^{-7}$ ($\epsilon_{\rm GPara} = 5 e^{-6}$) is used for all systems except Lorenz, which required a higher precision, so $\epsilon_{\text{Prob-GPara}} = 1e^{-9}$ ($\epsilon_{\rm GPara} = 5e^{-9}$) was taken. The number of samples is $n = 5000$ for all systems.
}
\label{tab:sim_res}
\end{table}

\subsubsection{Performance of Prob-GParareal}
\Cref{tab:sim_res} summarizes the performance of Prob-GParareal across various ODE systems evaluated using the metrics described in~\Cref{sec:prob_forecast}, and reports the runtimes $T$ and the number of iterations to converge $K_{\rm conv}$ of Prob-GParareal
and GParareal
for comparison. As reported in the acknowledgments, all simulations were executed on Warwick University High Performance Computing (HPC) facilities.

\begin{figure}[t]
    \centering
    \includegraphics[width=1\linewidth]{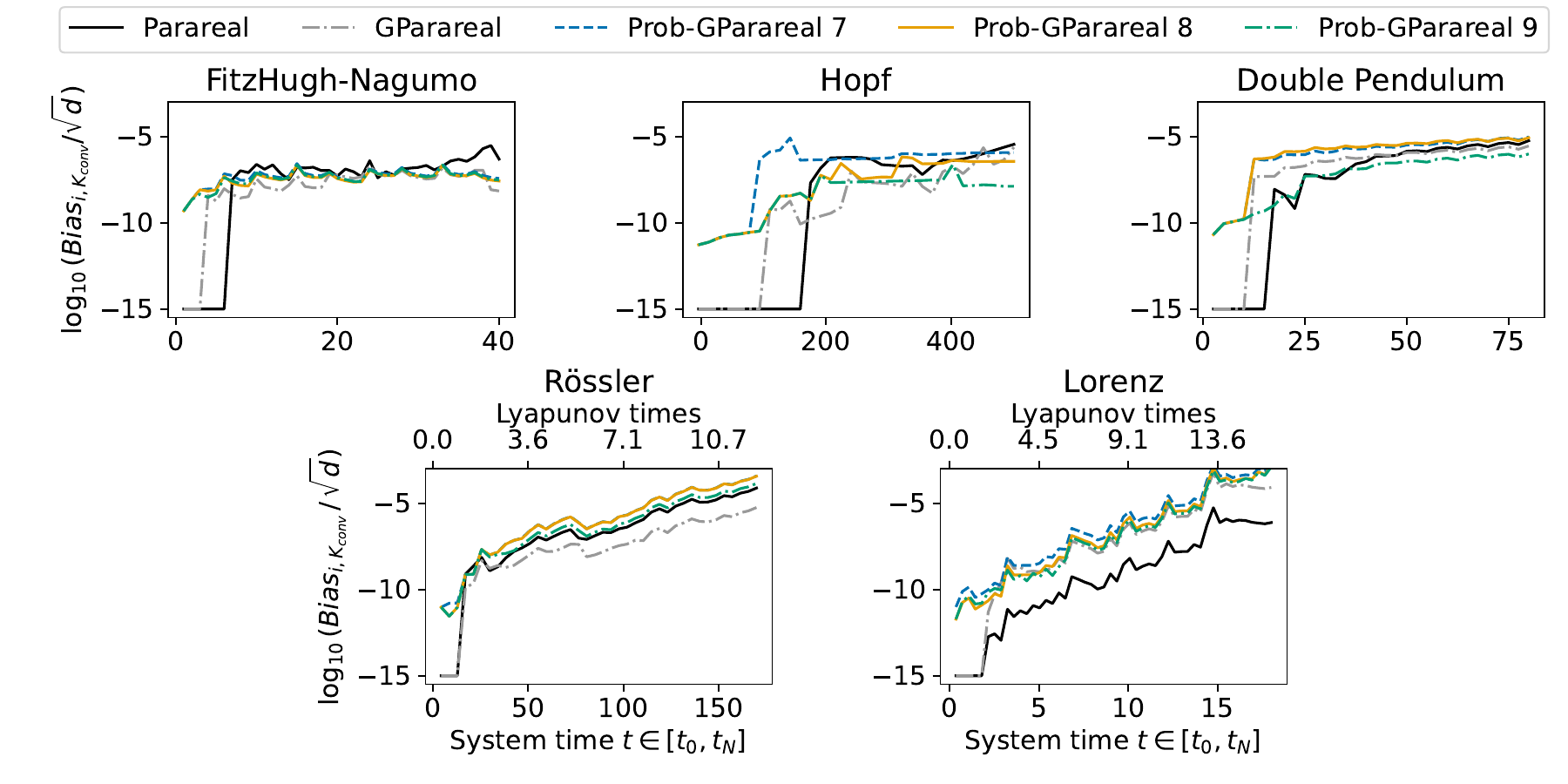}
    \caption{Performance comparison, evaluated using  the normalised bias ${\rm Bias}_{i,K_{\rm conv}}/\sqrt{d}$ (in $\log_{10}$), of the Parareal, GParareal and Prob-GParareal solutions with respect to the solution of the fine solver $\f$. Prob-GParareal is run using different values of $\epsilon$, with ``Prob-GParareal $l$'' in the legend referring to $\epsilon=1e^{-l}$, $l=7,8,9$. The Lyapunov times (for R\"ossler and Lorenz) are computed as the ratio between the system time and the Lyapunov time, as described in~\Cref{sec:chaotic_sys}. }
    \label{fig:rmse_comparison}
\end{figure}

The results demonstrate a good calibration and sharpness for all systems, with some performance reduction observed for Lorenz due to the chaotic nature of the system, as further discussed in Section \ref{sec:chaotic_sys}.
Since the metrics in~\Cref{tab:sim_res} are averaged over the entire time interval, they do not reveal the properties of the Prob-GParareal solution at individual time $t_i$. To this end, in~\Cref{fig:rmse_comparison}, we show the evolution of ${\rm Bias}_{i,k}$ normalized by $\sqrt{d}$, where $d$ is the dimension of the ODE, between the point solution provided by the fine solver $\f$, and that of Parareal, GParareal and Prob-GParareal. The normalization of the biases allows for a direct comparison of these metrics across models.
Since finding the value of the Prob-GParareal threshold $\epsilon$ that matches those of Parareal and GParareal is analytically challenging and system-dependent, particularly due to the different interpretations of the stopping thresholds $\epsilon$ in~\eqref{eq:para_stp_rule} and~\eqref{eq:pp_stp_rule}, we report in~\Cref{fig:rmse_comparison} the normalized bias for a range of values of the Prob-GParareal $\epsilon$ values. This shows that improved accuracy may be obtained by lowering such threshold. Further results on the impact of the threshold $\epsilon$ on the solution are given in~\Cref{app:n_analysis}. When looking at the normalized bias, we see that
the mean accuracy of Prob-GParareal is comparable to that of the other deterministic algorithms, with larger values for Lorenz and R\"ossler, the two chaotic systems.

\begin{figure}
    \centering
    \includegraphics[width=1\linewidth]{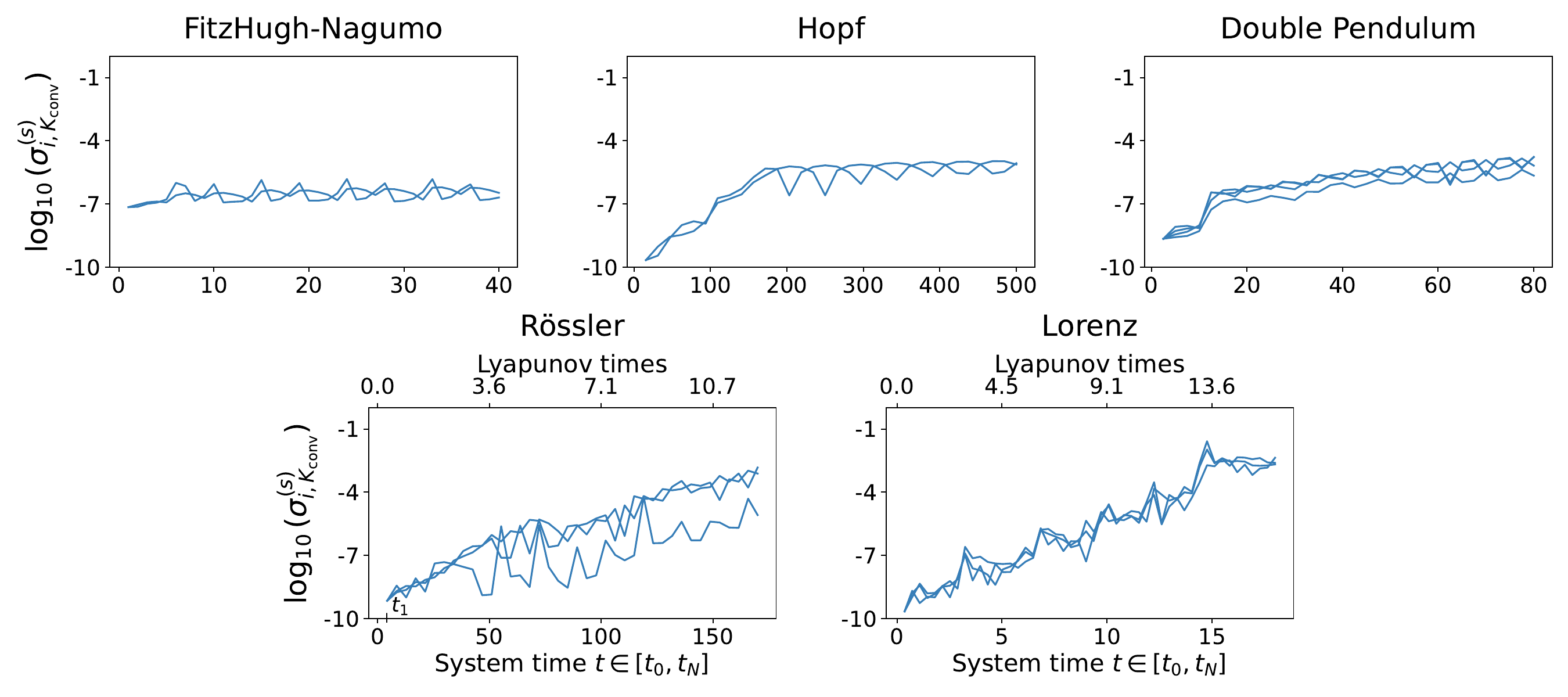}
    \caption{Evolution of the coordinate-wise standard deviation $\sigma_{i,K_{\rm conv}}^{(s)}$ (in $\log_{10}$), $s=1,\ldots, d$,
    of the converged Prob-GParareal solution $\mathcal{U}_{i,K_{\rm conv}}$
    for $i\in {0,\ldots, N}$, $t\in [t_0,t_N]$, with $N$ and $t_N$ which are
    system-specific, see \Cref{tab:simsetup} in~\Cref{app:setup}.
   The Lyapunov times (for R\"ossler and Lorenz) are computed as described in~\Cref{sec:chaotic_sys}. The simulation setup is as in~\Cref{tab:sim_res}.}
    \label{fig:std_evol}
\end{figure}
To better understand the different behavior across systems and characterize the Prob-GParareal solutions, in~\Cref{fig:std_evol}, we display the evolution of $\sigma_{i,k}^{(s)}$, the empirical coordinate-wise standard deviation of $\mathcal{U}_{i,K_{\rm conv}}$, across system time $t$ upon convergence. Note that Prob-GParareal shows steadily increasing uncertainty over time for the two chaotic systems, Lorenz and R\"ossler, which we explore further in~\Cref{sec:chaotic_sys}. A less pronounced effect is observed for the double pendulum, while the uncertainty stabilizes for FHN and Hopf.

\subsection{Focus: Chaotic systems}
\label{sec:chaotic_sys}
 As seen in~\Cref{tab:sim_res}, the performance for the Lorenz system is slightly worse than for the other systems in terms of scoring rules. Additionally, in~\Cref{fig:rmse_comparison} and \Cref{fig:std_evol}, we noticed a steady increase in the normalized bias and in the coordinate-wise standard deviation of the solution for both Lorenz and R\"ossler shown. This is an unavoidable consequence of the chaotic nature of the systems, specifically their sensitivity to initial conditions: two arbitrarily close trajectories diverge exponentially fast over time, with the divergence rate governed by the largest Lyapunov exponent~\citep{alligood1998chaos}. This increasing uncertainty limits the predictive capability of data-driven models, with reasonably accurate prediction horizons characterized in terms of the Lyapunov time, defined as the inverse of the largest Lyapunov exponent. These challenges are well known in the literature~\citep{pathak2021reservoir, pathak2017using, lu2018attractor, griffith2019forecasting, vlachas2020backpropagation}. In the figures, we show, when relevant, the Lyapunov times, defined as the ratio between the system time and the  Lyapunov time, to contextualize the prediction horizons. Using the estimates for the Lyapunov exponents in~\cite{sprott2003chaos}, we obtain a Lyapunov time equal to $14$  for R\"ossler, and $1.1$ for Lorenz.

\begin{figure}[t]
    \centering
    \includegraphics[width=1.0\linewidth]{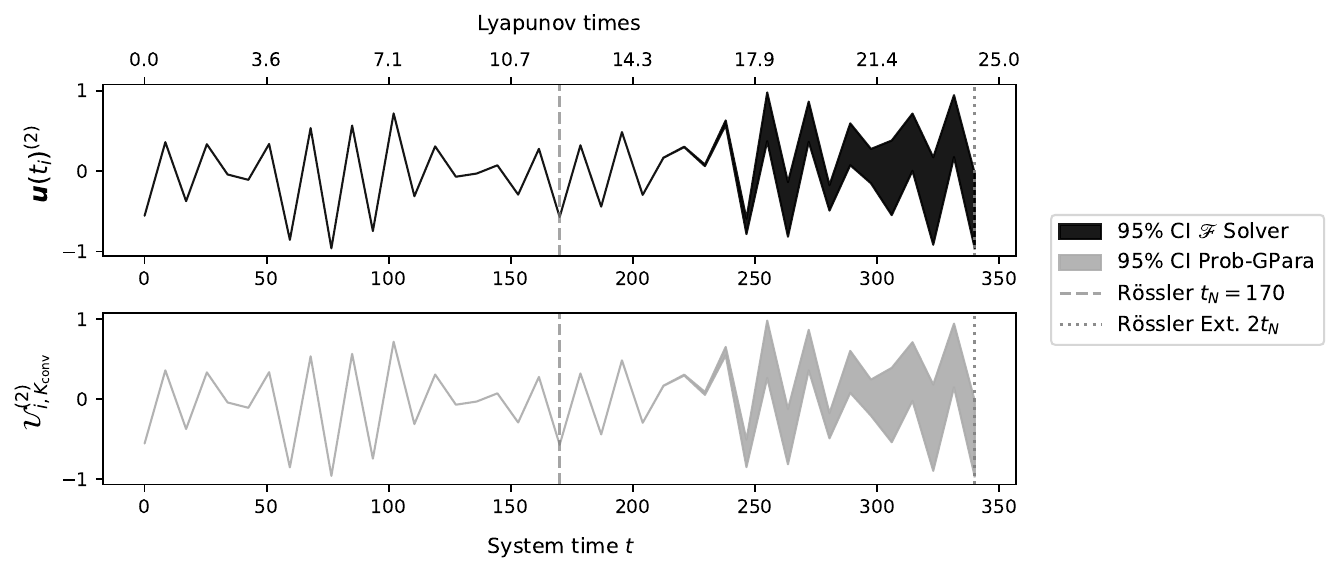}
    \caption{Impact of the initial condition on the solution of the chaotic R\"ossler system  evolved over a timespan of $t \in \left[0,340\right]$ (R\"ossler Ext. in the legend), twice longer than the timespan previously considered with $t_N=170$ in~\Cref{fig:rmse_comparison}  and \Cref{fig:std_evol}. Only the second coordinate (the $y$-coordinate) is shown. Top: empirical 95\% CI for the solution obtained via the fine solver $\f$, estimated using $1000$ initial conditions sampled from a multivariate Gaussian distribution centered at $\bs{u}_{(0)}$ with diagonal covariance matrix and identical standard deviation of $5 e^{-10}$. Bottom: Prob-GParareal 95\% CI obtained with $\epsilon=1e^{-9}$, and initial condition $\bs{u}_{0,0} = \bs{u}_{(0)}$. \mbox{The Lyapunov times are
    computed as described in~\Cref{sec:chaotic_sys}.}}
\label{fig:Rossler_explosion}
\end{figure}

In~\Cref{fig:Rossler_explosion}, top panel, we illustrate the impact of the initial condition on the evolution of the fine solver $\f$ for the 2nd coordinate of the R\"ossler system over $t \in \left[0,340\right]$, twice as much as the previously considered timespan.
We do this by sampling $1000$ initial conditions from a multivariate Gaussian distribution centered at $\bs{u}_{(0)}$ with diagonal covariance matrix and identical standard deviation of $5 e^{-10}$, comparable to that observed at interval $i=1$ (time $t_1$
),
and solve each IVP independently with $\f$. We adopt the same $\bs{u}_{(0)} = (0, -6.78, 0.02)$ as in previous works~\citep{pentland2023gparareal, nngparareal}. Note that the chosen $\bs{u}_{(0)}$ lies outside the R\"ossler attractor, and off-attractor transients may diverge differently before converging toward the invariant set, leading to more challenging numerical behavior. The exponential divergence after time 220 is noticeable, with an empirical 95\% CI for the $y$-coordinate of R\"ossler covering half the sample space by the end of the simulation.
 In~\Cref{fig:Rossler_explosion}, bottom, we present the corresponding 95\% CI generated by Prob-GParareal, which closely matches that on top.
Thus, contrary to what~\Cref{tab:sim_res} and Figures~\ref{fig:rmse_comparison} and
 \ref{fig:std_evol} alone might suggest, Prob-GParareal accounts for uncertainty equally well in both chaotic and non-chaotic systems.

\subsection{Impact of early algorithm termination}
\label{sec:early_stop}

\begin{figure}[t]
    \centering
    \includegraphics[width=1\linewidth]{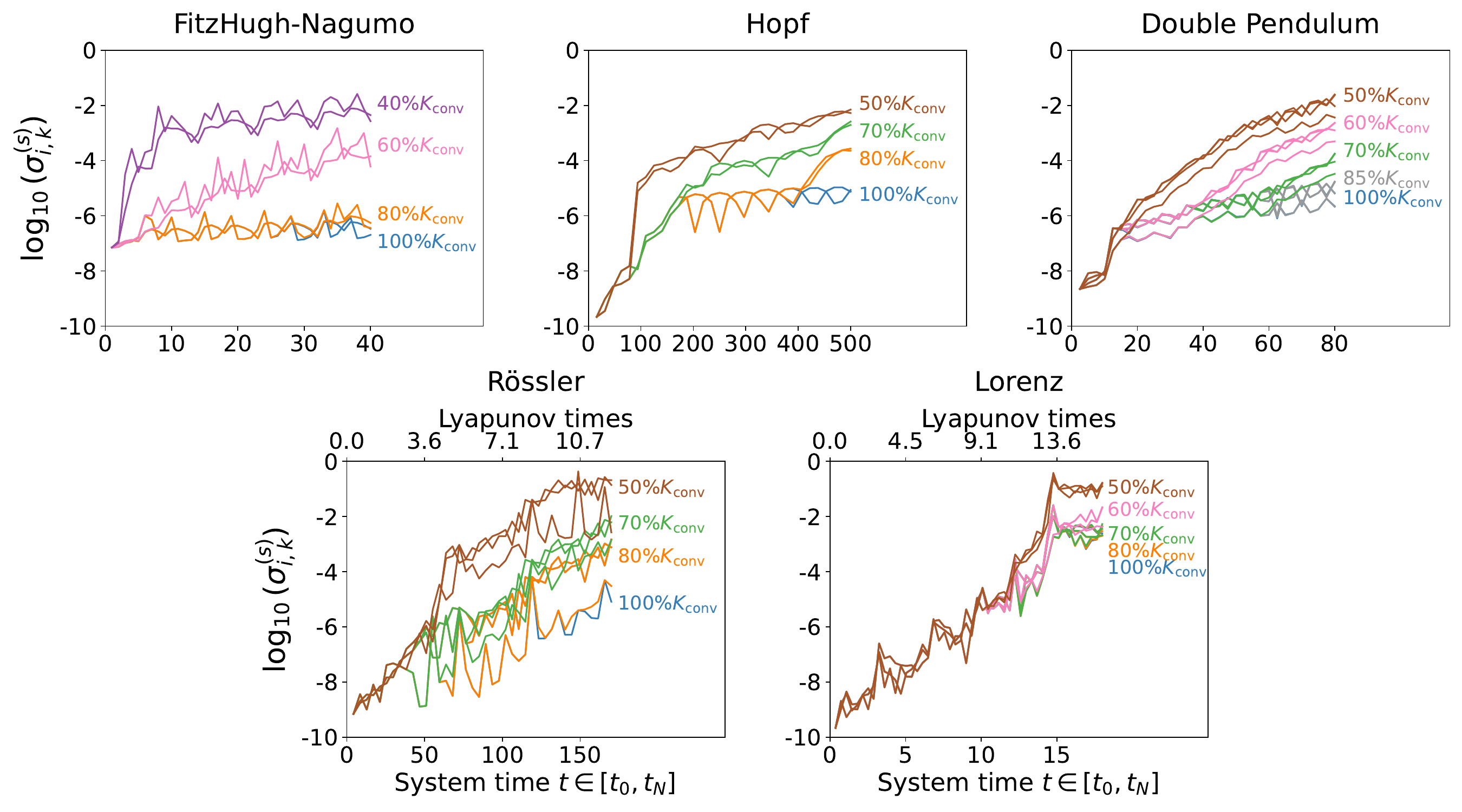}
    \caption{Evolution of the coordinate-wise standard deviation $\sigma_{i,k}^{(s)}$ (in $\log_{10}$), $s=1,\ldots, d$,
    of the Prob-GParareal solution $\mathcal{U}_{i,k}$ across system time $t$ for different phases of the algorithm, when $k=l\%K_{\rm conv}$ iterations to convergence are completed, with $K_{\rm conv}$ denoting the number of iterations to converge.
    The Lyapunov times are
    computed as described in~\Cref{sec:chaotic_sys}.
    The simulation setup is the same as that used in Table \ref{tab:sim_res}.}
   \label{fig:estop_std}
\end{figure}
\begin{figure}[t]
    \centering
    \includegraphics[width=1\linewidth]{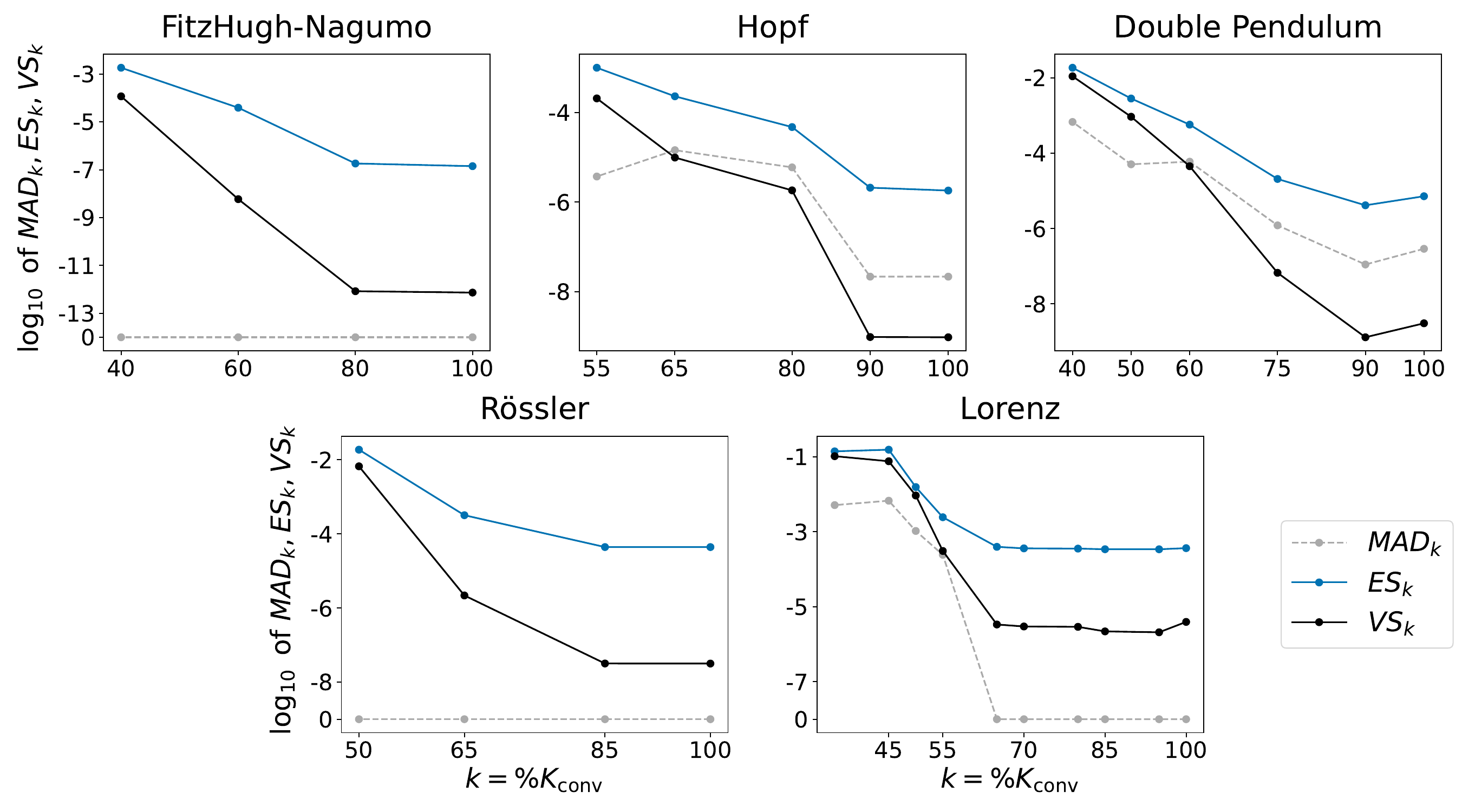}
    \caption{Evolution of ${\rm MAD}_k$, ${\rm ES}_k$, and ${\rm VS}_k$ (all in $\log_{10}$)
    of the Prob-GParareal solution $\mathcal{U}_{i,k}$ for different phases of the algorithm, i.e., when $k=l\%K_{\rm conv}$ iterations to convergence are completed, with $K_{\rm conv}$ denoting the number of iterations to converge. The simulation setup is the same as that used in Table \ref{tab:sim_res}.}
    \label{fig:EarlyTermination}
\end{figure}

As seen in Table \ref{tab:sim_res}, Prob-GParareal (and GParareal) converges with relatively low runtimes. However, if these simulations were to become more computationally intensive, the runtime would increase significantly, without a  reliable way to estimate completion time in advance. For instance, this would be the case when solving PDEs with thousands of spatial discretization points~\citep{randnet_parareal}.  In the worst-case scenario, Parareal-like algorithms ensure convergence within a time comparable to the sequential fine solver execution, which is often impractical. If simulations are stopped prior to convergence, say at time $t_i$, there are no guarantees or information on the error of unconverged intervals $t_{i+1}, \ldots, t_N$. In contrast, a probabilistic numerical solver enables UQ at \textit{any} point during execution, providing insight both at the conclusion of the algorithm and while running it. This is what we observe in~\Cref{fig:estop_std}, where we report the evolution of the coordinate-wise standard deviation of the solution $\mathcal{U}_{i,k}$ across system time, at different stages of the algorithm, with $
K_{\rm end}=l\%K_{\rm conv}$, $l\in [40,100]$
iterations to convergence completed when the algorithm is terminated, potentially prior to convergence at iteration $K_{\rm conv}$ (generally unknown before execution), i.e., $K_{\rm{end}}=K_{\rm stop}<K_{\rm conv}$ (early termination). As more Prob-GParareal iterations are carried out, and the algorithm approaches convergence, the uncertainty in the solution decreases.

The information carried by the UQ embedded in Prob-GParareal may also be used to set stopping criteria (e.g., based on the Wasserstein distance or the variance of the solution) and/or termination rules (e.g., defined in terms of the maximum allowed runtime).
The advantages of early termination  compared with traditional convergence are two-fold. From an algorithm design perspective, choosing a priori an optimal/suitable threshold $\epsilon$ for the Wasserstein distance can be challenging. However, by taking UQ into account, we may choose
it based on the variance of the Prob-GParareal solution, or use such variance to construct early stopping criteria. For example,
we may terminate the algorithm when the variance of the solution stops decreasing in consecutive iterations, as a sign that there is no further improvement in the learning of Prob-GParareal. For instance, by looking at \Cref{fig:estop_std}, we may stop Prob-GParareal for the FHN, Lorenz and double pendulum systems at $K
_\textrm{stop}=80\%K_{\rm conv}$, with similar results in terms of performance to the converged Prob-GParareal solution, as seen when looking at the score metrics (${\rm MAD}_k$, ${\rm ES}_k$, and ${\rm VS}_k$, in $\log_{10}$) in~\Cref{fig:EarlyTermination}. Interestingly, similar results are observed for different values of $\epsilon$ and $n$, as shown in~\Cref{fig:rob_ES} in~\Cref{app:n_analysis}.
This early termination of the algorithm results in a lower runtime, with a gain that depends on the model, as shown in~\Cref{fig:estop_runtime} in~\Cref{app:estop_runtime}. See also~\Cref{fig:slow_conv} and~\Cref{sec:prob_init_cond} for a further discussion on the impact of the variance of the solution on the convergence of the algorithm.

Alternatively, we may stop the algorithm based on the largest variance allowed over the solution timespan specified in the considered applications.
For example, suppose the user can tolerate a maximum coordinate-wise standard deviation of $\sigma_{i,k}^{(s)}=5e^{-4}$ in the probabilistic forecast. As shown in~\Cref{fig:estop_std}, FHN, Hopf, and R\"ossler would be terminated before convergence, at $K
_{\rm stop}=80\%K_{\rm conv}$, the double pendulum at $K
_{\rm stop}=70\%K_{\rm conv}$,
while Lorenz would continue until convergence, as its maximum (converged) $\sigma_{i,k}^{(s)}$ exceeds $5e^{-4}$. The performance of the early terminated solution in terms of score metrics is similar to the converged one, see Figure \ref{fig:EarlyTermination}, with advantages in terms of reduced runtime illustrated in~\Cref{fig:estop_runtime} in~\Cref{app:estop_runtime}.

The second advantage of early termination arises when the runtime of the algorithm exceeds the available computational budget. We may think, for example, of processes running on cloud infrastructure that are typically priced by usage. In such cases, forced termination ensures that the computations remain within budget constraints. The resulting forecast variance then reflects the uncertainty due to incomplete computation, which tends to be higher for intervals $i$ farther away from the initial condition, as shown in~\Cref{fig:estop_std}.

\subsection{Impact of random initial conditions on Prob-GParareal}
\label{sec:prob_init_cond}
We now investigate the impact of random initial conditions on the forecast and  convergence of Prob-GParareal. Let
$\bs{U}_{0,0}$ be multivariate Gaussian distributed, with mean $\bs{u}_0$
, and diagonal covariance matrix with identical diagonal entries $ \sigma_{\rm init}^2 \in \R^+$,
which we now assume strictly positive after having focused on $\sigma_{\rm init}=0$ (i.e. deterministic starting conditions) before. We explore values for the initial standard deviation $\sigma_{\rm init}$ ranging from $1e^{-6}$ to $1e^{-2}$. Larger values, such as $\sigma_{\rm init}=5e^{-2}$, lead to high uncertainty in the solution of the considered models, and are therefore excluded.  Although these values may appear small
, it is important to recall that the data have been normalized to lie within $[-1,1]^d$. Thus, on this scale, such standard deviations are not negligible.

\begin{figure}[t]
    \centering
\includegraphics[width=1\linewidth]{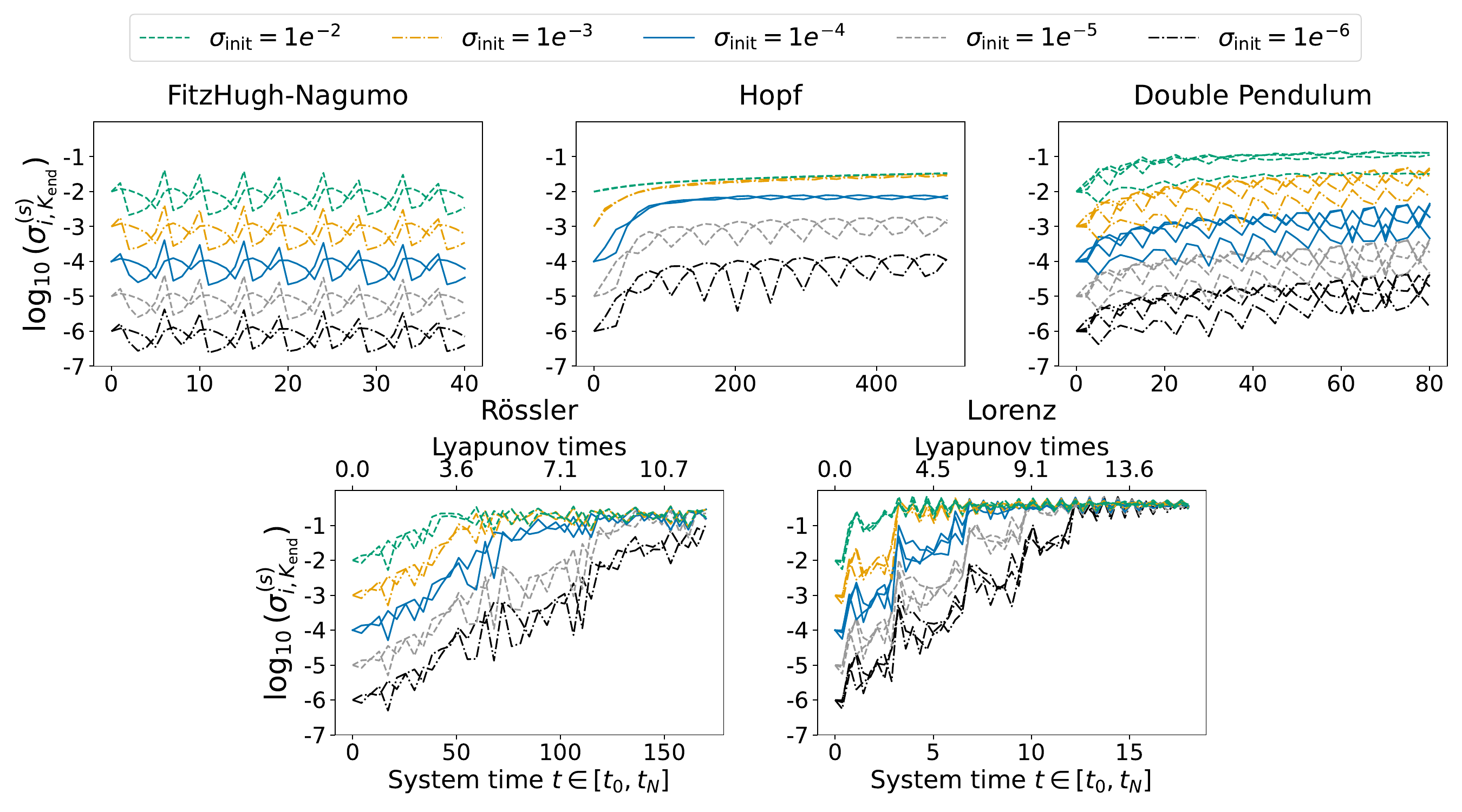}
    \caption{Impact of random initial conditions $\bs{U}_{0,0}$ on the coordinate-wise standard deviation $\sigma_{i,K_{\rm end}}^{(s)}, s=1,\ldots, d$ (in $\log_{10}$) of the converged Prob-GParareal solution $\bs{U}_{i,K_{\rm end}}$ for different systems. Here, $\bs{U}_{0,0}$ is a d-dimensional Gaussian distribution centered at $\bs{u}_{(0)}$ with diagonal covariance matrix with diagonal entries equal to $\sigma^2_{\textrm{init}}$, where $\sigma_{\textrm{init}}=1e^{-l}$, $l=2, 3, 4, 5, 6$. Simulations displaying nonconvergent behavior were stopped before convergence, at $K_{\rm end}=K_{\rm stop}$, with $K_{\rm stop}$ given in~\Cref{tab:simsetup} in~\Cref{app:setup}. The Lyapunov times are
    computed as described in~\Cref{sec:chaotic_sys}.}
    \label{fig:pp_initvar_std}
\end{figure}
The impact of random
initial conditions on the uncertainty of the Prob-GParareal solution is illustrated in~\Cref{fig:pp_initvar_std}.
The standard deviation of the final solution at $i=0$, that is,
$\sigma_{0,K_{\rm end}}^{(s)}$, is always higher than that in the deterministic case, reported in~\Cref{fig:std_evol}, and matches $\sigma_{\rm init}$, which sets a lower bound on the solution standard deviation across all considered systems. Moreover, the larger is $\sigma_{\rm init}$, the larger is the solution standard deviation,
resulting in stratified graphs. This behavior appears in both non-chaotic and chaotic systems: the former exhibit a relatively flat standard deviation trend over time, while the latter (here R\"ossler and Lorenz, and to a minor extent, the double pendulum) show the expected exponential growth in uncertainty (note the logarithmic scale on the $y$-axis).

\begin{figure}[t]
    \centering
\includegraphics[width=.9\linewidth]{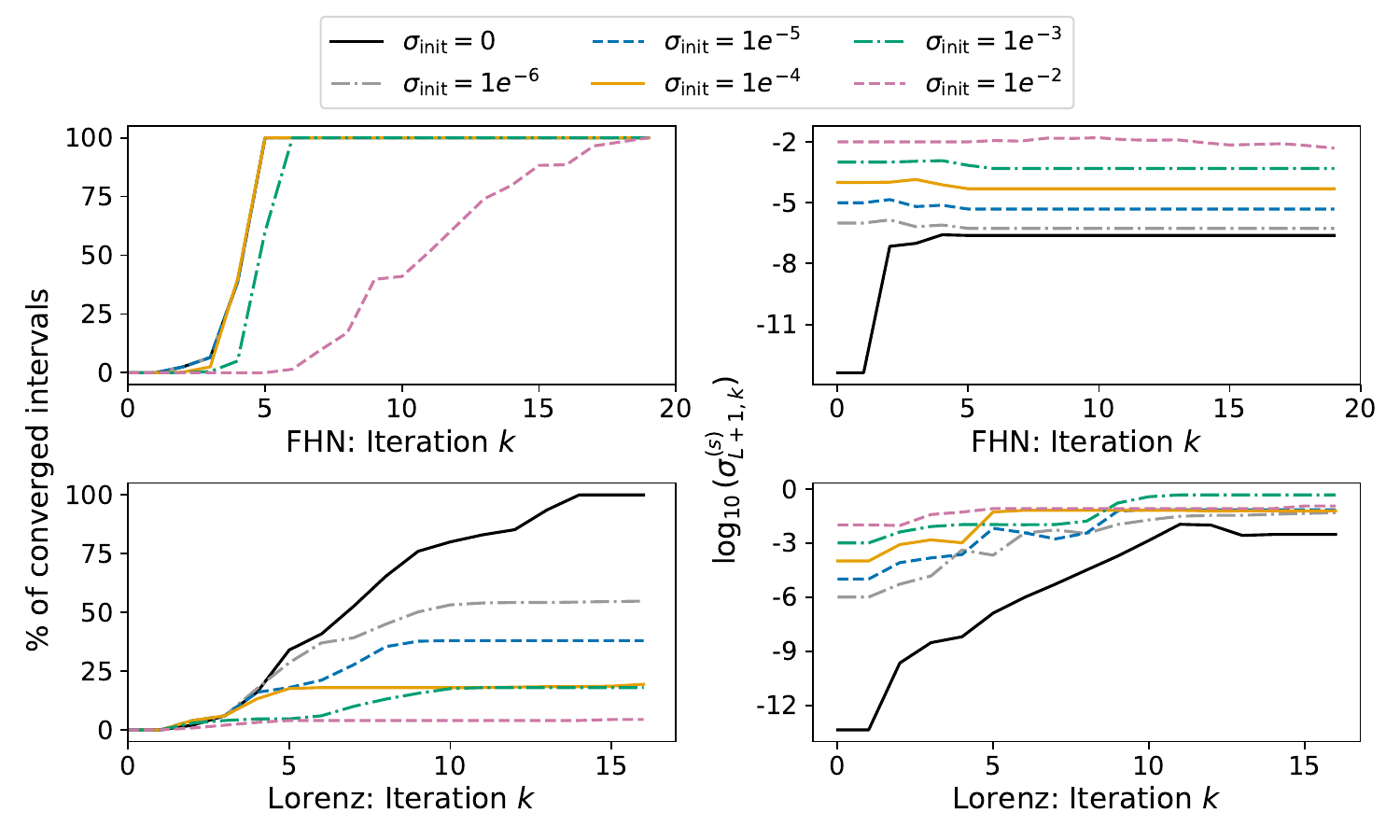}
    \caption{Impact of the initial standard deviation $\sigma_\text{init}$ of $\bs{U}_{0,0}$ on the Prob-GParareal convergence (left panels) and standard deviation (right panels) for the FHN (top panels) and Lorenz (bottom panels) systems. Left panels: percentage of converged intervals per iteration $k$ for different $\sigma_{\rm init}$ values. Right panels: standard deviation of $\mathcal{U}_{L+1,k}$, $\sigma_{L+1,k}^{(s)}$ (in $\log_{10}$), where $L+1$ is the first unconverged \mbox{interval. High variance at $L+1$ is associated with slow convergence.}}
    \label{fig:slow_conv}
\end{figure}

As expected, high standard deviations in the Prob-GParareal solution (e.g., $\sigma_{i,k}^{(s)} \in \left[0.05, 1\right]$, depending on the system)
are associated with slow or nonconvergent behavior of the algorithm. This can be seen in~\Cref{fig:slow_conv} for the FHN and Lorenz systems, where the left panels show the percentage of converged intervals across iterations, and the right panels the standard deviation $\sigma_{L+1,k}^{(s)}$ of the intermediate solution at iteration $k$, for the first unconverged interval $L+1$ (recall that $L$ denotes the number of converged intervals, see \Cref{sec:probpara_algo}). We choose to track the standard deviation for the first  $L+1$ interval as it tends to be the smallest across all the unconverged intervals $i=L+1,\ldots,N$. We note that when the coordinate-wise  standard deviation $\sigma_{L+1,k}^{(s)}$ reaches values around $5e^{-2}$ (from $k=1$ for FHN when $\sigma_{\rm init}=1e^{-2}$, at iteration $k=10$ for Lorenz over most values of $\sigma_{\rm init}$; right panels), the algorithm shows slow or nonconvergent behavior. This is reflected by the evolution of the percentage of converged intervals in the left panels, which for FHN increases to 100\% slowly, for $\sigma_{\rm init}=1e^{-2}$, while for Lorenz flattens around $k=10$ for most values of $\sigma_{\rm init}$. To save computation, we suggest stopping
the algorithm once this behavior is detected
, leading to an additional early termination criterion, similar to the first proposed in~\Cref{sec:early_stop}. Overall, we recommend choosing a starting condition $\bs{U}_{0,0}$ with $\sigma_\text{init}<\epsilon$ and  $\sigma_\text{init}\ll \epsilon$ for chaotic systems.

\section{Empirical results for Prob-nnGParareal}
\label{sec:nnprobpara}

In~\Cref{sec:comp_complx}, we showed that replacing the full GP with an nnGP
(as proposed in nnGParareal~\citealt{nngparareal}) yields substantial computational savings in the proposed probabilistic algorithm, Prob-nnGParareal. Importantly, using the nearest neighbors approximation has a limited impact, if any, on the predictive accuracy and uncertainty estimation of the algorithm, as shown in~\Cref{fig:nnppara_initcond} in~\Cref{app:Prob-nnGParareal}, where we replicate the analysis carried out in~\Cref{fig:pp_initvar_std},~\Cref{sec:prob_init_cond}, replacing the GP with an nnGP. While in some cases the use of nnGPs results in a slight increase in uncertainty over time compared to Prob-GParareal, the overall performance of the algorithm remains largely unaffected.

The reduced computational cost of Prob-nnGParareal (combined with an accuracy comparable to that of Prob-GParareal) makes it appropriate for solving PDEs. Here, we consider the Viscous Burgers equation~\citep{schmitt2018numerical}, a nonlinear one-dimensional PDE defined as
\begin{equation}
    \label{eq:burg}
         u_t = \nu u_{xx} - uu_x, \quad (x,t)\in \left[-L,L\right]
    \times \left[t_0,t_N\right],
\end{equation}
with initial condition $u(x,t_0)=u_{(0)}(x)$
, $x\in \left[-L,L\right], L>0$, and periodic boundary conditions
$u(-L,t)=u(L,t)$, $u_x(-L,t)=u_x(L,t)$, $t \in \left[t_0, t_N\right]$. We use the same setting and parameter values as in nnGParareal~\citep{nngparareal} for a direct comparison. Specifically, we take $L=1$, the diffusion coefficient $\nu=0.01$, and discretize the spatial domain using finite difference~\citep{fornberg1988generation} and equally spaced points $x_{l+1}=x_l+\Delta x$, with $\Delta x = 2L/d$ and $l=0,\ldots,d$, thus reformulating the PDE as a $d$-dimensional ODE system. We choose $N=d=128$, $u_{(0)}(x)=0.5(\cos (\frac{9}{2}\pi x)+1)$, $t_0=0$, and $t_N=5$. Moreover, we took $\g$ and $ \f$ to be Runge-Kutta of order 1 and 8, respectively, with a number of integration steps for numerical solvers over $[t_0, t_N]$ of $512$ for $\g$ and $5.12e^6$ for $\f$.

Note that in the present experiment both the fine and coarse propagators are applied to the same spatial discretization. More generally, however, the framework can accommodate different spatial discretizations for $\f$ and $\g$ by introducing suitable restriction and interpolation/prolongation operators between the corresponding spatial grids, as commonly done in Parareal methods with spatial coarsening and related multigrid-in-time methods~\citep{angel2021impact, howse2019parallel}. This would not fundamentally alter the structure of the algorithm, although the transfer operators would introduce additional approximation errors that should be accounted for in a refined analysis.

In~\Cref{tab:burges}, we compare the performance of Parareal, GParareal, nnGParareal, and Prob-nnGParareal, in terms of the number of iterations to converge $K_{\rm conv}$, computational cost of running $\g$, $\f$ and each algorithm, and the training cost $T_\text{model}$, together with an evaluation of each algorithm's performance with respect to the solution provided by the fine solver $\f$. Specifically, we  compute the bias and MSE for all the algorithms, and we additionally use the scoring rules for the probabilistic Prob-nnGParareal solution.
On the one hand, the runtime of Prob-nnGParareal is lower than its deterministic counterpart, nnGParareal, thanks to a lower $K_{\rm conv}$ to converge (with similar runtimes/costs per iteration). On the other hand, the solution accuracy, given in terms of both ${\rm Bias}_{k}$ and ${\rm MSE}_{k}$, is comparable.
However, such comparisons should be made with caution, as Prob-nnGParareal returns samples from a distribution rather than a single value, and, most importantly, the convergence criteria are inherently different, as the threshold $\epsilon$ controls the Wasserstein distance between probability distributions in Prob-(nn)GParareal, and the $L_\infty$ distance between vectors in
the other algorithms. Thus, the two $\epsilon$ are not immediately comparable, even if a  lower $\epsilon$ leads to higher accuracy, as shown in~\Cref{fig:rob_ES} in~\Cref{app:n_analysis} for Prob-GParareal.

\begin{table}[t]
    \centering
    {\footnotesize

\begin{tabular}{lccccc}
\hline
Algorithm & ${\rm Bias}_{K_{\rm conv}}$ & ${\rm MSE}_{K_{\rm conv}}$ & ${\rm MAD}_{K_{\rm conv}}$ & ${\rm VS}_{K_{\rm conv}}$ & ${\rm ES}_{K_{\rm conv}}$ \\
\hline
Fine solver $\f$ & $-$ & $-$ & $-$ & $-$ & $-$ \\
Parareal & $1.73e^{-07}$ & $7.12e^{-14}$ & $-$ & $-$ & $-$ \\
GParareal & $1.03e^{-07}$ & $3.21e^{-14}$ & $-$ & $-$ & $-$ \\
nnGParareal & $3.47e^{-07}$ & $1.33e^{-13}$ & $-$ & $-$ & $-$ \\
Prob-nnGPara & $3.23e^{-07}$ & $1.32e^{-12}$ & $3.53e^{-13}$ & $5.16e^{-10}$ & $3.65e^{-07}$ \\
\hline
\end{tabular}

\vspace{0.5em}

\begin{tabular}{lccccc}
\hline
Algorithm & $K_{\rm conv}$ & $T_\g$ & $T_\f$ & $T_{\rm model}$ & $T_{\rm alg}$ \\
\hline
Fine solver $\f$ & $-$ & $-$ & $-$ & $-$ & 13h 5m \\
Parareal & 10 & 0s & 6m & 0s & 1h 4m \\
GParareal & 6 & 0s & 6m & 1h 2m & 1h 39m \\
nnGParareal & 9 & 0s & 6m & 7m & 1h 3m \\
Prob-nnGPara & 5 & 3s & 6m & 3m & 33m \\
\hline
\end{tabular}
}
\caption{Performance of Parareal, GParareal, nnGParareal, and Prob-nnGParareal on Viscous Burgers' PDE~\eqref{eq:burg}, solved as a $d$-dimensional ODE, with $N=d=128$. $K_{\rm conv}$ is the number of iterations to convergence, $T_\g$ and $T_\f$ are the runtimes of $\g$ and $\f$ per iteration, respectively, while $T_{\rm model}$, and $T_{\rm alg}$ are the total cost of evaluating the correction function $f_c$ in \eqref{discrepancy}, and the total runtime of the algorithm, respectively.}
\label{tab:burges}
\end{table}

\section{Discussion}
\label{sec:conclusion}
In this paper, we have introduced Prob-(nn)GParareal, a probabilistic extension of the (nn)GParareal algorithm(s) capable of capturing and propagating the underlying numerical uncertainty.
This novel contribution addresses a general gap in probabilistic approaches in the PinT literature.
Unlike the work of~\cite{bosch2024parallel}, our approach does not rely on approximations for nonlinear ODE systems, and can be directly applied to any underlying numerical solver, facilitating its adoption in existing Parareal applications (e.g., for specific parabolic, hyperbolic or chaotic problems). UQg
is achieved by modeling the correction function with (nn)GPs,
and propagating the resulting posterior distribution through the (generally nonlinear) coarse solver $\g$ via Monte Carlo sampling.
The additional complexity introduced by Prob-(nn)GParareal is fully parallelizable, reducing the total cost to that of the corresponding deterministic version. Prob-GParareal also offers flexible resource management via early termination rules, supports probabilistic initial conditions, and shows good scalability in the number of processors by implementing recent advances based on nnGPs
\citep{nngparareal}, allowing the use of Prob-nnGParareal for PDE solutions.

Despite its several advantages, Prob-GParareal's ability to properly account for the numerical uncertainty depends on the assumptions that: 1) the fine solver $\f$ yields the true solution (which is common within the PinT literature); 2) the variance across the coordinates of the system can be modeled using an intrinsic coregionalization model~\citep{goovaerts1997geostatistics}, as discussed in~\Cref{sec:theoretical_construction}.
While these assumptions enable computational feasibility without requiring a linear vector field for the DE system, they may not always hold, leading to a bias in the mean and variance of the solution, respectively. The first assumption may be relaxed by either training local GPs to learn it, or using probabilistic solvers of it, allowing, in both cases, UQ for $\f$.
The GP covariance assumption can be relaxed by selecting more flexible GP models. In the most general multi-output case, this would require forming a covariance matrix of size $(Dd)\times(Dd)$, where $D$ is the size of the dataset and $d$ is the number of coordinates to be jointly modeled, with exact GP inference scaling as $\mathcal{O}(D^3d^3)$.
This may be feasible for nnGPs used on low-dimensional ODEs. Scaling to higher-dimensional ODEs and PDEs may require replacing the (nn)GPs with more scalable alternatives, as done for RandNet-Parareal in~\cite{randnet_parareal} for a deterministic algorithm. We defer the exploration of these problems to future research.

\acks{GG is funded by the Warwick Centre for Doctoral Training in Mathematics and Statistics. MT and LG acknowledge the financial support of the United Kingdom Engineering and Physical Sciences Research Council (EPSRC) grant EP/X020207/1. We are grateful to the Scientific Computing Research Technology Platform (SCRTP) at Warwick for the
provision of computational resources. All experiments were performed using the Warwick University HPC facilities on Dell PowerEdge C6420 compute nodes each with 2 x Intel Xeon Platinum 8268 (Cascade Lake) 2.9 GHz 24-core processors, with 48 cores per node and 192 GB DDR4-2933 RAM per node. Python code accompanying this manuscript is available at \href{https://github.com/Parallel-in-Time-Differential-Equations/ProbParareal}{https://github.com/Parallel-in-Time-Differential-Equations/ProbParareal}.}

\appendix

\section{Additional details on Gaussian process formulation}
\label{app:GP_details}
In this section, we provide the expressions for the GP posterior mean $\boldsymbol{\mu}_{\mathcal{D}_k}^{(s)}(\bs{u}') \in \mathbb{R}$ and posterior variance ${\sigma^{(s)}_{\mathcal{D}_k}}(\bs{u}')^2 \in \mathbb{R}^+$ introduced in~\Cref{sec:background}. These are given by
\begin{align}
    \mu^{(s)}_{\mathcal{D}_k}(\bs{u}')=& {K}({X}, \bs{u}')^\top \left({K}({X},{X}) + \sigma_{\rm reg}^2 \mathbb{I}_{Nk}\right)^{-1} {Y}_{(\cdot,s)}, \label{eq:gp_posterior_m}\\
    {\sigma^{(s)}_{\mathcal{D}_k}}(\bs{u}')^2 = &{K}_{\rm GP}(\bs{u}', \bs{u}') - {K}({X}, \bs{u}')^\top \left({K}({X},{X}) + \sigma_{\rm reg}^2 \mathbb{I}_{Nk}\right)^{-1} {K}({X}, \bs{u}'), \label{eq:gp_posterior_v}
\end{align}
where $X\in \mathbb{R}^{Nk\times d}$ denotes the matrix of inputs $\bs{u}_{i-1,j}$ and $Y\in \mathbb{R}^{Nk\times d}$ the matrix of outputs $f_c(\bs{u}_{i-1,j})$, taken from the dataset $\mathcal{D}_k$, with both matrices arranged by rows. The symbol $\mathbb{I}_{Nk}$ denotes the identity matrix of size $Nk$, while the vector ${K}({X}, \bs{u}')\in \mathbb{R}^{Nk}$ contains the covariances between every input in ${X}$ and the test point $\bs{u}'$, defined as \[
({K}({X},\bs{u}'))_{r}={K}_{\rm GP}({{X}_{(r, \cdot)}}^\top, \bs{u}' ), \quad r=1,\ldots, Nk,
\]
where ${{X}_{(r, \cdot)}}^\top$ denotes the transpose of the $r$th row of $X$. Similarly, the covariance matrix ${K}({X},{X})\in \mathbb{R}^{Nk\times Nk}$ is given by
\[({K}({X}, {X}))_{r,q} = {K}_{\rm GP}({{X}_{(r,\cdot)}} ^\top, {{X}_{(q,\cdot)}}^\top ), \quad r,q=1,\ldots, Nk.
\]
Here, as $K_{\rm GP}$ kernel, we use the Gaussian kernel $K_{\rm G}$, also known as radial basis function  or square-exponential kernel, defined as
\begin{equation}
\label{eq:kernel_Gaussian}
    {K}_{\rm G}(\bs{u},\bs{u}') = \sigma_{\rm o}^2 \exp\left(-\dfrac{\|\bs{u}-\bs{u}'\|^2}{\sigma_{\rm i}^2}\right), \quad \bs{u},\bs{u}' \in \mathbb{R}^d,
\end{equation}
where $\sigma_{\rm o}^2$ and $\sigma_{\rm i}^2$ represent the output and input length scales, respectively, and $\|\cdot \|$ denotes the Euclidean norm\footnote{Other choices of  kernels are also widely used, e.g. the Mat\'ern kernel in spatial statistics analysis~\citep{Matern1986}. For constants $\nu>0$ (smoothness parameter),   $\sigma_{\rm i}^2>0$ (input length scale),  $\sigma_{\rm o}^2 >0$ (output length scale), the Mat\'ern kernel $K_{\nu, \sigma_{\rm i},\sigma_{\rm o}}: \mathbb{R}^d \times \mathbb{R}^d \rightarrow \mathbb{R}$ is defined as
$$
{K}_{\nu, \sigma_{\rm i},\sigma_{\rm o}}\left(\bs{u}, \bs{u}^{\prime}\right)=\sigma_{\rm o}^2\frac{1}{2^{\nu-1} \Gamma(\nu)}\left(\frac{\sqrt{2 \nu}\left\|\bs{u}-\bs{u}^{\prime}\right\|}{\sigma_{\rm i}}\right)^\nu k_\nu\left(\frac{\sqrt{2 \nu}\left\|\bs{u}-\bs{u}^{\prime}\right\|}{\sigma_{\rm i}}\right), \quad \bs{u}, \bs{u}^{\prime} \in \mathbb{R}^d,
$$
where $\Gamma(\cdot)$ is the gamma function, and $k_\nu$ is the modified Bessel function of the second kind of order $\nu$.}.
The term $\sigma_{\rm reg}^2$ in~\eqref{eq:gp_posterior_m} is a regularization term, often referred to as jitter or nugget, which is used to improve the condition number of the covariance matrix during inversion. Note that the posterior variance in~\eqref{eq:gp_posterior_v} depends on the coordinate $s$, even though there is no explicit reference to $s$ in the expression. This is because the regularization term $\sigma_{\rm reg}^2$ and the input and output length scales ($\sigma_{\rm i}^2$ and $\sigma_{\rm o}^2$) form the hyperparameters of the $s$th GP, $\boldsymbol{\theta}^{(s)}:=({\sigma_{\rm i}^{(s)\; 2}}, {\sigma_{\rm o}^{(s)\; 2}}, {\sigma_{\rm reg}^{(s)\; 2}})$, where we add the superscript $s$ for clarity. Hence, $\boldsymbol{\theta}^{(s)}$ must be tuned independently for each of the $d$ coordinates. This is typically achieved through the numerical maximization of the marginal log-likelihood. Define $K_{\theta}^{(s)}
:=K^{(s)}(X,X)+\sigma_{\rm reg}^{(s)2}\mathbb{I}_{|\mathcal{D}_k|}$ and consider 
\[
\log p({Y}_{(\cdot,s)}|{X}, \boldsymbol{\theta}^{(s)}) \propto - {{Y}_{(\cdot,s)}}^\top {K_{\theta}^{(s)}}^{-1}{Y}_{(\cdot,s)} - \log {\rm det}(K_{\theta}^{(s)}),
\]
where $K^{(s)}(X,X)$, and hence ${K}^{(s)}_{\theta}$, depends on $\boldsymbol{\theta}^{(s)}$ through the kernel function ${K}_{\rm GP}$. For additional details on this process and the role of the regularization parameter $\sigma_{\rm reg}^2$, refer to~\cite{nngparareal}. Once the optimal hyperparameters are identified, the GPs are trained and can be used to make predictions.

\section{Proof of theoretical results}
In this section, we provide the proofs for the theorems and corollary presented in the manuscript. Before doing that, we recall some prior results needed for the proofs.

\subsection{Prior results from the literature}\label{pretheory}

We recall two results establishing the convergence of scalar-output GP posterior mean prediction (\Cref{th:GP_mean_bound} from~\citealt{wendland2004scattered}) and the rate of decay of the prediction variance (\citealt{wu1993local, kanagawa2018gaussian}), which we state here for the scalar-output GP used to model the $s$th coordinate of the correction function $f_c^{(s)}=(\f-\g)^{(s)}$. Starting from the variance decay, we report a two-part theorem, using the formulation of Theorem~5.4 in~\cite{kanagawa2018gaussian} in part {\bf (i)},
and Theorem 5.14 in~\cite{wu1993local} in part~{\bf (ii)}.
\begin{theorem}[\citealt{kanagawa2018gaussian,wu1993local}]
\label{th:posterior variance p smooth}
Let $K$ be a kernel on $\mathbb{R}^d$ and let ${\mathcal{H}}_{{K}}$ be the RKHS induced by it. Let $f_c^{(s)}=(\f-\g)^{(s)} \in {\mathcal{H}}_{{K}}$, $s=1,\ldots,d$, and let ${\sigma^{(s)}_{\mathcal{D}}}(\bs{u}')^2 \in \R$, $s=1,\ldots,d$, be the posterior variance of a scalar-output GP built on a dataset $\mathcal{D}$ to approximate $f_c^{(s)}
\in {\mathcal{H}}_{{K}}$, $s=1,\ldots,d$. Then:
\begin{description}
\item[(i)] If $K$ is such that its associated RKHS is norm-equivalent to $W^q_2(\mathbb{R}^d)$, the Sobolev space of order $q$,  then for any $\rho>0$, there exist constants $h_0>0$ and $\{C_s\}_{s=1}^d$, $C_s>0$, such that for any $\bs{u}' \in \mathbb{R}^d$ and any set of points $\mathcal{D}=\left\{\bs{u}_1, \ldots, \bs{u}_n\right\} \subset \mathbb{R}^d$ satisfying $h_{\rho, \mathcal{D}}(\bs{u}') \leq h_0$, it holds that ${\sigma^{(s)}_{\mathcal{D}}}(\bs{u}')^2 \leq C_s h_{\rho, \mathcal{D}}(\bs{u}')^{2 q-d}$, for $s=1,\ldots,d.$
\item[(ii)] If $K$ is infinitely smooth, then for any $\rho>0$ and $\alpha>0$, there exist constants $h_\alpha>0$ and $\{C_{\alpha,s}\}_{s=1}^d$, $C_{\alpha,s}>0$, all depending on $\alpha$, such that for any $\bs{u}' \in \mathbb{R}^d$ and any set of points $\mathcal{D}=\left\{\bs{u}_1, \ldots, \bs{u}_n\right\} \subset \mathbb{R}^d$ satisfying $h_{\rho, \mathcal{D}}(\bs{u}') \leq h_\alpha$, it holds that ${\sigma^{(s)}_{\mathcal{D}}}(\bs{u}')^2 \leq C_{\alpha,s} h_{\rho, \mathcal{D}}(\bs{u}')^{\alpha}$, for $ s=1,\ldots,d.$
\end{description}

\end{theorem}
\begin{remark}
The Gaussian kernel in~\Cref{app:GP_details} is infinitely smooth. By Corollary 10.48 in~\cite{wendland2004scattered}, the Mat\'ern kernel with smoothness parameter $\nu>0$ on $\mathbb{R}^d$ presented in~\Cref{app:GP_details} induces an RKHS $\mathcal{H}_K$ that is norm-equivalent to the Sobolev space $W^q_2(\mathbb{R}^d)$ with $q=\nu+d/2$. Note that functions in $\mathcal{H}_K$ are weak differentiable up to order $q$ and differentiable in the classical sense only up to $\nu$ (see Remark 2.11 in~\citealt{kanagawa2018gaussian}).
\end{remark}

\begin{theorem}[\cite{wendland2004scattered}, Theorem 11.4]
\label{th:GP_mean_bound}
    In the conditions of~\Cref{th:posterior variance p smooth}, the error between the correction function $f_c^{(s)}\in \mathcal{H}_{K} $ and the scalar-valued GP posterior mean $\mu^{(s)}_{\mathcal{D}}$, $s=1,\ldots, d$, estimated on $\mathcal{D}\subset \mathbb{R}^d$, satisfies
    \[ 
    |f_c^{(s)}(\bs{u}') - \mu^{(s)}_{\mathcal{D}}(\bs{u}')| \leq  {\sigma^{(s)}_{\mathcal{D}}}(\bs{u}') \;\| f_c^{(s)} \|_{\mathcal{H}_{{K}}}, \enspace {\rm for \, any}\enspace \bs{u}'\in \mathbb{R}^d.
    \]
\end{theorem}
We are now ready to prove  \Cref{prop:convergence} in~\Cref{app:pr:conv},  \Cref{th:probpara:var:bound} in~\Cref{app:pr:var_bound}, \Cref{prop:meanbound} and Corollary~\ref{cor:mean_full_expr} in~\Cref{app:pr:mean_bound}.

\subsection{Proof of \texorpdfstring{\Cref{prop:convergence}}{convergence}}
\label{app:pr:conv}
In this section, we prove~\Cref{prop:convergence} for the infinite smoothness case, case \textbf{(iii)}. Cases \textbf{(i)} and \textbf{(ii)} require minor modifications, which are detailed at the end of the Section.

\paragraph{Case \textbf{(iii)} Smoothness.}
    Consider the squared Wasserstein-2 distance between the true solution $\bs{u}(t_i)$ of~\eqref{eq:ode} at time $t_i$ and the distribution $P_{\bs{U}_{i,k}}$ of the Prob-GParareal solution at interval $i\in\{1,\ldots,N\}$ and iteration $k\in \mathbb{N}$ given by
\[
    e_{i,k} := W_2( \delta_{\bs{u}(t_i)}, P_{\bs{U}_{i,k}} )^2 = \inf_{ \gamma \in \Gamma( \delta_{\bs{u}(t_i)}, P_{\bs{U}_{i,k}} ) } \int _{\R^d \times \R^d} \| \bs{u}(t_i) - \bs{u}_{i,k} \|^2 d \gamma(\bs{u}(t_i), \bs{u}_{i,k}),
    \]
    where $\Gamma( \delta_{\bs{u}(t_i)}, P_{\bs{U}_{i,k}} )$ is the set of all couplings (joint distributions) with marginals $ \delta_{\bs{u}(t_i)}$ and $ P_{\bs{U}_{i,k}}$. Since $ \delta_{\bs{u}(t_i)}$ is a Dirac measure, the optimal coupling is unique and is given by the product measure (see~\citealt{villani2008optimal})  $    \gamma = \delta_{\bs{u}(t_i)} \otimes P_{\bs{U}_{i,k}}
$. Thus,
    \begin{align}
        e_{i,k} &= \int _{\R^d \times \R^d} \| \bs{u}(t_i) - \bs{u}_{i,k} \|^2 d (\delta_{\bs{u}(t_i)} \otimes P_{\bs{U}_{i,k}})(\bs{u}(t_i), \bs{u}_{i,k})\nonumber\\
         &=    \int_{\R^d \times \R^d} \| \bs{u}(t_i) - \bs{u}_{i,k} \|^2 \, d\delta_{\bs{u}(t_i)}\left(\bs{u}(t_i)\right) \, d P_{\bs{U}_{i,k}}(\bs{u}_{i,k})\nonumber\\
    &= \int_{\R^d} \| \bs{u}(t_i) - \bs{u}_{i,k} \|^2 \, dP_{\bs{U}_{i,k}}(\bs{u}_{i,k})= \mathbb{E}_{{\bs{U}_{i,k}}} \left[ \| \bs{u}(t_i) - \bs{U}_{i,k} \|^2 \right],\label{eik}
    \end{align}
    i.e., the $W_2^2$ distance equals the MSE of $\bs{u}(t_i)$, proving \eqref{eq:mse_w2}.
     We then seek to build and solve a recurrence relation for $e_{i,k}$. We first use the Prob-GParareal update rule~\eqref{eq:update_rule_probpara} and the assumption that the fine solver $\f$ is exact, and write
\begin{equation}
\label{eq:law of error}
    \bs{u}(t_i) - \bs{U}_{i,k} = \f(\bs{u}(t_{i-1}))- \g(\bs{U}_{i-1,k}) - \boldsymbol{Z}_{i,k}, \quad i=1,\ldots,N, \quad k \geq 1,
\end{equation}

where the conditional distribution of $\bs{Z}_{i,k}$ given $\bs{U}_{i-1,k}$ is $\boldsymbol{Z}_{i,k}|\bs{U}_{i-1, k} \sim$ \\$ \mathcal{N}_d(\boldsymbol{\mu}_{\mathcal{D}_k}\left(\bs{U}_{i-1, k}\right), \Sigma_{\mathcal{D}_k}(\bs{U}_{i-1, k}))$, see \eqref{eq:cond normal}, where $\bs{\mu}_{\mathcal{D}_k}$ and $\Sigma_{\mathcal{D}_k}$ are deterministic functions obtained by training a GP on $\mathcal{D}_{k}$, and for convenience, we write $\bs{Z}_{i,k}$ as
\begin{equation*}
    \boldsymbol{Z}_{i,k} = \bs{\mu}_{\mathcal{D}_k}(\boldsymbol{U}_{i-1,k}) + \boldsymbol{\xi}_{i,k}, \quad {\rm with} \enspace
    \quad \boldsymbol{\xi}_{i,k}|\bs{U}_{i-1,k}\sim \mathcal{N}_d\left( \mathbf{0}, \Sigma_{\mathcal{D}_k}(\boldsymbol{U}_{i-1,k}) \right).
   \end{equation*}

To derive a recurrence for errors $e_{i,k}$, $i=1,\ldots, N$, $k\in \mathbb{N}$, we add and subtract $\f(\bs{U}_{i-1,k})$, $\g(\bs{u}(t_{i-1}))$, and $\g(\bs{U}_{i-1,k})$ on the right-hand side of \eqref{eq:law of error}, and making use of the definition of $f_c$, we obtain the following decomposition:
    \begin{align}
        \boldsymbol{u}(t_i) - \boldsymbol{U}_{i,k} =& \underbrace{\left( f_c(\bs{U}_{i-1,k}) - \bs{\mu}_{\mathcal{D}_k}(\boldsymbol{U}_{i-1,k})\right)}_{Q} +
        \underbrace{(- \bs{\xi}_{i,k})}_{W} +
        \underbrace{\left( \g(\bs{u}(t_{i-1})) - \g(\bs{U}_{i-1,k}) \right)}_{R} \nonumber \\
        &+ \quad \underbrace{\left( f_c(\bs{u}(t_{i-1})) - f_c(\bs{U}_{i-1,k}) \right) }_{S},\label{error decomp}
    \end{align}
    for all $i=1,\ldots,N$ and $k\in \mathbb{N}$.
    By taking norm squared and applying the expectation operator on both sides of~\eqref{error decomp},
    we obtain the following bound for the error expression~\eqref{eik}:
    \begin{align*}
            e_{i,k}
            &=\mathbb{E}_{ {\boldsymbol{U}_{i-1,k}} } \left[\mathbb{E}_{ {\boldsymbol{U}_{i,k}|\bs{U}_{i-1,k}} } \left[ \| \boldsymbol{u}(t_i) - \boldsymbol{U}_{i,k} \|^2 \right] \right]\\
            &\leq
            4 \left(\mathbb{E}_{ {\boldsymbol{U}_{i-1,k}} } \left[ \| Q \|^2\right]+ \mathbb{E}_{ {\boldsymbol{U}_{i-1,k}} } \left[ \| W \|^2\right] + \mathbb{E}_{{\boldsymbol{U}_{i-1,k}} } \left[ \| R \|^2\right]+ \mathbb{E}_{{\boldsymbol{U}_{i-1,k}} }  \left[ \| S \|^2\right]\right).
    \end{align*}
    In this derivation, the first equality follows by the law of total expectation with the conditional expectation taken with respect to the conditional law in~\eqref{eq:pp_cond_distr}, while the inequality holds due to
   \begin{equation}
   \label{vector ineq}
    \left\| \sum_{j=1}^m \bs{v}_j \right\|^2 \leq m \sum_{j=1}^m \left\|\bs{v}_j\right\|^2,
    \end{equation}
    for any $\boldsymbol{v}_j$, $j=1,\ldots,m$, using the fact that the terms in~\eqref{error decomp} are measurable with respect to the conditional expectation. We continue by analyzing the terms on the right-hand side of this expression one by one. For the $R$ term, by Assumption~\ref{ass:3_beg}, we write
    \[
    \mathbb{E}_{{\boldsymbol{U}_{i-1,k}}} \left[ \| R \|^2\right]
        \leq L_{\g}^2 \mathbb{E}_{{\boldsymbol{U}_{i-1,k}}} \left[  \left\|   \boldsymbol{u}(t_{i-1}) - \boldsymbol{U}_{i-1,k}\right\|^2 \right]= L_{\g}^2 \,e_{i-1,k}.
    \]
    Next, for the $S$ term, note that Assumption~\ref{ass:2_beg}
holds by Remark~\ref{common_remark} under
Assumption~\ref{ass:1_beg}, and hence\[
\mathbb{E}_{{\boldsymbol{U}_{i-1,k}} } \left[ \| S \|^2\right]
\leq L_c^2\mathbb{E}_{\boldsymbol{U}_{i-1,k}}\left[\left\|\bs{u}(t_{i-1}) - \bs{U}_{i-1,k} \right\|^2\right]= L_c^2 \, e_{i-1,k}.
        \]
    For the  $Q$ term, we proceed as follows:
    \begin{align*}
\mathbb{E}_{{\boldsymbol{U}_{i-1,k}}} \left[ \| Q \|^2\right]
&= \mathbb{E}_{{\boldsymbol{U}_{i-1,k}}} \left[ \sum_{s=1}^d \left| f_c^{(s)}(\bs{U}_{i-1,k}) - \boldsymbol{\mu}^{(s)}_{\mathcal{D}_k}(\bs{U}_{i-1,k}) \right|^2 \right]\\
&\leq \mathbb{E}_{\boldsymbol{U}_{i-1,k}} \left[ \sum_{s=1}^d {\sigma^{(s)\; }_{\mathcal{D}_k}}(\bs{U}_{i-1,k})^2 \;\left\| f_c^{(s)} \right\|_{\mathcal{H}_{{K}}}^2 \right]\\
&\leq \max_{1\leq s\leq d}  \left\| f_c^{(s)} \right\|_{\mathcal{H}_{{K}}}^2 \sum_{s=1}^d\mathbb{E}_{\boldsymbol{U}_{i-1,k}} \left[   {\sigma^{(s)}_{\mathcal{D}_k}}(\bs{U}_{i-1,k})^2  \right]\\
&= \left\| f_c \right\|_{\infty, \mathcal{H}_K}^2 \sum_{s=1}^d\mathbb{E}_{\boldsymbol{U}_{i-1,k}} \left[    {\sigma^{(s)}_{\mathcal{D}_k}}(\bs{U}_{i-1,k})^2  \right],
\end{align*}
    where we used~\Cref{th:GP_mean_bound} in the first inequality and applied Definition~\ref{def:max RKHS norm} of the maximum norm in the last equality. Lastly, we apply~\Cref{th:posterior variance p smooth} part \textbf{(ii)} to bound posterior variances ${\sigma^{(s)}_{\mathcal{D}_k}}^2$, $s=1,\ldots,d$, and obtain
    \begin{align*}
\mathbb{E}_{{\boldsymbol{U}_{i-1,k}} }  \left[ \| Q \|^2\right]&\leq \|f_c \|_{\infty, \mathcal{H}_K}^2 \sum_{s=1}^d C_{s,\alpha} \mathbb{E}_{\boldsymbol{U}_{i-1,k}} \left[ h_{\rho, \mathcal{D}_k}(\bs{U}_{i-1,k})^{{\alpha}}\right]\\
&\leq C_{\alpha} \; d \; \| f_c \|_{\infty, \mathcal{H}_K}^2 \mathbb{E}_{\boldsymbol{U}_{i-1,k}} \left[ h_{\rho, \mathcal{D}_k}(\bs{U}_{i-1,k})^{\alpha}\right],
\end{align*}
with $C_{\alpha}=\max_{1\leq s\leq d} C_{s,\alpha}$.

    Finally, for the $W$ term, we use~\Cref{th:posterior variance p smooth} part \textbf{(ii)} and write:
    \begin{align*}
          \mathbb{E}_{{\boldsymbol{U}_{i-1,k}} }  \left[ \| W \|^2\right]&
          = \mathbb{E}_{{\boldsymbol{U}_{i-1,k}} } \left[ \operatorname{Tr} \left( \Sigma_{\mathcal{D}_k} ( \boldsymbol{U}_{i-1,k} ) \right )\right] =\sum_{s=1}^d\mathbb{E}_{{\boldsymbol{U}_{i-1,k}} } \left[    {\sigma^{(s)}_{\mathcal{D}_k}}(\bs{U}_{i-1,k}) ^2 \right]\\
          &\leq \sum_{s=1}^d C_{s,\alpha} \mathbb{E}_{{\boldsymbol{U}_{i-1,k}}} \left[  h_{\rho, \mathcal{D}_k}(\bs{U}_{i-1,k})^\alpha\right]\leq  d \; \max_{1\leq s\leq d} C_{s,\alpha} \mathbb{E}_{\boldsymbol{U}_{i-1,k}} \left[h_{\rho, \mathcal{D}_k}(\bs{U}_{i-1,k})^\alpha\right]\\
          &\leq d \; C_{\alpha} \mathbb{E}_{\boldsymbol{U}_{i-1,k}} \left[h_{\rho, \mathcal{D}_k}(\bs{U}_{i-1,k})^\alpha\right].
    \end{align*}
    Overall, we have the following bound for $e_{i,k}$
    \begin{align*}
        e_{i,k}  \leq &4 L_{\g}^2 \,e_{i-1,k} + 4 L_c^2 \, e_{i-1,k}   + 4d \; C_{\alpha} \mathbb{E}_{{\boldsymbol{U}_{i-1,k}}} \left[h_{\rho, \mathcal{D}_k}(\bs{U}_{i-1,k})^\alpha\right] \\
        & +4d \; C_{\alpha} \| f_c \|_{\infty,{\mathcal{H}_{{K}}}}^2 \mathbb{E}_{{\boldsymbol{U}_{i-1,k}}} \left[ h_{\rho, \mathcal{D}_k}(\bs{U}_{i-1,k})^{\alpha} \right], \quad i=1,\ldots, N, \enspace k\in \mathbb{N}.
    \end{align*}
    This upper bound can be written as the following recurrence
    \[
    e_{i,k} \leq a \, e_{i-1,k} + b_{i-1,k}, \quad i=1,\ldots, N, \enspace k\in \mathbb{N},
    \]
    with  $a = 4\left(L_{\g}^2 + L_c^2 \right)$ and
    \begin{align}
        b_{i-1,k} = 4d \, C_{\alpha} (1+\|f_c \|_{\infty,{\mathcal{H}_{{K}}}}^2) \mathbb{E}_{\bs{U}_{i-1,k}} \left[h_{\rho, \mathcal{D}_k}(\bs{U}_{i-1,k})^\alpha\right].\label{eq:b term}
    \end{align}
Unrolling the recurrence, we obtain
    \[
    e_{i,k} = W_2( \delta_{\boldsymbol{u}(t_i)}, P_{\boldsymbol{U}_{i,k}} )^2 \le\; a^i\,e_{0,k} \;+\;\sum_{j=1}^i a^{\,i-j}\,b_{j-1,k} \quad {\rm for }\enspace {\rm all} \enspace 1\leq i\leq N, \enspace k\in \mathbb{N}.
    \]
By~\eqref{eik},
\[
    e_{0,k}
    =
    \mathbb{E}_{\bs{U}_{0,k}}
    \left[
        \|\bs{u}(t_0)-\bs{U}_{0,k}\|^2
    \right],
\]
and since
$\mathbb{E}[\bs{U}_{0,k}]=\bs{u}_{(0)}=\bs{u}(t_0)$
and
${\rm Var}(\bs{U}_{0,k})=\Sigma_{0,k}$ by assumption,
\[
    e_{0,k}
    =
    \mathbb{E}_{\bs{U}_{0,k}}
    \left[
        \|\bs{U}_{0,k}-\mathbb{E}[\bs{U}_{0,k}]\|^2
    \right]
    =
    \operatorname{Tr}\!\left({\rm Var}(\bs{U}_{0,k})\right)
    =
    \operatorname{tr}(\Sigma_{0,k}).
\]

\medskip

For the remaining cases, the structure of the proof is unchanged, with minor modifications, which we briefly point out  below.

\paragraph{Case \textbf{(i)} Differentiability.}
The differentiability case differs from the infinite smoothness one in that  \Cref{th:posterior variance p smooth} does not apply, and is replaced by Assumption \ref{ass:Posterior variance decay}. In fact, under the conditions of Assumption \ref{ass:Posterior variance decay}, the $W$ and $Q$ terms are bounded using
\[
\sigma_{\mathcal{D}_k}^{(s)}(\bs{u}')^2 \leq C_{\beta, s} h_{\rho, \mathcal{D}_k}(\bs{u}')^\beta, \quad {\rm for}  \enspace s=1,\ldots,d.
\]
This leads to a similar expression for $b_{i-1,k}$ in~\eqref{eq:b term}, differing in the constant and the exponents of the fill distance:
\begin{align*}
b_{i-1,k} &= 4d \, C_{\beta} \left( 1+
              \| f_c \|_{\infty,{\mathcal{H}_{{K}}}}^2 \right)\mathbb{E}_{\bs{U}_{i-1,k}} \left[h_{\rho, \mathcal{D}_k}(\bs{U}_{i-1,k})^{\beta} \right],
\end{align*}
with $C_{\beta}=\max_{1\leq s\leq d}C_{\beta,s}$ with $\{C_{\beta,s}\}_{s=1}^d$, $C_{\beta,s}>0$, as in Assumption~\ref{ass:Posterior variance decay}.

\paragraph{Case \textbf{(ii)}  Sobolev norm-equivalence.} The only change required is the application of part {\bf(i)} of Theorem~\ref{th:posterior variance p smooth}, instead of  part {\bf(ii)} as for the infinite smoothness case. This leads to a different constant in the $b_{i-1,k}$ expression, and different exponent of the fill distance, namely
\[
b_{i-1,k} = 4d \, C \left( 1+
              \| f_c \|_{\infty,W^q_2}^2\right) \mathbb{E}_{\bs{U}_{i-1,k}}\left[h_{\rho, \mathcal{D}_k}(\bs{U}_{i-1,k})^{2q-d} \right],
\]
where $C=\max_{1\leq s\leq d}C_s$ with $\{C_s\}_{s=1}^d$, $C_s>0$, defined in part {\bf(i)} of Theorem~\ref{th:posterior variance p smooth}. $\blacksquare$

\subsection{Proof of \texorpdfstring{\Cref{th:probpara:var:bound}}{th}}
\label{app:pr:var_bound}
Similarly to the proof of~\Cref{prop:convergence} in~\Cref{app:pr:conv}, here we provide the proof of~\Cref{th:probpara:var:bound} for the infinite smoothness case, part \textbf{(iii)}. Parts \textbf{(i)} and \textbf{(ii)} require minor modifications, which are detailed at the end of the section.

\paragraph{Case ({\bf iii}) Smoothness.}
    Our goal is to upper bound the coordinate-wise variances of the Prob-GParareal solution $\bs{U}_{i+1,k}$ for all $i=1,\ldots, N$ and $k\in\mathbb{N}$, i.e., the diagonal elements of the covariance matrix $\rm{Var}(\bs{U}_{i+1,k})$, which we denote $\sigma_{i+1,k}^{(s),2}:=({\rm Var}(\bs{U}_{i+1,k}))^{(s)}, s=1,\ldots, d$, where for a matrix $S$ of dimension $d$, we write $S^{(s)}$ to denote the $s$th entry of its main diagonal.
 Then, we have
  \begin{eqnarray}        \nonumber\sigma^{(s),2}_{i+1,k}&=&\left(\mathbb{E}_{{\bs{U}_{i,k}}} \left[{\rm Var}_{{\bs{U}_{i+1,k}}|{\bs{U}_{i,k}}}\left(\boldsymbol{U}_{i+1,k}\right)\right]\right)^{(s)}+ \left({\rm Var}_{{\bs{U}_{i,k}}}\left(\mathbb{E}_{{\bs{U}_{i+1,k}}|{\bs{U}_{i,k}}}\left[\boldsymbol{U}_{i+1,k}\right]\right)\nonumber \right)^{(s)},\\
        &=&\left(\mathbb{E}_{\bs{U}_{i,k}}[\Sigma_{\mathcal{D}_k}(\boldsymbol{U}_{i,k})]\right)^{(s)} + \left({\rm Var}_{\bs{U}_{i,k}}\left((\g + \bs{\mu}_{\mathcal{D}_k})(\boldsymbol{U}_{i,k})\right)\right)^{(s)},\label{var_matrix}
    \end{eqnarray}
 where we used the  law of total covariance in the first equality, and the Prob-GParareal update rule \eqref{eq:update_rule_probpara} given by $\boldsymbol{U}_{i+1,k} = \g(\boldsymbol{U}_{i,k}) + \boldsymbol{Z}_{i+1,k}$, $i=1,\ldots, N, \enspace k\in \mathbb{N}$,
    with $\bs{Z}_{i+1,k} |\boldsymbol{U}_{i,k} \sim \mathcal{N}_d\left(\bs{\mu}_{\mathcal{D}_k}(\boldsymbol{U}_{i,k}),  \Sigma_{\mathcal{D}_k}(\boldsymbol{U}_{i,k}) \right)$, see \eqref{eq:cond normal} in the second equality.

     By applying part {\bf(ii)} in~\Cref{th:posterior variance p smooth} on the first term on \eqref{var_matrix}, we obtain the following bound for $s=1,\ldots, d$:
\begin{equation}
    \label{eq:pr_conv:1}
\left(\mathbb{E}_{\bs{U}_{i,k}}[\Sigma_{\mathcal{D}_k}(\boldsymbol{U}_{i,k})]\right)^{(s)} = \sigma^{(s)}_{\mathcal{D}_k}(\boldsymbol
    {U}_{i,k})^2 \leq
       C_{\alpha,s} \mathbb{E}_{\bs{U}_{i,k}}\left[ h_{\rho,\mathcal{D}_{k}}(\bs{U}_{i,k})^\alpha \right].\end{equation}
Denoting $g:=\g + \bs{\mu}_{\mathcal{D}_k}$, using the definition of variance, the second term on  \eqref{var_matrix} becomes
\begin{align*}
    ({\rm Var}_{\bs{U}_{i,k}}&\left(g(\boldsymbol{U}_{i,k})\right))^{(s)} =\mathbb{E}_{\bs{U}_{i,k}}\left[\left( g^{(s)}(\bs{U}_{i,k}) - \mathbb{E}_{\bs{U}_{i,k}}\left[g^{(s)}(\bs{U}_{i,k})\right]\right)^2\right]\\
        =& \,\mathbb{E}_{\bs{U}_{i,k}}\left[\left( g^{(s)}(\bs{U}_{i,k}) \mp  g^{(s)}\left(\mathbb{E}_{\bs{U}_{i,k}}\left[\bs{U}_{i,k}\right]\right)
        - \mathbb{E}_{\bs{U}_{i,k}}\left[g^{(s)}(\bs{U}_{i,k})\right]\right)^2\right]\\
        =\,&\mathbb{E}_{\bs{U}_{i,k}}\left[
        \left(
        g^{(s)}(\bs{U}_{i,k}) - g^{(s)}\left(\mathbb{E}_{\bs{U}_{i,k}}\left[\bs{U}_{i,k}\right]\right)
        \right)^2\right]\\
        &+ \mathbb{E}_{\bs{U}_{i,k}}\left[\left(
        g^{(s)}\left(\mathbb{E}_{\bs{U}_{i,k}}\left[\bs{U}_{i,k}\right]\right)
        - \mathbb{E}_{\bs{U}_{i,k}}\left[g^{(s)}(\bs{U}_{i,k})\right]\right)^2\right] \\
        &+ 2\,\mathbb{E}_{\bs{U}_{i,k}}\left[\left( g^{(s)}(\bs{U}_{i,k}) - g^{(s)}\left(\mathbb{E}_{\bs{U}_{i,k}}\left[\bs{U}_{i,k}\right]\right)\right) \left( g^{(s)}\left(\mathbb{E}_{\bs{U}_{i,k}}\left[\bs{U}_{i,k}\right]\right) - \mathbb{E}\left[g^{(s)}(\bs{U}_{i,k})\right]\right)\right]\\
        =\,& \mathbb{E}_{\bs{U}_{i,k}}\left[
        \left(
        g^{(s)}(\bs{U}_{i,k}) - g^{(s)}\left(\mathbb{E}_{\bs{U}_{i,k}}\left[\bs{U}_{i,k}\right]\right)
        \right)^2\right]\\
        &- \mathbb{E}_{\bs{U}_{i,k}}\left[\left(
        g^{(s)}\left(\mathbb{E}_{\bs{U}_{i,k}}\left[\bs{U}_{i,k}\right]\right)
        - \mathbb{E}_{\bs{U}_{i,k}}\left[g^{(s)}(\bs{U}_{i,k})\right]\right)^2\right]\\
        \leq \,& \mathbb{E}_{\bs{U}_{i,k}}\left[
        \left(
        g^{(s)}(\bs{U}_{i,k}) - g^{(s)}\left(\mathbb{E}_{\bs{U}_{i,k}}\left[\bs{U}_{i,k}\right]\right)
        \right)^2\right].
        \end{align*}
    In this derivation, we added and subtracted an intermediate term (second equality), calculated the square (third equality), and used the fact that the second multiplier in the cross-term is deterministic (fourth equality).
    Next, by replacing $g$ with its original definition on the right-hand side, and using inequality \eqref{vector ineq},
     we obtain
  \begin{align*}
    \left({\rm Var}\left(g(\boldsymbol{U}_{i,k})\right)\right)^{(s)}
        &\leq 2\underbrace{\mathbb{E}\left[\left(
  \g^{(s)}(\boldsymbol{U}_{i,k}) - \g^{(s)}(\mathbb{E}[\boldsymbol{U}_{i,k}])\right)^2\right]}_{\text{Term } T_1}+  2\underbrace{\mathbb{E}\left[\left(
  \bs{\mu}_{\mathcal{D}_k}^{(s)}(\boldsymbol{U}_{i,k}) -  \bs{\mu}_{\mathcal{D}_k}^{(s)}(\mathbb{E}[\boldsymbol{U}_{i,k}])\right)^2\right]}_{\text{Term } T_2}.
    \end{align*}
Assumption~\ref{ass:3_beg} implies
 \begin{align}
 \label{T1bound}
    T_1&\leq   L_{\g}^2 \; \mathbb{E}\left[\|\boldsymbol{U}_{i,k} - \mathbb{E}[\boldsymbol{U}_{i,k}]\|^2\right] =  L_{\g}^2 \operatorname{Tr}({\rm Var}(\boldsymbol{U}_{i,k}))=L_{\g}^2 \sum_{s=1}^d\sigma^{(s),2}_{i,k}.
  \end{align}
   For the $T_2$ term, we add and subtract intermediate terms $f_c^{(s)}(\boldsymbol{U}_{i,k})$ and $f_c^{(s)}(\mathbb{E}[\boldsymbol{U}_{i,k}])$, and obtain
  \begin{align*}
    T_2 &= \mathbb{E}\Big[\Big(
  \bs{\mu}_{\mathcal{D}_k}^{(s)}(\boldsymbol{U}_{i,k}) \pm
  f_c^{(s)}(\boldsymbol{U}_{i,k})
  \mp
  f_c^{(s)}(\mathbb{E}[\boldsymbol{U}_{i,k}])
  -  \bs{\mu}_{\mathcal{D}_k}^{(s)}(\mathbb{E}[\boldsymbol{U}_{i,k}])\Big)^2\Big]\\
  &\leq 3 \mathbb{E}\left[\left(
  \bs{\mu}_{\mathcal{D}_k}^{(s)}(\boldsymbol{U}_{i,k}) - f_c^{(s)}(\boldsymbol{U}_{i,k})\right)^2\right] + 3 \left( f_c^{(s)}(\mathbb{E}[\boldsymbol{U}_{i,k}]) -  \bs{\mu}_{\mathcal{D}_k}^{(s)}(\mathbb{E}[\boldsymbol{U}_{i,k}]) \right)^2 \\
  &\quad + 3 \mathbb{E}\left[\left(
  f_c^{(s)}(\boldsymbol{U}_{i,k}) - f_c^{(s)}(\mathbb{E}[\boldsymbol{U}_{i,k}]) \right)^2\right],
  \end{align*}
  where the last inequality follows again from \eqref{vector ineq}. Using~\Cref{th:GP_mean_bound} for the first and the second summand, and the fact that Assumption~\ref{ass:2_beg}
holds by Remark~\ref{common_remark} under
Assumption~\ref{ass:1_beg} for the third one, we get
\begin{align*}
    T_2\leq & {3 \mathbb{E}\left[{\sigma^{(s)}_{\mathcal{D}_k}}(\boldsymbol{U}_{i,k})^2 \;\right] \| f_c^{(s)} \|^2_{\mathcal{H}_{{K}}} + 3 {\sigma^{(s)}_{\mathcal{D}_k}}(\mathbb{E}\left[\boldsymbol{U}_{i,k}\right])^2 \| f_c^{(s)} \|^2_{\mathcal{H}_{{K}}} + 3 L_c^2 \mathbb{E}\left[\|\boldsymbol{U}_{i,k} - \mathbb{E}[\boldsymbol{U}_{i,k}]\|^2\right]} \\
    &\leq  3 C_{\alpha,s} \| f_c^{(s)} \|^2_{\mathcal{H}_{{K}}} \left\{\mathbb{E}\left[ h_{\rho,\mathcal{D}_{k}}(\bs{U}_{i,k})^{\alpha} \right]  +  h_{\rho,\mathcal{D}_{k}}(\mathbb{E}\left[\bs{U}_{i,k}\right])^{\alpha}\right\} + 3 L_c^2 \sum_{s=1}^d\sigma^{(s),2}_{i,k},
  \end{align*}
  where in the last inequality we used again part {\bf(ii)} in~\Cref{th:posterior variance p smooth} and analogous derivation to the one in \eqref{T1bound}.

  Finally,  using \eqref{eq:pr_conv:1} for the first summand on the right-hand side of \eqref{var_matrix},
  and terms $T_1$ and $T_2$ for the second summand, we obtain for \eqref{var_matrix}
  \begin{align*}
\sigma^{(s),2}_{i+1,k}
\leq& \, C_{\alpha,s} \mathbb{E}\left[ h_{\rho,\mathcal{D}_{k}}(\bs{U}_{i,k})^\alpha \right] + 2 L_{\g}^2 \sum_{s=1}^d\sigma^{(s),2}_{i,k}
\\&+ 6 C_{\alpha,s} \| f_c^{(s)} \|^2_{\mathcal{H}_{{K}}} \left\{\mathbb{E}\left[ h_{\rho,\mathcal{D}_{k}}(\bs{U}_{i,k})^{\alpha} \right]  +  h_{\rho,\mathcal{D}_{k}}(\mathbb{E}\left[\bs{U}_{i,k}\right])^{\alpha}\right\}  + 6 L_c^2 \sum_{s=1}^d\sigma^{(s),2}_{i,k}.
\end{align*}
Maximizing the variance across all components on both sides of the inequality,  we obtain a recursion for $\sigma_{i+1,k}^{{\rm max},2}$, defined in \eqref{max var}, as:
\begin{equation}
\label{max_recursion}
	\sigma^{{\rm max},2}_{{i+1,k}} \leq a \, \sigma^{{\rm max},2}_{{i,k}} + b_{i,k}, \quad i=0,\ldots,N-1, \enspace k\in \mathbb{N},
\end{equation}
where $a = 2 \; d(L_\g^2 + 3L_c^2)$ and 
    \begin{align*}
        b_{i,k} &= C_{\alpha} \mathbb{E}\left[ h_{\rho,\mathcal{D}_{k}}(\bs{U}_{i,k})^\alpha \right] + 6 C_{\alpha} \| f_c \|^2_{\infty,\mathcal{H}_{{K}}} \left\{\mathbb{E}\left[ h_{\rho,\mathcal{D}_{k}}(\bs{U}_{i,k})^{\alpha} \right]  +  h_{\rho,\mathcal{D}_{k}}(\bs{\mu}_{i,k})^{\alpha}\right\},
    \end{align*}
    with $C_{\alpha}=\max_{1\leq s\leq d} C_{\alpha,s}$. Unrolling the recurrence and re-indexing the sum, we obtain
    \[
    \sigma^{{\rm max},2}_{i,k} \le a^i\,\sigma^{{\rm max},2}_{0,k} + \sum_{j=1}^{\,i}a^{\,i-j}\,b_{j-1,k}, \quad i=1,\ldots,N, \enspace k\in \mathbb{N},
    \]
 as required.

\medskip

We now cover the remaining cases.

\paragraph{Case \textbf{(i)} Differentiability.} As seen in~\Cref{app:pr:conv}, the changes from the infinite smoothness case involve updating constant and exponents of the fill distance whenever either \Cref{th:posterior variance p smooth} was applied (note that \Cref{th:GP_mean_bound} depends on \Cref{th:posterior variance p smooth}), which is now replaced by Assumption \ref{ass:Posterior variance decay}, or by Assumption \ref{ass:2_beg}, which holds automatically with $L_c\leq\sqrt{d}\,(c_1+C_R\,\Delta t)\,\Delta t^{p+1}$
as in Remark~\ref{common_remark}. In particular, using Assumption \ref{ass:Posterior variance decay} (instead of \Cref{th:posterior variance p smooth} in~\eqref{eq:pr_conv:1}), we obtain
\begin{equation}
        \left(\mathbb{E}[\Sigma_{\mathcal{D}_k}(\bs{U}_{i,k})]\right)^{(s)} = \sigma^{(s)}_{\mathcal{D}_k}(\bs{U}_{i,k})^2 \leq
   C_{\beta,s} \mathbb{E}\left[ h_{\rho,\mathcal{D}_{k}}(\bs{U}_{i,k})^\beta \right].   \end{equation}
The other change is in the derivation of the $T_2$ term, which relied on both Assumptions \ref{ass:2_beg} and \ref{ass:Posterior variance decay}. Similarly to what shown in~\Cref{app:pr:conv}, only the constant and the exponents change, yielding
\begin{align*}
    T_2\leq &   3 C_{\beta,s} \| f_c^{(s)} \|^2_{\mathcal{H}_{{K}}} \left\{\mathbb{E}\left[ h_{\rho,\mathcal{D}_{k}}(\bs{U}_{i,k})^{\beta} \right]  +  h_{\rho,\mathcal{D}_{k}}(\mathbb{E}\left[\bs{U}_{i,k}\right])^{\beta}\right\}  + 3 L_c^2 \sum_{s=1}^d\sigma^{(s),2}_{i,k}.
  \end{align*}
Overall, we obtain that there exists constant $C_{\beta}>0$ such that
\begin{align*}
b_{i,k} &= C_{\beta} \mathbb{E}\left[ h_{\rho,\mathcal{D}_{k}}(\bs{U}_{i,k})^\beta \right] + 6 C_{\beta} \| f_c \|^2_{\infty,\mathcal{H}_{{K}}} \left\{\mathbb{E}\left[ h_{\rho,\mathcal{D}_{k}}(\bs{U}_{i,k})^{\beta} \right]  +  h_{\rho,\mathcal{D}_{k}}(\boldsymbol{\mu}_{i,k})^{\beta}\right\},
\end{align*}
where $C_{\beta}=\max_{1\leq s\leq d}C_{\beta,s}$ with $\{C_{\beta,s}\}_{s=1}^d$, $C_{\beta,s}>0$, as in Assumption~\ref{ass:Posterior variance decay}.

\paragraph{Case \textbf{(ii)}  Sobolev norm-equivalence.} The only change required here is the application of part {\bf(i)} of Theorem~\ref{th:posterior variance p smooth} instead of  part {\bf(ii)} as for the infinite smoothness case. This leads to a different constant in the $b_{i,k}$ expression, and different exponent of the fill distance
\begin{align*}
b_{i,k} &= C \mathbb{E}\left[ h_{\rho,\mathcal{D}_{k}}(\bs{U}_{i,k})^{2q-d} \right] + 6 C \| f_c \|^2_{\infty,W^q_2} \left\{\mathbb{E}\left[ h_{\rho,\mathcal{D}_{k}}(\bs{U}_{i,k})^{2q-d} \right]  +  h_{\rho,\mathcal{D}_{k}}(\boldsymbol{\mu}_{i,k})^{2q-d}\right\},
\end{align*}
where $C=\max_{1\leq s\leq d}C_s$ with $\{C_s\}_{s=1}^d$, $C_s>0$, defined in part {\bf(i)} of Theorem~\ref{th:posterior variance p smooth}. $\blacksquare$

\subsection{Proof of \texorpdfstring{\Cref{prop:meanbound}}{th}}
\label{app:pr:mean_bound}
We provide the proof of \Cref{prop:meanbound}, part \textbf{(iii)}. Parts \textbf{(i)} and \textbf{(ii)} follow the same strategy as detailed in~\Cref{app:pr:var_bound}, and are therefore omitted. Explicit values for the coefficients $a$ and $b_{i,k}$ for these cases are given in  \Cref{prop:meanbound}. The proof for Corollary~\ref{cor:mean_full_expr} is provided at the end of the section.

\paragraph{Case ({\bf iii}) Smoothness.}
Although the following derivations are demonstrated for some fixed interval $i$ and iteration $k$, they hold true for any $i\in \{1,\ldots, N\}$ and $k\in \mathbb{N}$ . We start by recalling that, according to the introduced notation,
    \begin{align}
    \label{error to bound}
        \left\| \bs{\mu}_{i+1,k}- \boldsymbol{u}_{i+1,k}^{\rm GPara} \right\| &=  \left\| \mathbb{E}\left[ \bs{U}_{i+1,k} \right]- \left(\g + \bs{\mu}_{\mathcal{D}_k}\right)(\boldsymbol{u}_{i,k}^{\rm GPara})\right\|.
    \end{align}
Using the law of total expectation and the Prob-GParareal update rule \eqref{eq:update_rule_probpara}-\eqref{eq:cond normal}, we obtain
    \[
    \begin{aligned}
        \mathbb{E}[\bs{U}_{i+1,k}]&=\mathbb{E}_{\bs{U}_{i,k}} [\mathbb{E}_{\bs{U}_{i+1,k}|\bs{U}_{i,k}}[\bs{U}_{i+1,k}]]=\mathbb{E}[(\g+\bs{\mu}_{\mathcal{D}_k})(\bs{U}_{i,k})],
        \end{aligned}
    \]
By the hypotheses of the theorem, $(\g+\bs{\mu}_{\mathcal{D}_k})\in C^2$, so we expand it around $\mathbb{E}\left[ \bs{U}_{i,k} \right ] \in \mathbb{R}^d$ using the second-order Taylor expansion as follows:
    \begin{align}
    \label{eq:Taylor}
    (\g+\bs{\mu}_{\mathcal{D}_k})(\bs{U}_{i,k}) = (\g+\bs{\mu}_{\mathcal{D}_k}) \left(\mathbb{E}\left[ \bs{U}_{i,k} \right ]\right) + J_{(\g+\bs{\mu}_{\mathcal{D}_k})}\left(\mathbb{E}\left[ \bs{U}_{i,k} \right ]\right) \left(\bs{U}_{i,k} - \mathbb{E}\left[ \bs{U}_{i,k} \right ]\right) + \bs{R}_{i,k}.
    \end{align}
    In this expansion, $J_{(\g+\bs{\mu}_{\mathcal{D}_k})}$ is the Jacobian matrix of $(\g+\bs{\mu}_{\mathcal{D}_k})$ and $\bs{R}_{i,k}\in\mathbb{R}^d$ is defined as
        \[
    \bs{R}_{i,k}^{(s)} = \frac{1}{2} \left(\bs{U}_{i,k}-\mathbb{E}\left[ \bs{U}_{i,k} \right ]\right)^\top H_{(\g+\bs{\mu}_{\mathcal{D}_k})^{(s)}}(\bs{\zeta}_{i,k})     \left(\bs{U}_{i,k}-\mathbb{E}\left[ \bs{U}_{i,k} \right ]\right), \enspace s=1,\ldots d,
    \]
    where $H_{(\g+\bs{\mu}_{\mathcal{D}_k})^{(s)}}$ denotes the Hessian matrix of $(\g+\bs{\mu}_{\mathcal{D}_k})^{(s)}$ evaluated at $\bs{\zeta}_{i,k} \in \mathbb{R}^d$ on the line segment between $\bs{U}_{i,k}$ and $\mathbb{E}\left[ \bs{U}_{i,k} \right ]$.

    By taking expectation with respect to $\bs{U}_{i,k}$ on both sides of \eqref{eq:Taylor}, the term involving the Jacobian vanishes and one writes
    \begin{equation}
        \label{eq:proof:mn_err_1}
        \mathbb{E}\left[(\g+\bs{\mu}_{\mathcal{D}_k})(\bs{U}_{i,k}) \right]= (\g+\bs{\mu}_{\mathcal{D}_k})(\mathbb{E}\left[ \bs{U}_{i,k} \right]) + \mathbb{E}\left[ \bs{R}_{i,k} \right].
    \end{equation}
    We now further analyze the remainder $\bs{R}_{i,k}$:
   \begin{align}
    \left|\mathbb{E}\left[ \bs{R}_{i,k}^{(s)} \right] \right|&\leq
    \mathbb{E}\left[ \left| \bs{R}_{i,k}^{(s)} \right| \right]
    = \frac{1}{2}\mathbb{E} \left[ \left| \left(\bs{U}_{i,k}
    -\mathbb{E}\left[ \bs{U}_{i,k} \right]\right)^\top H_{(\g+\bs{\mu}_{\mathcal{D}_k})^{(s)}}(\bs{\zeta}_{i,k})
    \left(\bs{U}_{i,k}-\mathbb{E}\left[ \bs{U}_{i,k} \right]\right)
    \right| \right] \nonumber\\
    &\leq \frac{1}{2}\mathbb{E} \left[\left\|H_{(\g+\bs{\mu}_{\mathcal{D}_k})^{(s)}}(\bs{\zeta}_{i,k}) \right\|
    \left\| \bs{U}_{i,k}-\mathbb{E}\left[ \bs{U}_{i,k}
    \right]\right\|^2 \right]
    \leq \frac{M_s}{2} \, \mathbb{E}\left[ \left\| \bs{U}_{i,k}
    -\mathbb{E}\left[ \bs{U}_{i,k} \right] \right\|^2\right] \nonumber \\
    &= \frac{M_s}{2} \operatorname{Tr}\left( {\rm Var}
    \left( \bs{U}_{i,k}\right)\right)
    \leq \frac{M_s\,  d}{2} \, \sigma^{{\rm max},2}_{{i,k}},
    \label{eq:proof:2}
\end{align}
for all $s = 1,\ldots,d$, where we used that $|\bs{x}^\top A\bs{x}| \leq \|A\|\,\|\bs{x}\|^2$
for any $\bs{x}\in\R^d$ and symmetric matrix $A$ (second inequality), the assumption on
the boundedness of the spectral norm of the Hessian (third inequality), and the definition
of variance and~\eqref{max var} (last inequality).

We now  obtain
\begin{align}
\left\|\mathbb{E}\left[\bs{R}_{i,k}\right]\right\|
    &\leq \sqrt{d} \left\|\mathbb{E}\left[\bs{R}_{i,k}\right]\right\|_\infty \leq  \sqrt{d} \max_{1\leq s \leq d} \left|\mathbb{E}\left[ \bs{R}_{i,k}^{(s)} \right] \right| \nonumber \\
    & \leq \sqrt{d}  \max_{1\leq s \leq d} \frac{M_s d}{2}  \sigma^{{\rm max},2}_{{i,k}} = \frac{M  d^{3/2}}{2} \sigma^{{\rm max},2}_{{i,k}},
    \label{eq:proof:R_vector}
\end{align}
where we use that $M = \max_{1\leq s\leq d} M_s$.

      Using \eqref{eq:proof:mn_err_1} and \eqref{eq:proof:R_vector} in \eqref{error to bound}, we obtain
      \begin{align}
    \label{error to bound 3}
        \left\| \bs{\mu}_{i+1,k}- \boldsymbol{u}_{i+1,k}^{\rm GPara} \right\| \leq&  \left\| (\g+\bs{\mu}_{\mathcal{D}_k})(\mathbb{E}\left[ \bs{U}_{i,k} \right])  - \left(\g + \bs{\mu}_{\mathcal{D}_k}\right)(\boldsymbol{u}_{i,k}^{\rm GPara})\right\|+\|\mathbb{E}[ \bs{R}_{i,k} ]\|\nonumber\\
        \leq &  \left\| (\g+\bs{\mu}_{\mathcal{D}_k})(\mathbb{E}\left[ \bs{U}_{i,k} \right]) - \left(\g + \bs{\mu}_{\mathcal{D}_k}\right)(\boldsymbol{u}_{i,k}^{\rm GPara})\right\| + \frac{M \, d^{3/2}}{2} \, \sigma^{{\rm max},2}_{{i,k}} \nonumber \\
         \leq & \left\| \g(\mathbb{E}\left[ \bs{U}_{i,k} \right])   -  \g(\boldsymbol{u}_{i,k}^{\rm GPara})\right\|
         +
        \left\| \bs{\mu}_{\mathcal{D}_{k}}(\mathbb{E}\left[ \bs{U}_{i,k} \right ])   -  \bs{\mu}_{\mathcal{D}_{k}}(\boldsymbol{u}_{i,k}^{\rm GPara})\right\|\nonumber \\ &+ \frac{M \, d^{3/2}}{2}  \sigma^{{\rm max},2}_{{i,k}},
    \end{align}
    where we used the triangle inequality in the second and third inequalities. For the second summand on the right-hand side of \eqref{error to bound 3} we can use the same approach as for term $T_2$ in the proof of~\Cref{th:probpara:var:bound} (see~\Cref{app:pr:var_bound}), that is
    \[
    \begin{aligned}
       & \| \bs{\mu}_{\mathcal{D}_{k}}(\mathbb{E}\left[ \bs{U}_{i,k} \right ])-  \bs{\mu}_{\mathcal{D}_{k}}(\boldsymbol{u}_{i,k}^{\rm GPara})\|\\
       &\leq  \| \bs{\mu}_{\mathcal{D}_k}(\mathbb{E}\left[ \bs{U}_{i,k} \right ]) - f_c(\mathbb{E}\left[ \bs{U}_{i,k} \right ]) \| + \| f_c(\boldsymbol{u}_{i,k}^{\rm GPara}) - \bs{\mu}_{\mathcal{D}_{k}}(\boldsymbol{u}_{i,k}^{\rm GPara}) \| + \| f_c(\mathbb{E}\left[ \bs{U}_{i,k} \right ]) - f_c(\boldsymbol{u}_{i,k}^{\rm GPara}) \|.
        \end{aligned}
    \]
We now notice that by~\Cref{th:GP_mean_bound} one can bound the first summand on the right-hand side of this inequality. More precisely,
\[
    \begin{aligned}
        \| \bs{\mu}_{\mathcal{D}_k}(\mathbb{E}\left[ \bs{U}_{i,k} \right ]) - f_c(\mathbb{E}\left[ \bs{U}_{i,k} \right ]) \|^2 &\leq \sum_{s=1}^d  | \bs{\mu}^{(s)}_{\mathcal{D}_k}(\mathbb{E}\left[ \bs{U}_{i,k} \right ]) - f_c^{(s)}(\mathbb{E}\left[ \bs{U}_{i,k} \right ]) |^2 \nonumber \\
       &\leq \| f_c \|^2_{\infty,{\mathcal{H}_{{K}}}} \sum_{s=1}^d  \sigma^{(s)}_{\mathcal{D}_k} (\mathbb{E}\left[ \bs{U}_{i,k} \right ])^2\leq \| f_c \|^2_{\infty,{\mathcal{H}_{{K}}}} \sum_{s=1}^d C_{\alpha,s} h_{\rho, \mathcal{D}_k}(\mathbb{E}\left[ \bs{U}_{i,k} \right]) ^\alpha\\
       & \leq \| f_c \|^2_{\infty,{\mathcal{H}_{{K}}}} d C_{\alpha} h_{\rho, \mathcal{D}_k}(\mathbb{E}\left[ \bs{U}_{i,k} \right]) ^\alpha,
        \end{aligned}
    \]
with $C_{\alpha}=\max_{1\leq s\leq d} C_{\alpha,s}$,
and the second summand can be bounded analogously. For the third one, we use \eqref{ass:2}. We therefore write that
\[
    \begin{aligned}
        \| \bs{\mu}_{\mathcal{D}_{k}}(\mathbb{E}\left[ \bs{U}_{i,k} \right ])-  \bs{\mu}_{\mathcal{D}_{k}}(\boldsymbol{u}_{i,k}^{\rm GPara})\|\leq&      \sqrt{C_{\alpha} d} \| f_c \|_{\infty,{\mathcal{H}_{{K}}}} h_{\rho, \mathcal{D}_k}(\mathbb{E}\left[ \bs{U}_{i,k} \right]) ^{\alpha/2}\\&+ \sqrt{C_{\alpha} d} \| f_c \|_{\infty,{\mathcal{H}_{{K}}}} h_{\rho, \mathcal{D}_k}(\boldsymbol{u}_{i,k}^{\rm GPara}) ^{\alpha/2}  + L_c \| \mathbb{E}\left[\boldsymbol{U}_{i,k}\right]- \boldsymbol{u}_{i,k}^{\rm GPara}\|.
        \end{aligned}
    \]
Using this bound in \eqref{error to bound 3}, and noticing that the second summand  on its right-hand side can be bounded using Assumption~\ref{ass:3_beg}, we get
    \begin{align*}
        \left\| \bs{\mu}_{i+1,k}- \boldsymbol{u}_{i+1,k}^{\rm GPara} \right\| \leq &  \frac{M \, d^{3/2}}{2} \, \sigma^{{\rm max},2}_{{i,k}} + L_\g \| \mathbb{E}\left[\bs{U}_{i,k}\right]- \boldsymbol{u}_{i,k}^{\rm GPara}\| + L_c \| \mathbb{E}\left[\boldsymbol{U}_{i,k}\right]- \boldsymbol{u}_{i,k}^{\rm GPara}\| \\
        &+\sqrt{C_\alpha d} \| f_c \|_{\infty,{\mathcal{H}_{{K}}}} \left(h_{\rho, \mathcal{D}_k}(\mathbb{E}\left[ \bs{U}_{i,k} \right]) ^{\alpha/2} + h_{\rho, \mathcal{D}_k}(\boldsymbol{u}_{i,k}^{\rm GPara})^{\alpha/2}\right)\nonumber\\
        \leq &  \frac{M \, d^{3/2}}{2}  \sigma^{{\rm max},2}_{{i,k}} + (L_\g + L_c) \left\| \bs{\mu}_{i,k}- \boldsymbol{u}_{i,k}^{\rm GPara} \right\|   \\
        &+\sqrt{C_\alpha d} \| f_c \|_{\infty,{\mathcal{H}_{{K}}}} \left(h_{\rho, \mathcal{D}_k}(\bs{\mu}_{i,k}) ^{\alpha/2} + h_{\rho, \mathcal{D}_k}(\boldsymbol{u}_{i,k}^{\rm GPara})^{\alpha/2}\right).
    \end{align*}
  Notice that this expression is a recursive relation which could be written more compactly as:
    \[
    \| \bs{\mu}_{i+1,k}- \boldsymbol{u}_{i+1,k}^{\rm GPara} \| \leq a \| \bs{\mu}_{i,k}- \boldsymbol{u}_{i,k}^{\rm GPara} \|  + b_{i,k}, \quad i=1,\ldots,N, \enspace k\in \mathbb{N},
    \]
    where  $a = L_\g + L_c$ and
    \begin{align*}
        b_{i,k} &= \frac{M \, d^{3/2}}{2} \, \sigma^{{\rm max},2}_{{i,k}} + \sqrt{C_\alpha d} \| f_c \|_{\infty,{\mathcal{H}_{{K}}}}  \left(h_{\rho, \mathcal{D}_k}(\bs{\mu}_{i,k}) ^{\alpha/2} + h_{\rho, \mathcal{D}_k}(\boldsymbol{u}_{i,k}^{\rm GPara})^{\alpha/2}\right).
    \end{align*}
    Unrolling the recurrence from $\|\bs{\mu}_{0,k} -
\bs{u}_{0,k}^{\rm GPara}\|=0$ and re-indexing, we obtain
\[
\|\bs{\mu}_{i,k}- \boldsymbol{u}_{i,k}^{\rm GPara}\|
\leq \sum_{j=1}^{i} a^{i-j} b_{j-1,k}
\quad \text{for all } 1\leq i\leq N, \enspace k\in \mathbb{N},
\]
which proves the claim. $\blacksquare$

\paragraph{Proof of Corollary~\ref{cor:mean_full_expr}.}

The proof is given for case \textbf{(iii)}. Cases  \textbf{(i)} and \textbf{(ii)} can be proven analogously.
Let $a$ and $b_{i,k}$ be defined as in \eqref{a def var} and in part \textbf{(iii)} in~\Cref{th:probpara:var:bound}, respectively. We use $\tilde a$ to denote $a$ defined in \eqref{a def mean} and $\tilde b_{i,k}$  to denote $b_{i,k}$ in part \textbf{(iii)} of~\Cref{prop:meanbound}, respectively. For convenience, we recall the latter below
\begin{align*}
\tilde b_{l,k} &= \frac{M \, d^{3/2}}{2} \, \sigma^{{\rm max},2}_{{l,k}} + \sqrt{C_\alpha d} \| f_c \|_{\infty,{\mathcal{H}_{{K}}}} \left(h_{\rho, \mathcal{D}_k}(\boldsymbol{\mu}_{l,k}) ^{\alpha/2} + h_{\rho, \mathcal{D}_k}(\bs{u}_{l,k}^{\rm GPara})^{\alpha/2}\right),
\end{align*}
with $l=0,\ldots, N-1$, $k \in \mathbb{N}$.

Next, we substitute into this expression the variance bound \eqref{bound variance} in~\Cref{th:probpara:var:bound}, namely,
\begin{equation*}
\sigma^{{\rm max},2}_{l,k} \le a^l\,\sigma^{{\rm max},2}_{0,k} + \sum_{j=1}^{\,l}a^{\,l-j}\,b_{j-1,k},
\end{equation*}
and obtain the following inequality:
$$
\begin{aligned}
\tilde b_{l,k}
\le &\frac{M \, d^{3/2}}{2} \Bigl(a^l\,\sigma^{\max,2}_{0,k}
   \;+\;\sum_{j=1}^l a^{\,l-j}\,b_{j-1,k}\Bigr)
  \\
  &+
   \sqrt{C_\alpha d} \| f_c \|_{\infty,{\mathcal{H}_{{K}}}} \left(h_{\rho, \mathcal{D}_k}(\boldsymbol{\mu}_{l,k}) ^{\alpha/2} + h_{\rho, \mathcal{D}_k}(\bs{u}_{l,k}^{\rm GPara})^{\alpha/2}\right).
\end{aligned}
$$
Using this expression, we now rewrite the mean error
bound~\eqref{mean bound} in~\Cref{prop:meanbound} as follows:
\begin{align}
\|\boldsymbol{\mu}_{i,k}-\bs{u}_{i,k}^{\rm GPara}\|
&\le
\sum_{j=1}^i \tilde{a}^{\,i-j}\,\tilde{b}_{j-1,k}
\nonumber\\
&\le
\sum_{j=1}^i \tilde{a}^{\,i-j}
\Bigl[
\frac{M\,d^{3/2}}{2}
\Bigl(a^{j-1}\,\sigma^{\max,2}_{0,k}
+\sum_{m=1}^{j-1}a^{\,j-1-m}\,b_{m-1,k}\Bigr)
\nonumber\\
&\quad
+ \sqrt{C_\alpha d}\,\|f_c\|_{\infty,\mathcal{H}_K}
\bigl(h_{\rho,\mathcal{D}_k}(\boldsymbol{\mu}_{j-1,k})^{\alpha/2}
+h_{\rho,\mathcal{D}_k}(\bs{u}_{j-1,k}^{\rm GPara})^{\alpha/2} \bigr)
\Bigr]
\nonumber\\
&=
\frac{M\,d^{3/2}}{2}
\left(
\underbrace{
\sigma^{\max,2}_{0,k}
\sum_{j=1}^i\tilde{a}^{\,i-j}a^{j-1}
}_{\text{Term }T_1}
+
\underbrace{
\sum_{j=1}^i\sum_{m=1}^{j-1}
\tilde{a}^{\,i-j}a^{\,j-1-m}\,b_{m-1,k}
}_{\text{Term }T_2}
\right)
\nonumber\\
&\quad
+\,\sqrt{C_\alpha d}\,\|f_c\|_{\infty,\mathcal{H}_K}
\sum_{j=1}^i\tilde{a}^{\,i-j}
\bigl(h_{\rho,\mathcal{D}_k}(\boldsymbol{\mu}_{j-1,k})^{\alpha/2}
+h_{\rho,\mathcal{D}_k}(\bs{u}_{j-1,k}^{\rm GPara})^{\alpha/2}\bigr).
\label{error expr}
\end{align}
Whenever $a\neq\tilde{a}$, we can further develop the terms
$T_1$ and $T_2$ as follows. Term $T_1$ can be written as
\[
T_1
= \sigma^{\max,2}_{0,k}
\sum_{j=1}^i\tilde{a}^{\,i-j}a^{j-1}
= \sigma^{\max,2}_{0,k}\,\tilde{a}^{i-1}
\sum_{j=1}^i\Bigl(\frac{a}{\tilde{a}}\Bigr)^{j-1}
= \sigma^{\max,2}_{0,k}\,
\frac{\tilde{a}^i - a^i}{\tilde{a} - a}.
\]
while for $T_2$ we get
\[
T_2
=\sum_{m=1}^{i-1}b_{m-1,k}
\sum_{j=m+1}^{i}\tilde{a}^{\,i-j}a^{\,j-1-m}
=\frac{1}{\tilde{a}-a}
\sum_{m=1}^{i-1}b_{m-1,k}
\bigl(\tilde{a}^{\,i-m}-a^{\,i-m}\bigr).
\]
Substituting $T_1$ and $T_2$ into~\eqref{error expr} yields,
for all $i=1,\ldots,N$ and $k\in\mathbb{N}$,
\begin{align*}
\|\boldsymbol{\mu}_{i,k}-\bs{u}_{i,k}^{\rm GPara}\|
&\le
\frac{M\,d^{3/2}}{2(\tilde{a}-a)}
\left(
\sigma^{\max,2}_{0,k}
\bigl(\tilde{a}^i - a^i\bigr)
+\sum_{j=1}^{i-1}b_{j-1,k}
\bigl(\tilde{a}^{\,i-j}-a^{\,i-j}\bigr)
\right)
\\
&\quad
+\,\sqrt{C_\alpha d}\,\|f_c\|_{\infty,\mathcal{H}_K}
\sum_{j=1}^i\tilde{a}^{\,i-j}
\bigl(
h_{\rho,\mathcal{D}_k}(\boldsymbol{\mu}_{j-1,k})^{\alpha/2}
+h_{\rho,\mathcal{D}_k}(\bs{u}_{j-1,k}^{\rm GPara})^{\alpha/2}
\bigr),
\end{align*}
as required.

It remains to consider the case $a=\tilde a$. Starting again from~\eqref{error expr}, the terms $T_1$ and $T_2$ can then be evaluated directly as
\[
T_1
=
\sigma^{\max,2}_{0,k}
\sum_{j=1}^i a^{\,i-j}a^{j-1}
=
\sigma^{\max,2}_{0,k}\, i a^{i-1},
\]
and
\[
T_2
=
\sum_{m=1}^{i-1}b_{m-1,k}
\sum_{j=m+1}^{i}a^{\,i-j}a^{\,j-1-m}
=
\sum_{m=1}^{i-1}b_{m-1,k}(i-m)a^{\,i-m-1}.
\]
Substituting these expressions into~\eqref{error expr} gives
\begin{align*}
\|\boldsymbol{\mu}_{i,k}-\bs{u}_{i,k}^{\rm GPara}\|
&\le
\frac{M\,d^{3/2}}{2}
\left(
\sigma^{\max,2}_{0,k}\, i a^{i-1}
+\sum_{j=1}^{i-1}b_{j-1,k}(i-j)a^{\,i-j-1}
\right)
\\
&\quad
+\,\sqrt{C_\alpha d}\,\|f_c\|_{\infty,\mathcal{H}_K}
\sum_{j=1}^i a^{\,i-j}
\bigl(
h_{\rho,\mathcal{D}_k}(\boldsymbol{\mu}_{j-1,k})^{\alpha/2}
+h_{\rho,\mathcal{D}_k}(\bs{u}_{j-1,k}^{\rm GPara})^{\alpha/2}
\bigr).
\end{align*}
Equivalently, the equality case is obtained from the expression for $a\neq\tilde a$ by replacing each factor
\[
\frac{\tilde a^q-a^q}{\tilde a-a}
\]
with its limiting value $q a^{q-1}$, for $q\geq 1$.  $ \blacksquare$

\section{Robustness to \texorpdfstring{$n$}{n} and effect of \texorpdfstring{$\epsilon$}{epsilon} on the Prob-GParareal performance}
\label{app:n_analysis}
\begin{figure}[ht]
    \centering
    \includegraphics[width=1\linewidth]{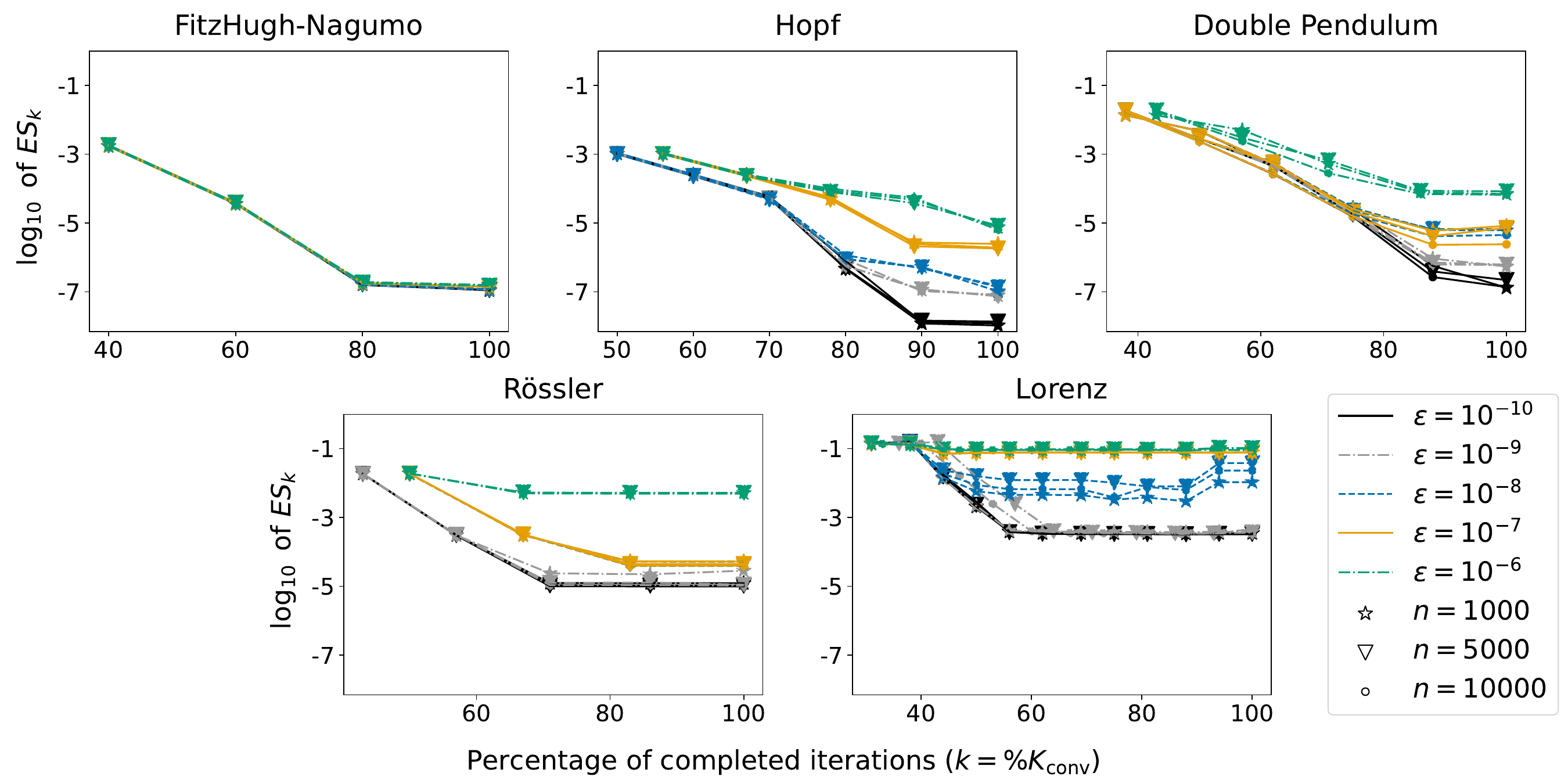}
    \caption{Impact of the tolerance level $\epsilon$ and the number of samples
    $n$ on the Prob-GParareal performance (${\rm ES}_k$, in $\log_{10}$) across systems. The x-axes show the percentage of completed iterations relative to $K_{\rm conv}$, the number of iterations to converge.
    All results are averaged over ten independent Prob-GParareal runs.}
    \label{fig:rob_ES}
\end{figure}
\begin{figure}[t]
    \centering    \includegraphics[width=1\linewidth]{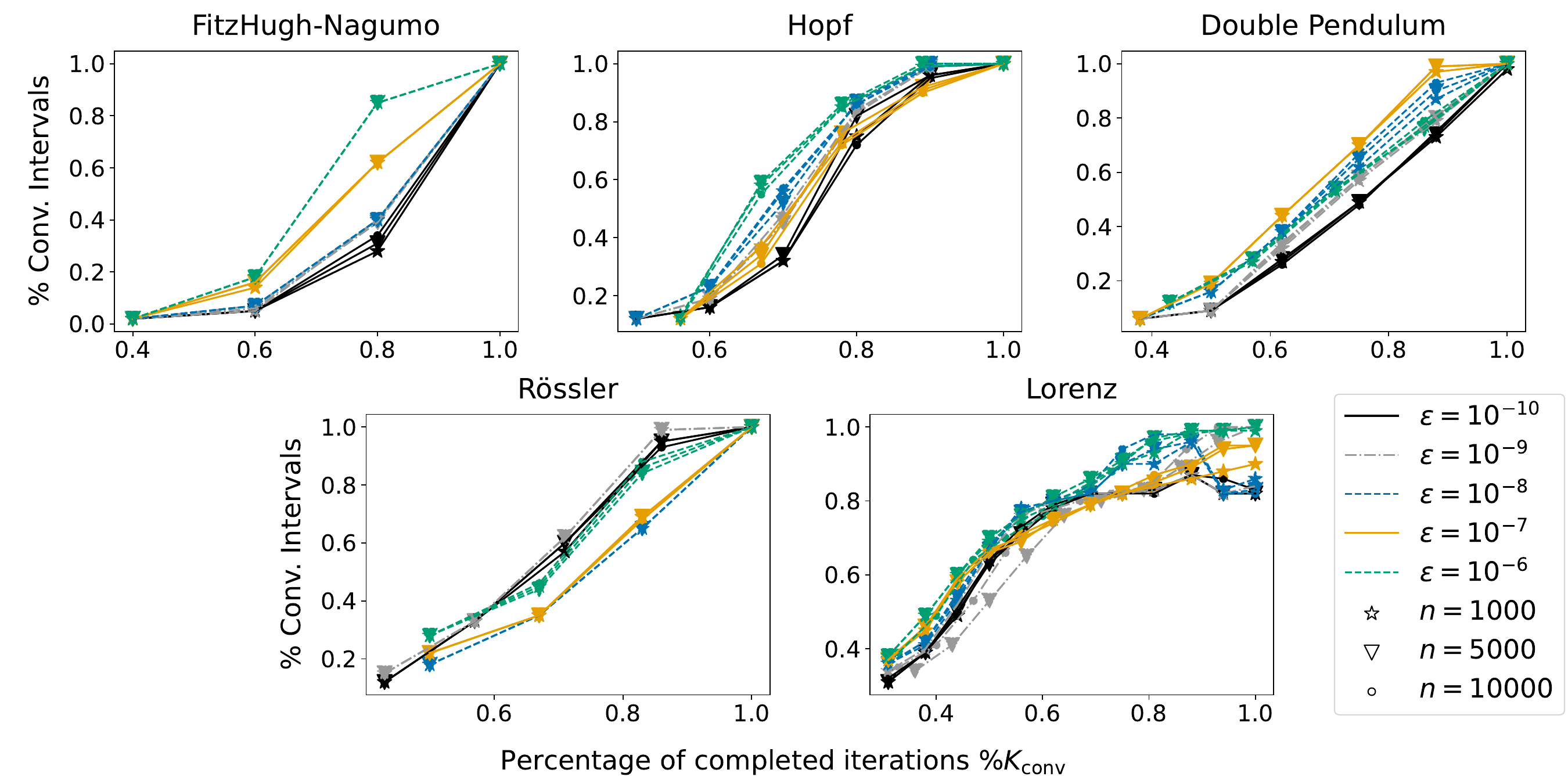}
    \caption{Impact of the tolerance level $\epsilon$ and the number of observations $n$ on the Prob-GParareal convergence. The x-axis shows the percentage of completed iterations relative to $K_{\rm conv}$, the iterations required to converge (generally unknown). The y-axis displays the percentage of converged intervals at iteration $k = \%K_{\rm conv}$. All results are averaged over ten independent Prob-GParareal runs.}
    \label{fig:rob_conv}
\end{figure}
\begin{figure}[t]
    \centering
    \includegraphics[width=1\linewidth]{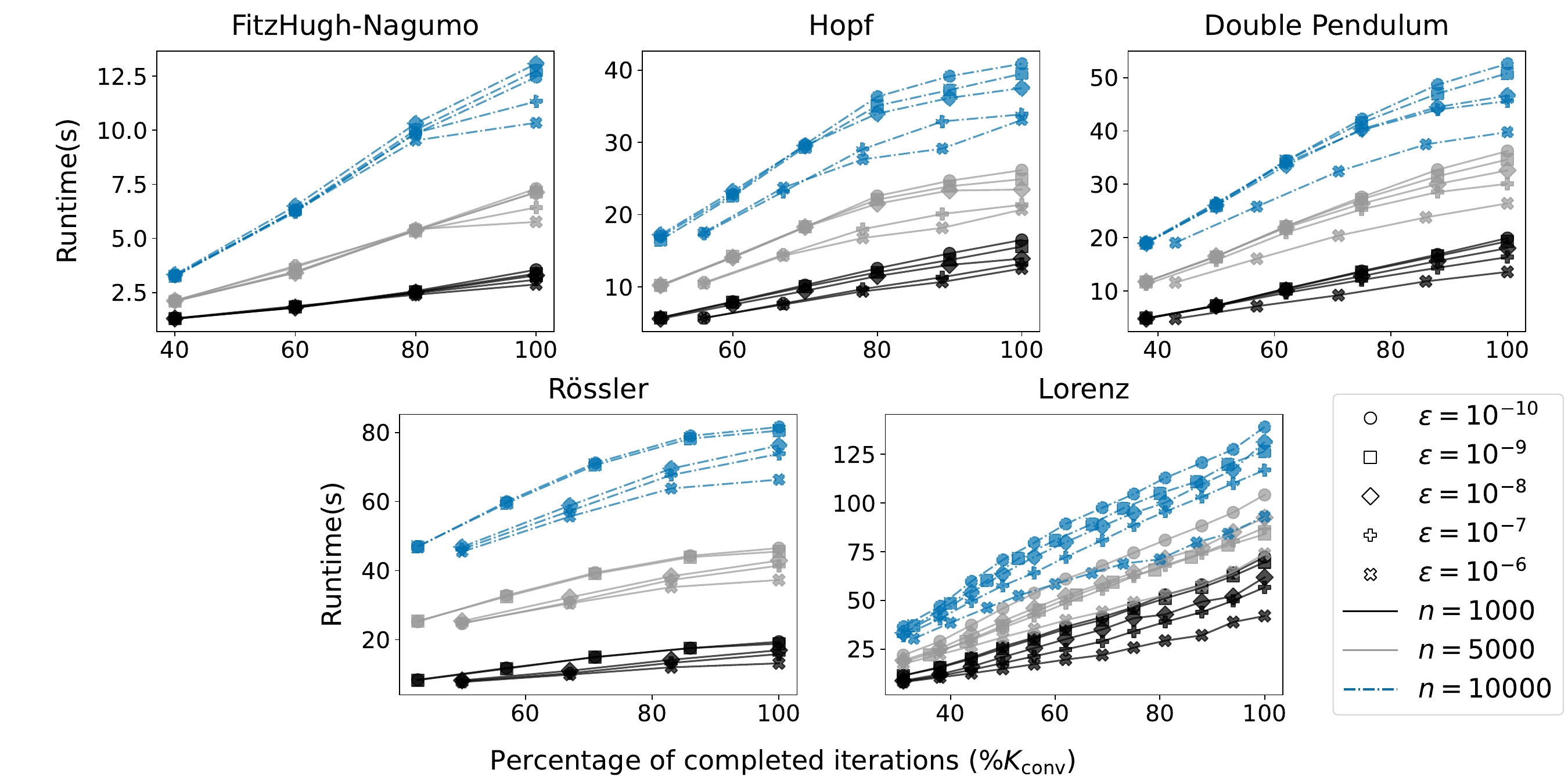}
    \caption{Impact of the tolerance level $\epsilon$ and the number of samples
    $n$ on the empirical Prob-GParareal runtime.  The x-axes show the percentage of completed iterations relative to $K_{\rm conv}$, the number of iterations to converge.
    All results are averaged over ten independent Prob-GParareal runs.}
    \label{fig:rob_runtime}

\end{figure}

In this section, we examine the impact of $n$ (the number of draws used to represent the probabilistic solution at each interval $i$) and $\epsilon$ (the tolerance threshold used to establish convergence in~\eqref{eq:pp_stp_rule}) on the accuracy (\Cref{fig:rob_ES}),  convergence (\Cref{fig:rob_conv}) and runtime (\Cref{fig:rob_runtime}) of the Prob-GParareal algorithm.
All results are averaged over ten independent runs using the same setup as in~\Cref{sec:early_stop}, with $\sigma_{\rm init} = 0$. Smaller thresholds generally lead to more accurate results but slightly increased runtimes due to additional iterations required for convergence (results not shown), with an expected more notable impact for more expensive evaluations of $\f$.

\Cref{fig:rob_ES} highlights the impact of $n$ and $\epsilon$ on the quality of the forecast. For clarity of presentation, we only report the energy score (ES), although similar patterns are observed for the other metrics. In this figure, the gray colors indicate different values of $\epsilon$, while the marker symbols represent $n$. The most notable difference occurs in $\epsilon$, with lower threshold values yielding better performance. The effect is system dependent, though: while FHN is unaffected by the choice of $\epsilon$, Lorenz, being chaotic, demonstrates poor performance with $\epsilon = 10^{-7}$, as it fails to capture the distribution's characteristics adequately. Generally, the $\epsilon$ values around $10^{-7}$ or $10^{-8}$ provide sufficient performance for most cases. However, the lower is $\epsilon$, the longer the algorithm may take to converge,
as shown in~\Cref{fig:rob_conv}, with an impact on the runtime as well, see \Cref{fig:rob_runtime}. Although this effect is not particularly pronounced here, it may become significant in real-world applications with costly fine solvers, where each additional iteration is expensive. The impact of $n$ on convergence is marginal, while it affects the algorithm runtime, as shown in~\Cref{fig:rob_runtime}.

\section{Fill distance analysis}
\label{app:fill_dist}
In this section, we present empirical evidence on the evolution of the local fill distance, which was introduced in~\Cref{sec:th_convergence} to quantify dataset quality and establish bounds on the Wasserstein distance and on the variance of the solution.

\begin{figure}[t]
    \centering    \includegraphics[width=1\linewidth]{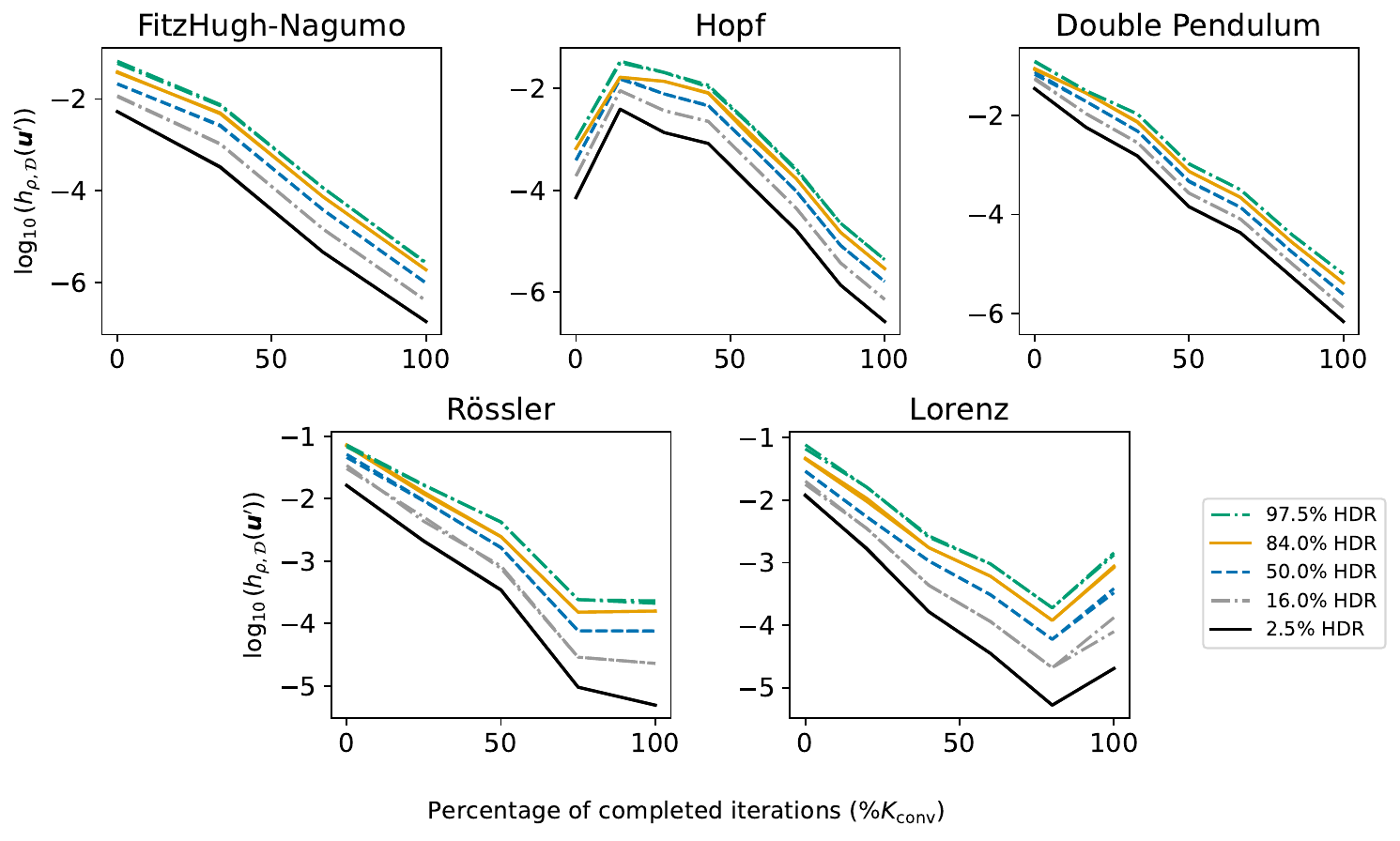}
    \caption{Evolution of the fill distance evaluated at representative observations placed on the boundary of the $\alpha$-HDR. Fill distance values are averaged over intervals $i$ and plotted as a function of the percentage of completed iterations for $\sigma_{\textrm{init}}=0$. }
    \label{fig:fill_dist}
\end{figure}

To compute the fill distance at iteration $k$, we evaluate $h_{\rho,\mathcal{D}_k}(\bs{u}')$ at every point $\bs{u}' = \bs{u}_{i,k}^{(j)}$ for $i=1,\ldots,N$ and $j=1,\ldots,n$. Recall that for a constant $\rho>0$, the local fill distance  at $\bs{u}' \in \mathcal{U}$ is defined by
$$
h_{\rho, \mathcal{D}}(\bs{u}'):=\sup _{\bs{u} \in B_{\rho}(\bs{u}')} \min _{\bs{u}_i \in \mathcal{D}}\left\|\bs{u}-\bs{u}_i\right\|,
$$
where $B_\rho(\bs{u}')\in \mathbb{R}^d$ denotes a ball of radius $\rho>0$ around $\bs{u}'$, with $\rho$ chosen as the smallest radius which ensures that the ball contains at least one observation. Rather than evaluating $h_{\rho,\mathcal{D}_k}$ at every $\bs{u}' \in \mathcal{U}$, we select representative points of the sample $\mathcal{U}_{i,k}$. In one-dimensional settings, quantiles provide a natural choice. For multivariate data, various generalizations of quantiles exist (see~\cite{cai2010multivariate} and references therein). Here, we use the highest density regions (HDR,~\citealt{hyndman1996computing}), defined as the smallest regions that contain $\alpha\%$ of the probability mass. For example, for multivariate Gaussians, HDRs correspond to hyperellipsoids. For each interval $i$ and $\alpha$-HDR, we take two observations from $\mathcal{U}_{i,k}$ that lie on the HDR boundary or are closest to it in the Euclidean sense, evaluate the fill distance at these points, and then average them across intervals $i$. By repeating this process across iterations, we observe how the fill distance changes for points located near or far from the bulk of the data. Recall that the dataset at iteration $k$ is updated with the observations $
\left(\overline{\bs{u}}_{i,k-1}, f_c(\overline{\bs{u}}_{i,k-1}) \right)$, $ i=1,\ldots,N$,
so we expect the points closer to the mean $\overline{\bs{u}}_{i,k-1}$ to be better represented,
with lower fill distances. This
aligns with the empirical results shown in~\Cref{fig:fill_dist}, where we observe an exponentially fast decrease in the local fill distance over the iterations.

\section{Impact of
early termination on the algorithm
runtime}
\label{app:estop_runtime}
\begin{figure}[t]
    \centering
    \includegraphics[width=0.7\linewidth]{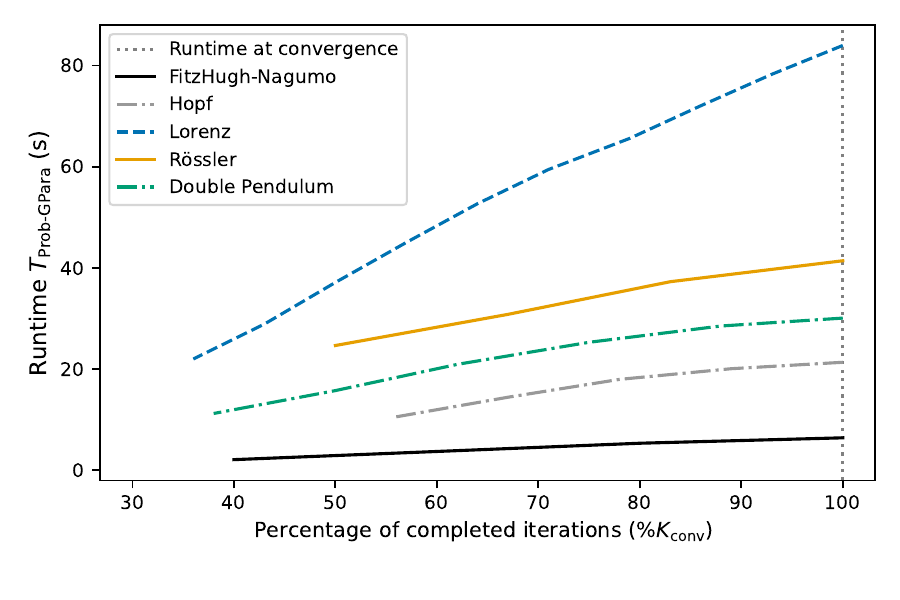}
    \caption{Effect of early termination of Prob-GParareal on its runtime, in seconds. All results are averaged over ten independent runs of the Prob-GParareal algorithm.}
    \label{fig:estop_runtime}
\end{figure}

In this section, we investigate the computational cost savings obtained by stopping Prob-GParareal prior to convergence. In~\Cref{fig:estop_runtime}, we report the algorithm runtime (in seconds) as a function of the percentage of completed iterations to convergence ($\%K_{\rm conv}$). Reducing the Prob-GParareal execution by even one iteration leads to one less parallel application of the fine solver $\f$, and fewer (nn)GP operations. Hence, the impact is system dependent, with more expensive $\f$ having the largest effect.

\section{Additional results on Prob-nnGParareal}\label{app:Prob-nnGParareal}
\begin{figure}[t]
    \centering
    \includegraphics[width=1\linewidth]{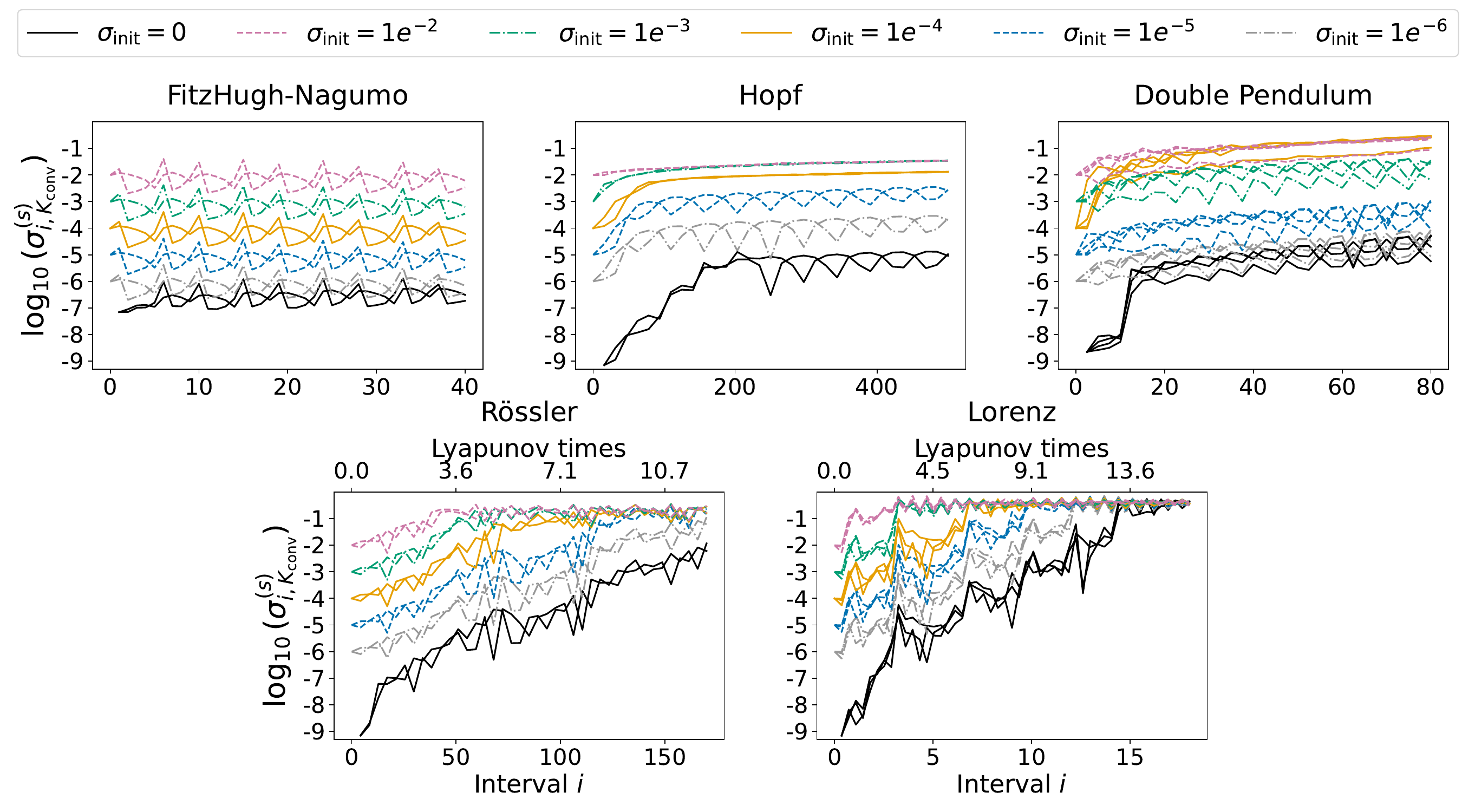}
    \caption{Impact of random initial conditions $\bs{U}_{0,0}$ on the coordinate-wise standard deviation of the converged Prob-nnGParareal solution $\mathcal{U}_{i,K_{\rm conv}}$ (with $m=15$ nearest neighbors) across intervals $i$ for different systems, using the same settings as~\Cref{fig:pp_initvar_std}, to favor a comparison with the Prob-GParareal results.}
    \label{fig:nnppara_initcond}
\end{figure}

In~\Cref{fig:nnppara_initcond}, we report the impact of random initial conditions $\bs{U}_{0,0,}$ on the coordinate-wise standard deviation of the converged Prob-nnGParareal solution $\mathcal{U}_{i,K_{\rm conv}}$ across intervals $i$ for different systems. We sample the initial state as $\bs{U}_{0,0} \;\sim\; \mathcal{N}\!\bigl(\bs{u}_{(0)},\,\sigma^2_{\textrm{init}} \mathbb{I}_d\bigr)$, where $\sigma_{\textrm{init}}\in \{0,e^{-2},e^{-3},e^{-4},e^{-5},e^{-6} \}$,
is the standard deviation of all coordinates. This setup mimics the experiments in~\Cref{fig:pp_initvar_std},~\Cref{sec:prob_init_cond} for Prob-GParareal, and is used here to assess the impact of switching from GPs to nnGPs. While in some cases the use of nnGP results in a slight increase in uncertainty over time compared to Prob-GParareal, the overall algorithm performance remains largely unaffected.

\section{Empirical results setup}
\label{app:setup}

In Table \ref{tab:simsetup}, we summarize the experimental setup used to obtain the results reported in the main text. The system definitions and parameters follow~\cite{nngparareal}, while all implementation details, including solver configurations and evaluation scripts, are available in the accompanying code repository. Unless otherwise stated, Prob-GParareal is run with a fixed number of samples \(n=5000\), and convergence is determined using the stopping criterion in~\eqref{eq:pp_stp_rule} with tolerance \(\epsilon = 10^{-8}\). All reported results are averaged over ten independent runs with different random seeds.
\begin{table}[th!]
\centering
{
\footnotesize
\begin{tabular}{lcccccc}
    \toprule
    System & $d$ & System Parameters & $[t_0,t_N]$ & $N$ & $K_{\rm stop}$ & $\bs{u}_{(0)}$ \\
    \midrule
    FHN & 2 & $a=b=0.2,\;c=3$ & $[0,40]$ & 40 & 9  & $(-1,1)$ \\
    R\"ossler & 3 & $a=b=0.2,\;c=5.7$ & $[0,170]$ & 40 & 14 & $(0,-6.78,0.02)$ \\
    Double Pend. & 4 & None & $[0,80]$ & 32 & 12 & $(-0.5,0,0,0)$ \\
    Hopf & 2 & None & $[-20,500]$ & 32 & 12 & $(0.1,0.1)$ \\
    Lorenz & 3 & $(\gamma_1,\gamma_2,\gamma_3)=(10,28,8/3)$ & $[0,18]$ & 50 & 16 & $(-15,-15,20)$ \\
    R\"ossler Ext. & 3 & $a=b=0.2,\;c=5.7$ & $[0,340]$ & 40 & 14 & $(0,-6.78,0.02)$ \\
    \bottomrule
\end{tabular}

\vspace{1em}

\begin{tabular}{lcccc}
    \toprule
    System & $\g$ & $\f$ & $\g$ steps & $\f$ steps \\
    \midrule
    FHN & RK2 & RK4 & 160 & $1.6e^5$ \\
    R\"ossler & RK1 & RK4 & $9e^4$ & $4.5e^7$ \\
    Double Pend. & RK1 & RK8 & 3104 & $2.17e^5$ \\
    Hopf & RK1 & RK8 & 2048 & $1.7e^5$ \\
    Lorenz & RK4 & RK4 & $3e^2$ & $2.25e^4$ \\
    R\"ossler Ext. & RK1 & RK4 & $9e^4$ & $4.5e^7$ \\
    \bottomrule
\end{tabular}

}
    \caption{Description of the simulation setup used to obtain the experimental results. The system equations and corresponding parameters are provided in~\cite{nngparareal}. Here, $d$ represents the system dimension, $[t_0, t_N]$ the evolution timespan $t \in \left[t_0, t_N\right]$, $N$ the number of intervals, $K_{\rm stop}$ the maximum number of iterations before the execution is stopped, $\bs{u}_{(0)}$ the deterministic initial condition, and $\f$ and $\g$ are the fine and coarse solvers, respectively, with `RK$p$' indicating a Runge-Kutta method of order $p$. Finally, `$\f$ steps' refers to the number of integration steps performed by the fine solver over $[t_0, t_N]$, where a higher count implies greater accuracy. The same applies for `$\g$ steps'. R\"ossler Ext. refers to the R\"ossler system over an extended solution timespan, namely $t \in \left[0,340\right]$, twice as much as the previously considered value of $t_N=170$. 
    Additional implementation details, including the stopping criterion, number of samples, and random seeds, are provided in the accompanying code repository.
    }
    \label{tab:simsetup}
\end{table}

\clearpage

\section{Empirical assessment of bound sharpness}\label{AppendixH}
\label{sec:bound_validation}
\begin{figure}[t]
\centering
\includegraphics[width=\linewidth]{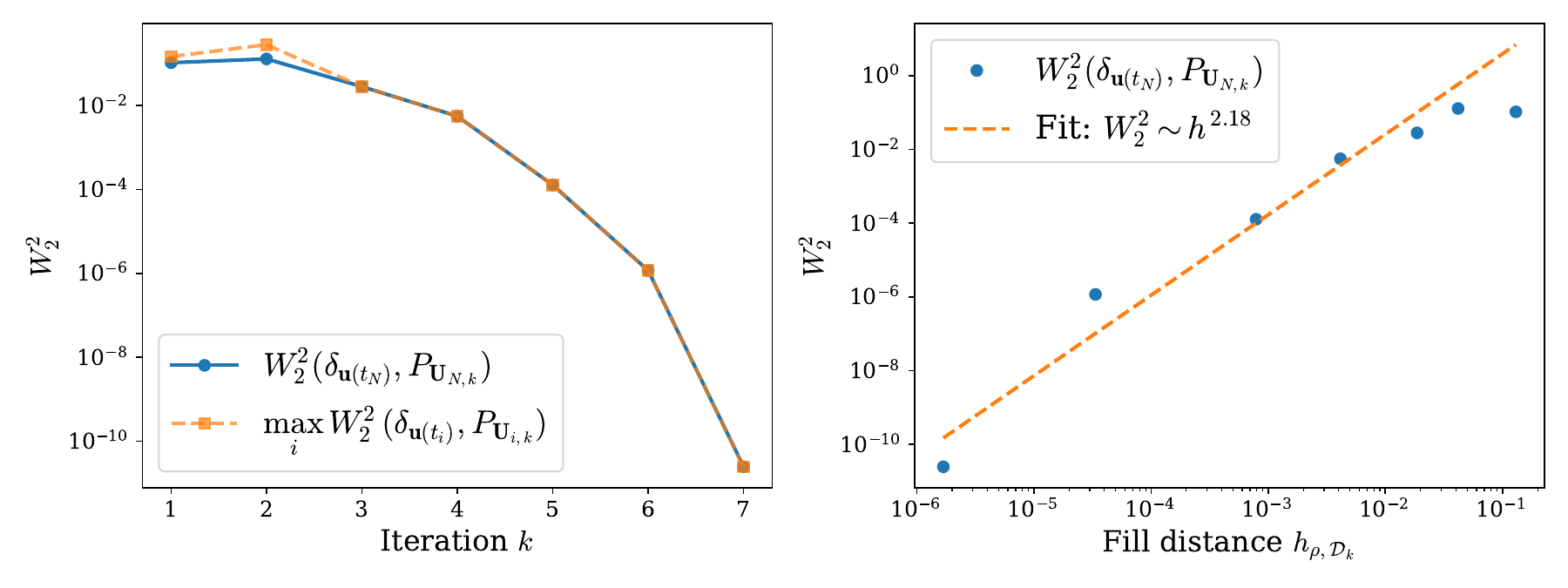}
\caption{
Left: decay of the empirical Wasserstein error
$W_2(\delta_{\bs{u}(t_N)}, P_{\bs{U}_{N,k}})^2$
as a function of the iteration number $k$, together with the maximum error over time
$\max_i W_2(\delta_{\bs{u}(t_i)}, P_{\bs{U}_{i,k}})^2$.
Right: empirical Wasserstein error at final time versus the local fill distance
$h_{\rho,\mathcal{D}_k}$ on a log-log scale. Each point corresponds to one iteration.
A power-law relation is observed with slope approximately equal to $2$, indicating that
$W_2^2 \sim h_{\rho,\mathcal{D}_k}^{\eta}$ with $\eta \approx 2$ over the range of iterations considered.
}
\label{fig:fill_distance_validation}
\end{figure}

In this section, we provide an empirical assessment of the bounds in \Cref{prop:convergence} by comparing the Wasserstein error $W_2 \bigl(\delta_{\bs{u}(t_i)}, P_{\bs{U}_{i,k}}\bigr)^2$ with the local fill distance $h_{\rho,\mathcal{D}_k}$ appearing in the coefficients~\(b_{l,k}\). We consider the Double Pendulum system and evaluate both quantities across Parareal iterations. We focus on the Wasserstein error at the final time \(t_N\), namely $W_2 \bigl(\delta_{\bs{u}(t_N)}, P_{\bs{U}_{N,k}}\bigr)^2$.
As shown in \Cref{fig:fill_distance_validation} (left), this quantity decays rapidly with the iteration number \(k\) and closely tracks the maximum error across time intervals \(i\). The reference solution \(\bs{u}(t_i)\) is obtained by propagating the initial condition $\bs{u}_{(0)}$ with the fine solver $\mathcal{F}$. For each iteration \(k\), the Prob-GParareal solution is represented by a particle approximation \(P_{\bs{U}_{i,k}}\) with \(n=10^4\) samples. The quantity
\[
W_2 \bigl(\delta_{\bs{u}(t_i)}, P_{\bs{U}_{i,k}}\bigr)^2
=
\mathbb{E}\!\left[\|\bs{u}(t_i) - \bs{U}_{i,k}\|^2\right]
\]
is estimated via Monte Carlo using the available particles. The local fill distance $h_{\rho,\mathcal{D}_k}$ is evaluated as described in \Cref{app:fill_dist}, averaging over quantiles.
To directly assess the dependence on $h_{\rho,\mathcal{D}_k}$ predicted by the bound, \Cref{fig:fill_distance_validation} (right) reports the empirical error $W_2 \bigl(\delta_{\bs{u}(t_N)}, P_{\bs{U}_{N,k}}\bigr)^2$ as a function of the fill distance on a log-log scale. A  power-law relationship of the form
\begin{equation*}
W_2 \bigl(\delta_{\bs{u}(t_N)}, P_{\bs{U}_{N,k}}\bigr)^2
\;\approx\;
C h_{\rho,\mathcal{D}_k}^{\eta},
\end{equation*}
with an estimated exponent \(\eta \approx 2\) is observed.
Since we employ a Gaussian kernel, the relevant theoretical regime is case~(iii) of \Cref{prop:convergence}. In this case, the bound is consistent with
arbitrarily algebraic $h_{\rho, \mathcal{D}_k}^\alpha$ for $\alpha>0$ and does not prescribe a specific exponent $\alpha$. The observed value $\eta\approx 2$ should therefore be
interpreted as an effective algebraic rate over the finite range of fill
distances explored in the experiment, rather than as a sharp theoretical
exponent. As expected for worst-case estimates, the theoretical constants are
conservative and the bound is not intended to be tight in magnitude.
Nevertheless, the experiment supports the interpretation that the convergence
of Prob-GParareal is governed by the decay of the local fill distance.

\section{Empirical validation of the Prob-GParareal computational complexity}
\label{app:cost_model_validation}
In this section, we provide an empirical validation of the Prob-(nn)GParareal computational complexity derived in \Cref{sec:comp_complx}. To do so, we extend the  runtime comparison in \Cref{fig:rob_runtime} by adding, for each system, the runtime predicted by the cost model for a representative configuration with \(\epsilon = 10^{-8}\) and \(n=5000\). The theoretical curves are obtained by evaluating the expression in \Cref{sec:comp_complx} at the corresponding values of \(N\), \(n\), and \(K_{\rm conv}\), up to proportionality constants estimated from the data.

\begin{figure}[t]
    \centering
    \includegraphics[width=\linewidth]{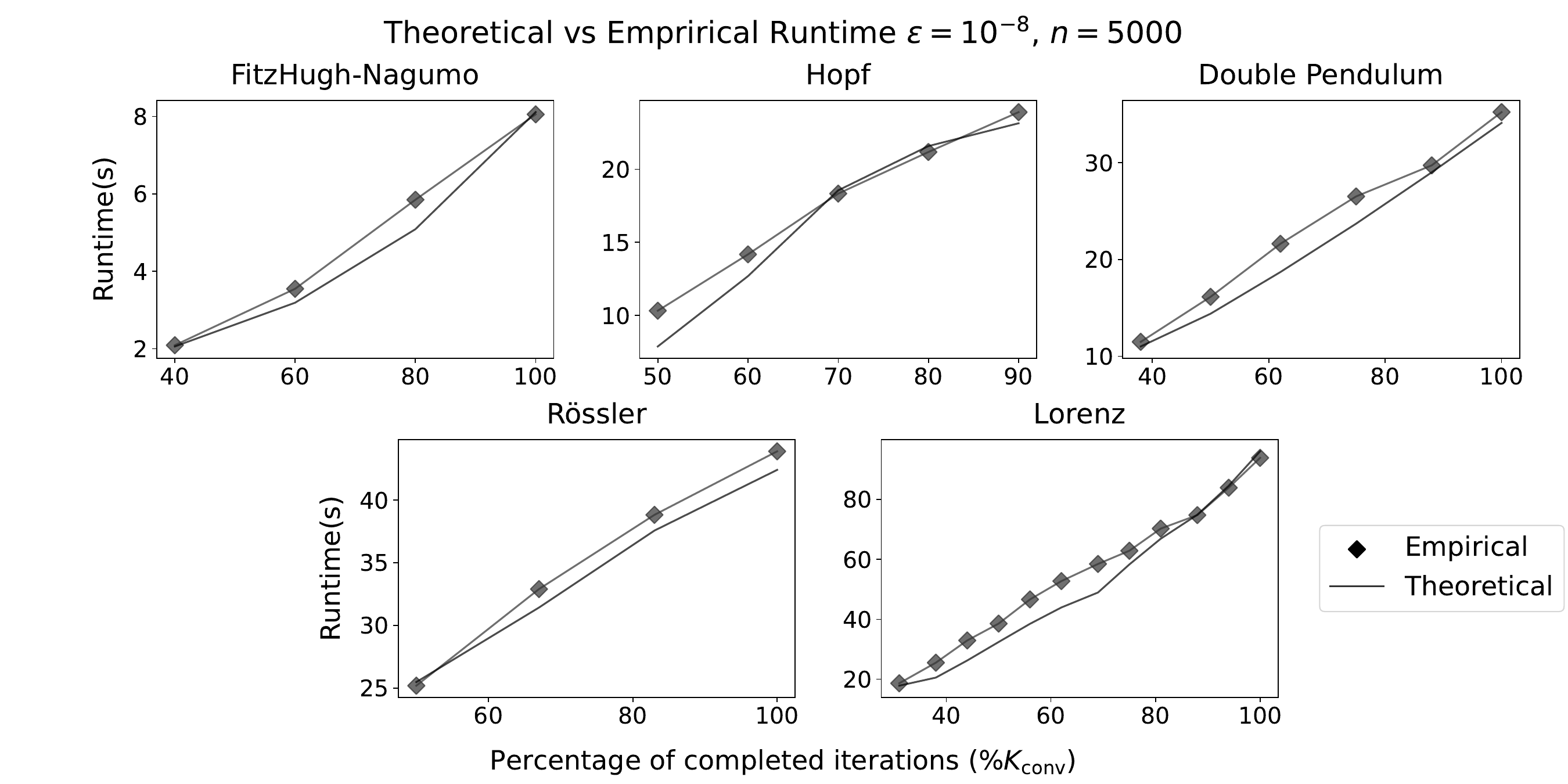}
    \caption{Comparison between empirical wall-clock time and runtime predicted by the cost model for Prob-GParareal across systems. The x-axis shows the percentage of completed iterations relative to \(K_{\rm conv}\). Empirical runtimes are averaged over ten independent runs, while the theoretical costs are obtained by evaluating the cost model from \Cref{sec:comp_complx} at the corresponding values of \(N\), \(n\), and \(K_{\rm conv}\).}    \label{fig:cost_model_validation}
\end{figure}

\Cref{fig:cost_model_validation} compares the resulting theoretical runtime with the empirical wall-clock time averaged over ten independent runs. Overall, the model captures the observed trends across systems and reproduces the increase in computational cost as the number of completed iterations grows. The small irregularities visible in some theoretical curves reflect the discrete structure of the cost model, in particular its dependence on the number of converged intervals and hence on the number of coarse solver evaluations, together with the nonlinear contribution of the model-cost term. While the agreement is not exact, which is expected since the analysis neglects serial overheads, the comparison provides empirical support for the derived theoretical cost model as a predictor of practical runtime behavior.

\begin{figure}[t]
    \centering
    \includegraphics[width=\linewidth]{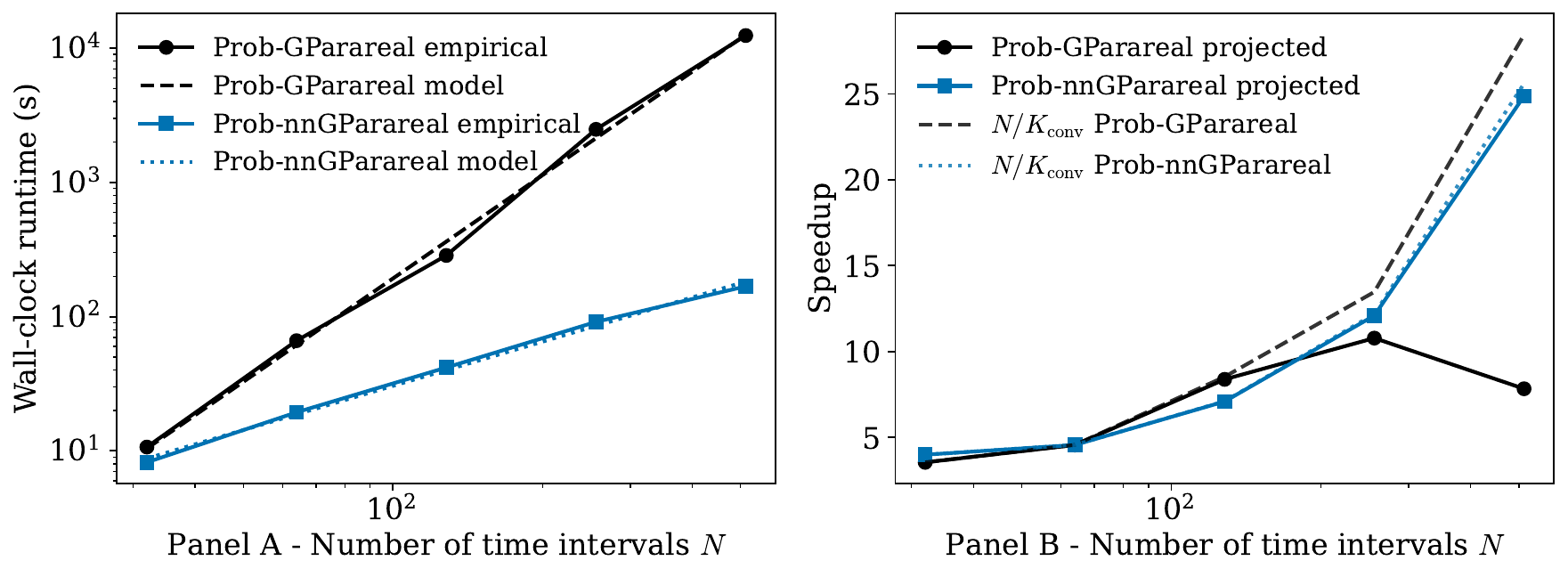}
    \caption{Empirical validation of the Prob-(nn)GParareal cost model with respect to the number of time intervals \(N\).
Panel A: empirical wall-clock runtime (solid lines with markers) together with the scaling predicted by the cost model from \Cref{sec:comp_complx} for Prob-GParareal (dashed lines) and Prob-nnGParareal (dotted lines).
Panel B: corresponding speedup as a function of \(N\). The nnGP variant closely follows the  ideal scaling \(N/K_{\rm conv}\), while the full GP variant exhibits reduced efficiency for large \(N\) due to the higher cost of the correction step.}    \label{fig:runtime_scaling_validation}
\end{figure}
We further validate the cost model by examining the dependence of the runtime on the number of time intervals \(N\). \Cref{fig:runtime_scaling_validation}, Panel A, shows the empirical wall-clock runtime for both Prob-GParareal and Prob-nnGParareal as \(N\) increases, together with the scaling predicted by the cost model. All experiments are conducted on a fixed problem setup (same ODE, solver configuration, tolerance \(\epsilon\), and number of samples \(n\)), varying only the number of time intervals \(N\); runtimes are measured as total wall-clock time. As before, the theoretical curves are obtained by evaluating the dominant terms in \Cref{sec:comp_complx}, up to proportionality constants estimated from the data. For Prob-GParareal, the model predicts a superlinear growth driven by the cubic dependence of the GP training cost on the dataset size, which is reflected in the empirical scaling. In contrast, Prob-nnGParareal exhibits a significantly milder dependence on \(N\), consistent with the reduced complexity induced by the nearest-neighbour approximation. Additionally, \Cref{fig:runtime_scaling_validation}, Panel B, reports the corresponding speedup as a function of \(N\). Prob-nnGParareal closely follows the  ideal scaling \(N/K_{\rm conv}\), maintaining a substantial speedup as \(N\) increases. In contrast, Prob-GParareal exhibits a progressive loss of efficiency for large \(N\), as the cubic GP training cost eventually dominates the runtime. Overall, these results confirm that the cost model captures both the scaling of the runtime and the resulting parallel efficiency, and highlight the importance of the nnGP approximation for achieving scalable performance as the number of time intervals increases.

\section{Comparison with \texorpdfstring{\citet{bosch2024parallel}}{Bosch et al. (2024)} parallel-in-time probabilistic ODE solver}
\label{app:bosch_comparison}

In this section, we provide a comparison with the parallel-in-time probabilistic ODE solver of \citet{bosch2024parallel}. Their method formulates probabilistic ODE solving as a time-parallel iterated extended Kalman smoother (IEKS), combining a global first-order linearization of the nonlinear observation model with a time-parallel Kalman filtering and smoothing step. For affine problems, the resulting Gaussian state estimation problem can be solved exactly, while for nonlinear vector fields the method relies on the IEKS approximation.

\begin{figure}[t]
    \centering
    \includegraphics[width=\linewidth]{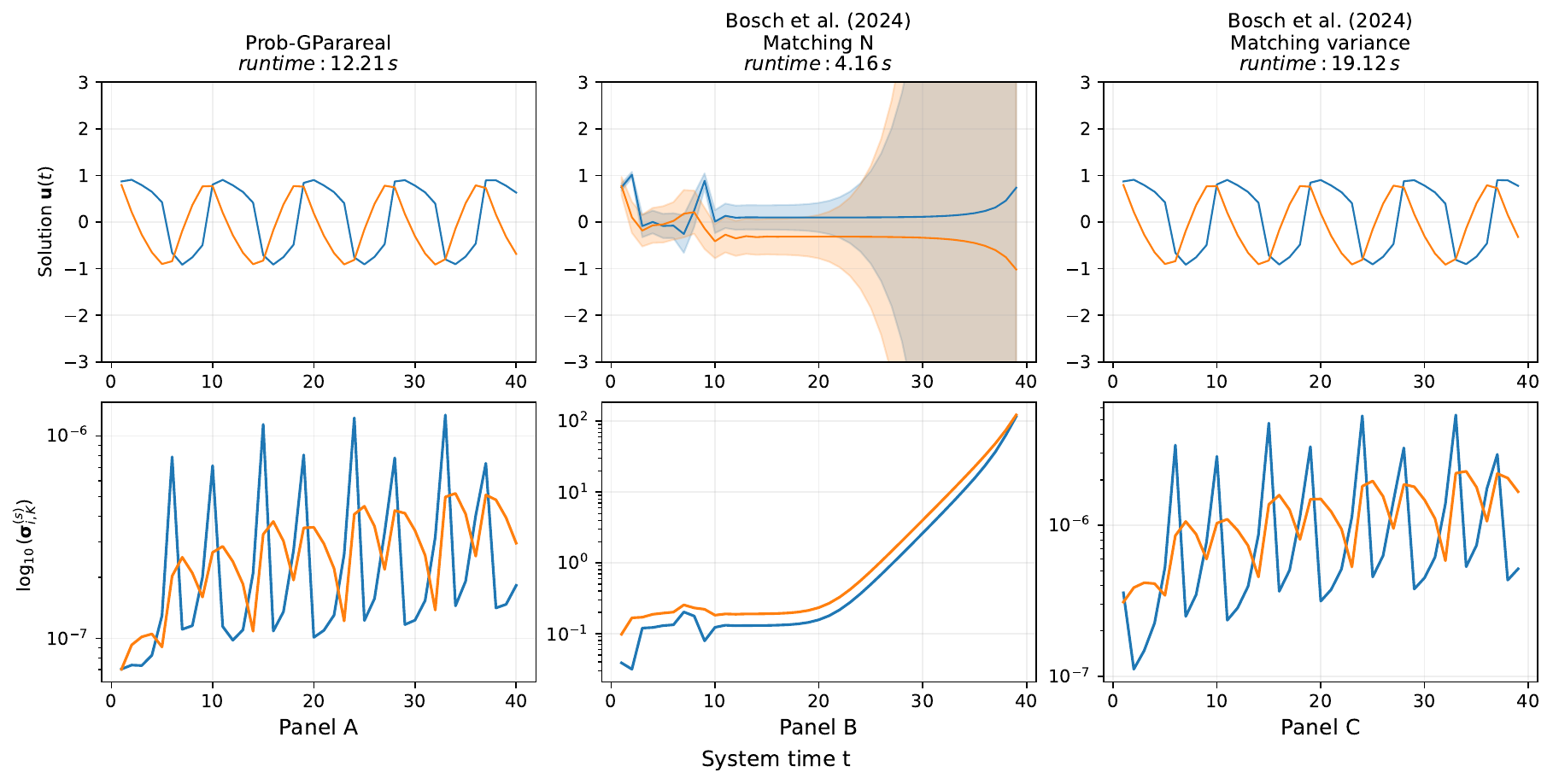}
    \caption{Comparison between Prob-GParareal and the parallel-in-time probabilistic ODE solver of \citet{bosch2024parallel} for the FitzHugh-Nagumo system. Columns (A), (B), and (C) correspond respectively to Prob-GParareal, the baseline run on a grid with matching number of time intervals, and the baseline run on a denser grid (5000 points). The first row shows the posterior mean (together with \(\pm 2\) standard deviation bands for \cite{bosch2024parallel}) for each coordinate of the system $(d=2)$, while the second row reports the marginal posterior standard deviations. The runtime of each method is shown in the column title.}
    \label{fig:bosch_comparison}
\end{figure}

We consider the FHN system and compare Prob-GParareal with the solver of \citet{bosch2024parallel} in two representative settings. In the first, the baseline is run on a time grid matching the number of time intervals used by Prob-GParareal, i.e., \(N=32\). In the second, it is run on a substantially denser time grid (5000 points), chosen so that the resulting posterior standard deviations of the two algorithms are of comparable magnitude in the displayed solution. For each method, shown in \Cref{fig:bosch_comparison}, we report the posterior mean, the associated uncertainty bands, and the observed runtime. The matching-\(N\) configuration (Panel B) is approximately three times faster than Prob-GParareal (Panel A), but yields substantially larger posterior uncertainty, with average marginal standard deviations several orders of magnitude higher. In contrast, when the baseline is run on a denser grid (Panel C), the resulting uncertainty becomes comparable to that of Prob-GParareal, at the cost of an increase in runtime by a factor of approximately \(1.5\). In terms of solution accuracy, both Prob-GParareal (Panel A) and the denser-grid baseline (Panel C) achieve similar performances, while the matching-\(N\) configuration (Panel B) exhibits a larger error.

Some cautionary words. This comparison should be interpreted as contextual rather than definitive. The two methods are based on different modeling assumptions: \citet{bosch2024parallel} place a prior directly on the continuous-time trajectory and perform inference through IEKS, whereas Prob-GParareal models the Parareal correction and propagates uncertainty through the coarse solver by sampling. Accordingly, the aim here is not to establish superiority of one method over the other, but to provide a baseline for context. As is well known in the PinT literature, comparisons across methods with different underlying principles are inherently challenging and therefore uncommon \citep{randnet_parareal}. Nonetheless, presenting both approaches on the same problem offers useful insight into the practical trade-offs in runtime, solution accuracy, and uncertainty quantification.

\bibliography{sample}

\end{document}